\documentclass[aps,prc,showpacs,twocolumn,nofootinbib,superscriptaddress,preprintnumbers]{revtex4-1}

\usepackage{graphicx}
\usepackage{dcolumn}
\usepackage{bm}
\usepackage{amsmath,amssymb}
\usepackage{color}
\usepackage{subfigure}

\newcommand{\bra}[1]{\langle \, #1 \, |}
\newcommand{\ket}[1]{| \, #1 \, \rangle}
\newcommand{\kket}[1]{\, #1 \, \rangle}
\newcommand{\rts}{\sqrt{s}}
\newcommand{\re}{\text{Re}}
\newcommand{\im}{\text{Im}}
\newcommand{\KN}{{\bar{K}N}}
\newcommand{\pS}{{\pi\Sigma}}
\newcommand{\pL}{{\pi\Lambda}}
\newcommand{\eq}{\text{equiv}}
\newcommand{\eff}{\text{eff}}
\newcommand{\delV}{\text{equiv,$\Delta V$}}

\newcommand{\eqdel}{\text{equiv},\delta}
\newcommand{\eqg}{\text{equiv},g}
\newcommand{\eqY}{\text{equiv,Y}}

\newcommand{\g}{\text{G}}
\newcommand{\del}{\partial}
\newcommand{\hP}{\hat{P}}
\newcommand{\hQ}{\hat{Q}}
\newcommand{\hH}{\hat{H}}
\newcommand{\hV}{\hat{V}}

\newcommand{\chiral}{SU(3)${}_{L}\times$SU(3)${}_{R}$ }

\begin{document}

\preprint{YITP-18-35}


\title{Construction of a local $\bar{K}N$-$\pi\Sigma$-$\pi\Lambda$ potential and composition of the $\Lambda(1405)$}


\author{Kenta~Miyahara}
\email[]{miyahara.kenta.62r@st.kyoto-u.ac.jp}
\affiliation{Department of Physics, Graduate School of Science, Kyoto University, Kyoto 606-8502, Japan}

\author{Tetsuo~Hyodo}
\affiliation{Yukawa Institute for Theoretical Physics, Kyoto University, Kyoto 606-8502, Japan}

\author{Wolfram~Weise}
\affiliation{Physik-Department, Technische Universit\"at M\"unchen, 85748 Garching, Germany}


\date{\today}

\begin{abstract}                           
A $\KN$-$\pS$-$\pL$ coupled-channel potential is constructed on the basis of chiral SU(3) dynamics. Several matching conditions are introduced to formulate an equivalent local potential that reproduces the coupled-channel scattering amplitudes resulting from chiral \chiral meson-baryon effective field theory. In contrast to a previously constructed effective single-channel $\KN$ potential, the explicit treatment of the $\pS$ channel yields a natural description of the low-mass pole as part of the two-pole structure of the $\Lambda(1405)$ resonance. The energy dependence of the potential can now be parametrized with a minimum of polynomial orders. To study the properties of the $\Lambda(1405)$ as a quantum-mechanical quasibound state, we derive the normalization condition of its wave function generated by the energy-dependent coupled-channel potential, using the Feshbach projection method. This framework provides an improved understanding of this system from the viewpoint of the compositeness of hadrons. With the properly normalized wave function, we demonstrate and confirm that the high-mass pole of the $\Lambda(1405)$ is dominated by the $\KN$ component.
\end{abstract}



\maketitle

\section{Introduction}  \label{sec:intro}  

An active research area of hadron physics is the search for exotic baryons, systems with one unit of baryon number that do not fit into the traditional scheme of ordinary three-quark states. A prominent candidate in this category has always been the $\Lambda(1405)$\,\cite{Patrignani:2016xqp,Hyodo:2011ur,Kamiya:2016jqc}. Notorious difficulties in understanding the $\Lambda(1405)$ within the frame of standard quark models\,\cite{Isgur:1978xj} have stimulated considerations toward a possibly more complex structure.

 A successful picture began to emerge many decades ago\,\cite{Dalitz:1959dn,Dalitz:1960du,Dalitz:1967fp} when the $\Lambda(1405)$ was treated as a $\KN$ quasibound state embedded in the $\pS$ continuum, using a coupled-channel approach combined with a vector meson exchange potential model. Later developments in Refs.\,\cite{Kaiser:1995eg,Oset:1998it,Oller:2000fj,Lutz:2001yb} established such a framework from the viewpoint of low-energy QCD as an effective meson-baryon field theory with spontaneously (and explicitly) broken chiral \chiral symmetry. Several examples of more recent theoretical evidence support this picture. In a lattice QCD simulation\,\cite{Menadue:2011pd,Hall:2014uca}, the strange quark contribution to the magnetic form factor of the $\Lambda(1405)$ is shown to vanish when approaching physical quark masses, in qualitative contrast to expectations from a simple $uds$ constituent-quark model. The spatial structure of the $\Lambda(1405)$ is studied by evaluating its form factors\,\cite{Sekihara:2008qk,Sekihara:2010uz}, utilizing finite-volume effects\,\cite{Sekihara:2012xp}, and analyzing the $\KN$ wave function\,\cite{Dote:2014ema,Miyahara:2015bya}. In all cases, the spatial size of the dominant $\KN$ component is found to be unusually large and of a magnitude indicating a hadronic molecular picture of the $\Lambda(1405)$. A further criterion comes from evaluating the compositeness of hadrons\,\cite{Hyodo:2011qc,Aceti:2012dd,Hyodo:2013nka,Sekihara:2014kya,Guo:2015daa,Kamiya:2015aea,Sekihara:2015gvw,Kamiya:2016oao,Sekihara:2016xnq,Tsuchida:2017gpb,Oller:2017alp}, a concept generalizing the wave function renormalization constant\,\cite{Weinberg:1965zz,Baru:2003qq}. Recent studies of the compositeness of the $\Lambda(1405)$ reveal once again that its structure is dominated by the $\KN$ component\,\cite{Sekihara:2012xp,Sekihara:2014kya,Garcia-Recio:2015jsa,Guo:2015daa,Kamiya:2015aea,Molina:2015uqp,Kamiya:2016oao,Tsuchida:2017gpb}.

The attractive $\KN$ interaction underlying the picture of the $\Lambda(1405)$ as a two-body quasibound state, with a nominal binding energy of 27 MeV, has motivated a multitude of studies concerning the possible existence of antikaon-nuclear quasibound systems ($\bar{K}$-nuclei)\,\cite{Nogami:1963xqa}. In particular, predictions of deeply bound states of some antikaonic nuclei were made in Refs.\,\cite{Akaishi:2002bg,Yamazaki:2002uh}, based on $g$-matrix calculations with optical potentials derived from phenomenological meson-baryon two-body interactions. These studies were followed by more elaborate and accurate few-body calculations, either using variational approaches\,\cite{Yamazaki:2007cs,Dote:2008in,Dote:2008hw,Barnea:2012qa,Ohnishi:2017uni} or solving Faddeev equations\,\cite{Shevchenko:2006xy,Shevchenko:2007ke,Ikeda:2007nz,Ikeda:2010tk,Revai:2014twa}. For the $\KN N$ prototype system, all of these computations agree qualitatively about the existence of a quasibound state with spin parity $J^{P}=0^{-}$ and isospin $I=1/2$ in the energy window between $\KN N$ and $\pS N$ thresholds. However, the predicted binding energy $B_{\KN N}$ and decay width $\Gamma_{\KN N}$ vary over a wide range depending on the type of potential used and its extrapolation below $\KN$ threshold. 

An important empirical condition at threshold is imposed by the measurement of the energy shift and width of the kaonic hydrogen $1S$ state, performed by the SIDDHARTA Collaboration\,\cite{Bazzi:2011zj,Bazzi:2012eq}. An accurate value of the complex $K^-p$ scattering length was deduced from these data through the improved Deser formula\,\cite{Meissner:2004jr}. Thanks to this strong constraint the uncertainties of theoretical subthreshold extrapolations of $\bar{K}$-nucleon amplitudes have been significantly reduced\,\cite{Ikeda:2011pi,Ikeda:2012au}. Energy-dependent interactions based on chiral SU(3) dynamics and subject to this constraint generally produce modest binding, $B_{\KN N} \sim $ 15--35 MeV, together with widths $\Gamma_{\KN N}\sim$ 30--50 MeV. On the other hand, purely phenomenological, energy-independent potentials tend to give much stronger binding. The still existing theoretical uncertainties of $\bar{K}$-nuclear calculations are primarily rooted in the subthreshold behavior of the $\KN$ two-body interaction. In particular, the energy region around the $\Lambda(1405)$ is governed by the dynamics of the $\pS$ and $\pL$ channels and their coupling to the subthreshold $\KN$ system. 

Experimental searches for a $\KN N$ bound state have been actively pursued in recent years\,\cite{Agnello:2005qj,Bendiscioli:2007zza,Yamazaki:2010mu,Ichikawa:2014ydh,Tokiyasu:2013mwa,Hashimoto:2014cri,Agakishiev:2014dha,Sada:2016nkb},
although a fully conclusive and consistent answer to the quest for a $\bar{K}$-nuclear bound state has not been reached. Recently the J-PARC~E15 experiment observed a peak structure in the $\Lambda p$ invariant mass distribution of the $K^- {}\,^3\text{He}\to\Lambda p\,n$ reaction, interpreted in terms of $B_{\KN N} = 15\pm 7\pm12$ MeV and $\Gamma_{\KN N}= 110\pm18\pm27$ MeV\,\cite{Sada:2016nkb}. In view of the strong broadening of the observed signal, questions remain, however, concerning, e.g., the role of final state interactions and related reaction mechanisms (see also discussion in Ref.\,\cite{Sekihara:2016vyd}). Further E15 investigations with improved statistics and including a measurement of the $``K^-pp"\rightarrow \pi\Sigma N$ decay channels are being performed to clarify the situation\,\cite{Sakuma:2017muh}.

Motivated by these recent developments, two of the present authors have constructed a quantitatively reliable $\KN$ single-channel potential constrained by the SIDDHARTA data\,\cite{Miyahara:2015bya}. This complex and energy-dependent effective potential is particularly suitable for applications in few-body calculations. It follows the strategy described in Ref.\,\cite{Hyodo:2007jq} where a realistic model based on chiral SU(3) dynamics has been developed that succeeds in reproducing the available $K^-p$ cross sections and the SIDDHARTA data with $\chi^2/\text{d.o.f}\simeq1$\,\cite{Ikeda:2011pi,Ikeda:2012au}. Chiral SU(3) dynamics\,\cite{Kaiser:1995eg,Oset:1998it,Oller:2000fj,Lutz:2001yb,Hyodo:2011ur} is a nonperturbative coupled-channel extension of chiral \chiral perturbation theory. It is designed to extrapolate $\KN$ amplitudes reliably into the subthreshold region not directly accessible by $\KN$ scattering experiments. A characteristic feature of chiral SU(3) dynamics in the $\KN$-$\pi\Sigma$ coupled channels with isospin $I=0$ is the appearance of two resonance poles corresponding to the $\Lambda(1405)$ in the scattering amplitude\,\cite{Jido:2003cb,Hyodo:2011ur,Kamiya:2016jqc,Patrignani:2016xqp}. In Ref.\,\cite{Miyahara:2015bya}, the equivalent local $\KN$ potential has been constructed to reproduce this two-pole structure of the amplitude in the complex energy plane. 

An instructive recent example for the application of this $\KN$ potential near threshold is the high-precision three-body calculation of the $1S$ energy shift and width of kaonic deuterium\,\cite{Hoshino:2017mty}. The same potential has also been applied in computations of $\bar{K}$ nuclei up to seven-body systems using an accurate few-body technique: the stochastic variational method with a correlated Gaussian basis\,\cite{Ohnishi:2017uni}. In the $\KN N$ system, a relatively small binding energy of 25--28 MeV is found. This binding energy increases as one adds more nucleons, and it reaches 70--80 MeV in the seven-body systems. However, in view of the fact that the $\pS$ threshold lies roughly 100 MeV below the $\KN$ threshold, an explicit treatment of the $\pS$ channel is certainly necessary for a more detailed analysis of such deeply bound states. In fact, even in the $\KN N$ system, the importance of treating the $\pS$ channel explicitly has been pointed out in Refs.\,\cite{Shevchenko:2007ke,Ikeda:2008ub}. 

The present work extends the previous construction of the $\KN$ single-channel  potential\,\cite{Hyodo:2007jq,Miyahara:2015bya} to a multichannel local potential with explicit treatment of the $\KN$-$\pS$-$\pL$ coupled channels. The framework is again chiral SU(3) dynamics with inclusion of the SIDDHARTA constraint\,\cite{Ikeda:2011pi,Ikeda:2012au}. Some issues inherent in coupled-channel scattering require special attention. For example, in contrast to the complex $\KN$ single-channel potential, with its imaginary part reflecting the open $\pS$ and $\pL$ channels, the coupled-channel potential is given in matrix form with real elements representing the interactions in the $\KN$, $\pS$, and $\pL$ channels and their couplings.

This newly constructed potential can then be used to analyze the structure of $\Lambda(1405)$ by evaluating the wave functions of the two-body eigenstates. However, the normalization of the wave function is not straightforward. The coupled-channel potential is energy dependent. For such potentials, it is known that the standard normalization condition and the rules for computing expectation values are not valid\,\cite{Lepage:1997cs,Lepage:1977gd,Caswell:1978mt,Sazdjian:1986qn,Formanek2004,2003quant.ph.12148Z,Benchikha:2013mba}. Furthermore, the $\Lambda(1405)$ is an unstable state, and therefore the boundary condition for the eigenstate inevitably makes the system non-Hermitian, even for a real potential. Hence, we are going to establish a method for treating a non-Hermitian system with energy-dependent potential, based on the Feshbach projection method. This formulation provides a natural interpretation of the wave-function normalization condition and the compositeness of the state under consideration\,\cite{Hyodo:2011qc,Aceti:2012dd,Hyodo:2013nka,Sekihara:2016xnq}.

This paper is organized as follows: In Sec.~\ref{sec:local_pot}, we develop the scheme for deriving the coupled-channel local potential equivalent to chiral SU(3) dynamics. The direct comparison with the interaction kernel of chiral SU(3) dynamics determines the strengths of the equivalent local potential. The energy dependence of the potential strengths is parametrized in each channel with a minimal set of polynomial orders. The explicit construction of the realistic $\KN$-$\pS$-$\pL$ potential follows in Sec.~\ref{sec:results_const_pot}, mainly focusing on the $I=0$ channel in which the $\Lambda(1405)$ appears. This matrix potential reproduces the original scattering amplitudes resulting from chiral SU(3) dynamics in the complex energy plane, including the poles relevant to the structure of the $\Lambda(1405)$. In Sec.~\ref{sec:L1405}, we derive the normalization condition for the wave functions of non-Hermitian systems resulting from an energy-dependent potential. With this formalism, the compositeness and the spatial structure of the $\Lambda(1405)$ are analyzed. The paper closes with a summary in Sec.~\ref{sec:summary}.

\section{Construction scheme for the coupled-channel potential} \label{sec:local_pot}  

This section introduces the procedures for constructing a local meson-baryon potential with explicit treatment of coupled channels, generalizing the single-channel case in Refs.\,\cite{Kaiser:1995eg,Hyodo:2007jq}. The aim is to generate a coupled-channel potential such that the solution of the Schr\"odinger equation equivalently reproduces the scattering amplitudes of chiral SU(3) dynamics which, in turn, reproduce all available $\KN$ scattering data. To this end, we derive the relation between the strengths of the coupled-channel potential and the interaction kernel in chiral SU(3) dynamics with several matching conditions. Finally, an explicit form of the parametrized potential is given for practical use. 

\subsection{Chiral SU(3) dynamics for meson-baryon scattering} \label{subsec:ch_u}

The analysis of the $\Lambda(1405)$ baryon resonance requires a nonperturbative calculation of the two-body scattering amplitude. In chiral SU(3) dynamics\,\cite{Kaiser:1995eg,Oset:1998it,Oller:2000fj,Lutz:2001yb}, the coupled-channel $T$ matrix $T_{ij}$, with the indices $i$ and $j$ denoting the relevant meson-baryon channels, is computed by resumming tree-level amplitudes, $V_{ij}$, derived from leading orders of chiral \chiral meson-baryon effective field theory (EFT). The strategy and framework is analogous to the chiral EFT treatment of the nuclear force\,\cite{Machleidt:2011zz,Epelbaum:2008ga}. The Bethe-Salpeter equation for the $S$-wave $T$ matrix at the center-of-mass energy $\rts$ is
\begin{align}
T_{ij}(\rts) &= V_{ij}(\rts) + \sum_k V_{ik}(\rts)\, G_k(\rts)\, T_{kj}(\rts) \notag \\
&= \left( \left[ \{ V(\rts)\}^{-1}-G(\rts) \right]^{-1} \right)_{ij},
\label{eq:BS_algebraic}
\end{align}
where $G_i(\rts)$ is the meson-baryon loop function in channel $i$. With the convention in Ref.\,\cite{Hyodo:2011ur}, the corresponding meson-baryon scattering amplitudes, $F_{ij}$, are given by
\begin{align}
F_{ij}(\rts) &=-\frac{\sqrt{M_iM_j}}{4\pi\rts}T_{ij}(\rts),
\end{align}
with the baryon masses $M_i$ and $M_j$ in channels $i$ and $j$, respectively. 

This framework has been applied to the $\Lambda(1405)$ system in many studies\,\cite{Oset:2001cn,Hyodo:2002pk,Hyodo:2003qa,Borasoy:2004kk,Borasoy:2005ie,Borasoy:2006sr,Ikeda:2011pi,Ikeda:2012au,Guo:2012vv,Mai:2012dt,Mai:2014xna}. In the present investigation, we adopt the model of Refs.\,\cite{Ikeda:2011pi,Ikeda:2012au}, with interaction kernels up to next-to-leading-order (NLO) terms in chiral perturbation theory. In this model, the free parameters are the low-energy constants of the NLO Lagrangian and the subtraction constants in the meson-baryon loop functions. These parameters are fixed by fits to the following experimental data of the low-energy $\KN$ system:
\begin{itemize}
\item[(1)] $K^-p$ elastic and inelastic cross sections\,\cite{Abrams:1965zz, Sakitt:1965kh, Kim:1965zzd, Csejthey-Barth:1965izu, Mast:1975pv, Bangerter:1980px, Ciborowski:1982et, Evans:1983hz},
\item[(2)] branching ratios at $K^-p$ threshold\,\cite{Ikeda:2011pi,Ikeda:2012au}, and
\item[(3)] the energy shift and width of kaonic hydrogen from the SIDDHARTA measurements\,\cite{Bazzi:2011zj,Bazzi:2012eq} and the deduced $K^- p$ scattering length\,\cite{Meissner:2004jr}.
\end{itemize}

The result of the $\chi^{2}$ fit is $\chi^2/\text{d.o.f}=0.96$. The amplitudes thus determined are the basis for the quantitative discussion in the following sections. The precise threshold constraint provided by the SIDDHARTA measurements limits the theoretical uncertainties of the subthreshold extrapolations of the amplitudes to about 20\%. 

For practical applications of the potential in few-body calculations, it is useful to reduce the number of coupled channels from the full model space in the original amplitude. This channel reduction can be exactly performed as shown in Refs.\,\cite{Hyodo:2007jq,Miyahara:2015bya}. Here we explicitly include the channels that are open and active below the $\KN$ threshold, namely $\KN$, $\pS$, and $\pL$. In the following, $V_{ij}$ stands for the effective interaction suitably constructed within the set of these active channels. Other channels with thresholds at higher energies ($\eta\Lambda$, $\eta\Sigma, K\Xi$) are ``integrated out." Their effects are absorbed through the additional term in the interaction kernel. We also use isospin-averaged masses to avoid the splitting of threshold energies for isospin multiplets. This turns out to be a well-justified approximation. Deviations appear only in the near-threshold region as shown in Ref.\,\cite{Miyahara:2015bya}. 

\subsection{Construction of the equivalent potential}  \label{subsec:equiv_pot}

Consider now the equivalent potential, $V^\eq_{ij}$, to be used in the Schr\"odinger equation,
\begin{align}
\left[ -\frac{\nabla^2}{2\mu_i}\delta_{ij} + \Delta M_i\, \delta_{ij} + V^\eq_{ij}(\bm{r},E) \right]\psi_{j}(\bm{r}) = E \psi_{i}(\bm{r})~.
\label{eq:Sch_psi}
\end{align}
The nonrelativistic two-body energy is
\begin{align}
 E=\sqrt{s}-m_{\bar{K}}-M_{N}=\sqrt{s}-M_i-m_i+\Delta M_i~, 
\end{align}
where $m_i$ and $M_i$ are the meson and baryon masses in channel $i$. 
The mass difference in that channel, measured from the reference energy at $\KN$ threshold, is
\begin{align}
\Delta M_i \equiv m_i+M_i - (m_{\bar{K}}+M_N)~.
\end{align}
The kinetic energy term involves the reduced mass, 
\begin{align}
\mu_i\equiv {m_{i}M_{i}\over m_{i}+M_{i}}~. 
\end{align}
The two-body wave function in channel $i=1,\dots,N$ is denoted by $\psi_i(\bm{r})$.

The scattering solution of Eq.\,\eqref{eq:Sch_psi} is subject to the boundary condition for incoming waves. By choosing the incident channel with index $j$, the asymptotic form of the wave function with angular momentum $l=0$ in channel $i$ for a given energy $E$ is related to the coupled-channels $S$ matrix, $S_{ij}$, as
\begin{align}
\left.r\psi_{i,j}^{l=0}(r)\right|_{r\to \infty}
& \propto
e^{-ik_{i}r}\delta_{ij}
-\sqrt{\frac{\mu_{i}k_{j}}{\mu_{j}k_{i}}}\,
S_{ij}(E)\,
e^{ik_{i}r} , \label{eq:asymptotic} \\
k_{i}& =\sqrt{2\mu_{i}(E-\Delta M_{i})} .
\end{align}
The detailed derivation of this wave function from the coupled-channel Schr\"odinger equation~\eqref{eq:Sch_psi} is given in Appendix\,\ref{app:wf}.
The $s$-wave scattering amplitude is then obtained as
\begin{align}
F_{ij}^{\eq}(E)
& =\frac{S_{ij}(E)-\delta_{ij}}{2i\sqrt{k_{i}k_{j}}} .
\end{align}

Our aim is to construct the equivalent potential $V^\eq_{ij}$ such that $F^\eq_{ij}(E)$ reproduces the original amplitudes $F_{ij}(\rts)$:
\begin{align}
F^{\eq}_{ij}(E) = F_{ij}(\rts)\,.
\label{eq:matching}
\end{align}
One expects that $V^\eq_{ij}$ is related to the original $V_{ij}$, but one should note that the scattering equations which use these potentials as input are different. For first orientation, consider identifying $V^\eq_{ij}$ with the Fourier transform of $V_{ij}$. As the interaction kernels in Refs.\,\cite{Ikeda:2011pi,Ikeda:2012au} are momentum independent, this Fourier transform gives a potential proportional to a $\delta$ function in coordinate space:
\begin{align}
V^{\eqdel}_{ij}(\bm{r},E) &= \delta^3(\bm{r})\, N_{ij}(\rts)\, V_{ij}(\rts)\,,
\label{eq:Veq_eff_deltar}
\end{align}
where $N_{ij}$ is a kinematic factor that accounts for the difference between scattering equations, determined in Born approximation for Eq.\,\eqref{eq:matching} as in Ref.\,\cite{Kaiser:1995eg}. In chiral SU(3) dynamics, the Born approximation of the amplitude $F_{ij}$ is
\begin{align}
F_{ij}(\rts) &=-\frac{\sqrt{M_iM_j}}{4\pi\rts}T_{ij}(\rts) \overset{\mathrm{Born}}{\approx} -\frac{\sqrt{M_iM_j}}{4\pi\rts}V_{ij}(\rts)\,.
\label{eq:FCh_Born}
\end{align}
On the other hand, the equivalent potential in Eq.\,\eqref{eq:Veq_eff_deltar} gives
\begin{align}
F^{\eqdel}_{ij}(E) &=-(2\pi)^2\sqrt{\mu_i\mu_j} \ 
T_{ij}^{\eqdel}(E) \notag \\
\overset{\mathrm{Born}}{\approx}& -\frac{\sqrt{\mu_i\mu_j}}{2\pi}\int d^3r \ e^{-i(\bm{k}_i-\bm{k}_j)\cdot\bm{r}} \ V^{\eqdel}_{ij}(\bm{r},E) \notag \\
&=-\frac{\sqrt{\mu_i\mu_j}}{2\pi} N_{ij}(\rts)\,V_{ij}(\rts)\,,
\label{eq:Fsch_Born}
\end{align}
where $\bm{k}_i$ is the meson momentum in channel $i$ in the center-of-mass frame. Comparing Eqs.~\eqref{eq:FCh_Born} and \eqref{eq:Fsch_Born}, $N_{ij}$ is determined as%
\footnote{In previous works\,\cite{Hyodo:2007jq,Miyahara:2015bya,Miyahara:2015uya}, a semirelativistic form of the flux factor $N_{ij}$ in  Ref.\,\cite{Kaiser:1995eg} has been used, whereas Eq.~\eqref{eq:flux_factor} should have been used in order to be consistent with the Schr\"odinger equation~\eqref{eq:Sch_psi}. The difference between these flux factors is absorbed by the adjustment term $\Delta V$ (see Sec.~\ref{subsec:parameterization}) so that the parametrized form of the potential is consistent with the (nonrelativistic) Schr\"odinger equation.}
\begin{align}
N_{ij}(\rts) &= \frac{1}{2}\sqrt{\frac{M_iM_j}{s\mu_i\mu_j}}\,.
\label{eq:flux_factor}
\end{align}

Although the Fourier transform of $V_{ij}(\rts)$ formally provides a $\delta$-type potential in coordinate space, this is not an exact correspondence because ultraviolet divergences are tamed by regularizing the loop functions $G_i$ in chiral SU(3) dynamics. The equivalent potential correspondingly involves finite-range distributions which replace $\delta^3(\bm{r})$ in Eq.\,(\ref{eq:Veq_eff_deltar}). The physical interpretation is as follows. Contact terms and subtraction constants associated with the regularization of loops in chiral SU(3) dynamics reflect physics at high-energy scales not treated explicitly in (low-energy) chiral EFT. The complementary coordinate-space potential does not resolve details of the corresponding short-distance physics, which are then encoded in conveniently  parametrized finite-range distributions. These distributions can be thought of as representing length scales characteristic of short-range effects such as vector meson exchange and finite-size meson-baryon vertex form factors. 

Expressing the spatial distributions of the potential by functions $g_{ij}(r)$, we rewrite the potential as
\begin{align}
V^{\eqg}_{ij}(\bm{r},E) &= g_{ij}(r)\, N_{ij}(\rts)\, V_{ij}(\rts)\,.
\label{eq:Vequiv_full}
\end{align}
The normalization of $g_{ij}(r)$ is determined as follows. First, we impose the condition that the diagonal parts of the amplitudes in Born approximation coincide with each other at each threshold:
\begin{align}
\left. F^{\eqg}_{ii}(E=\Delta M_i) \right|_{\text{Born}} = \left. F_{ii}(\rts=M_i+m_i) \right|_{\text{Born}}\,,
\label{eq:diag_Born_thre}
\end{align}
which implies for the diagonal component in channel $i$
\begin{align}
\int d^3r\,g_{ii}(r) &= 1\,.
\label{eq:g_diag}
\end{align}
The range parameters in the off-diagonal distributions $g_{ij}(r)$  with $i\neq j$ should be determined by the diagonal parts since the regularization in chiral SU(3) dynamics is performed for the diagonal loop function in each channel. Motivated by the separable form of the regulator function, we make the ansatz of a ``geometric mean'' of the diagonal parts:
\begin{align}
g_{ij}(r) = \left[ g_{ii}(r)g_{jj}(r) \right]^{1/2}\,.
\label{eq:g_offdiag}
\end{align}
In practice, Gaussian distributions are used which are convenient for few-body calculations. With normalization conditions specified by Eqs.~\eqref{eq:g_diag} and \eqref{eq:g_offdiag}, an explicit form of the spatial distribution is
\begin{align}
g_{ij}(r) &= \frac{e^{-r^2(1/2b_i^2 + 1/2b_j^2)}}{
(\pi b_{i}b_{j})^{3/2}
}\,,
\end{align}
where $b_i$ represents the potential range in the diagonal channel $i$.%
\footnote{An alternative prescription for the off-diagonal distribution is $g_{ij}(r)=e^{-r^2/b_{ij}^2}/(\pi^{3/2}b_{ij}^3)$ with  $b_{ij}=(b_{i}+b_{j})/2$ which satisfies $\int d^3r\ g_{ij}(r)=1$. We have checked that the results of the scattering amplitudes change only marginally with this prescription.}

The resulting equivalent potential becomes
\begin{align}
V^{\eqg}_{ij}(\bm{r},E) &= \frac{e^{-r^2(1/2b_i^2+1/2b_j^2)}}{2(\pi b_ib_j)^{3/2}}\sqrt{\frac{M_iM_j}{s\mu_i\mu_j}} \ V_{ij}(\rts)\,.
\label{eq:Vequiv_eff}
\end{align}
In the next step, we examine the original condition\,\eqref{eq:matching} with this potential. The condition for the Born approximation amplitudes, Eq.\,\eqref{eq:diag_Born_thre}, can be satisfied by any range parameters $b_{i}$ under the normalization\,\eqref{eq:g_diag}, whereas the nonperturbative scattering amplitude $F^{\eqg}_{ij}(E = \Delta M_i; \{b_i\})$, determined by the asymptotic behavior of the wave function, depends on the range parameters in all channels. It is thus required that the equivalent potential should reproduce the diagonal amplitudes $F_{ii}$ of the full chiral SU(3) dynamics calculation at the threshold energies of each channel $i$: 
\begin{align}
F^{\eqg}_{ii}(E = \Delta M_i; \{b_i\}) = F_{ii}(\rts=M_i+m_i)\, ,
\label{eq:requirement_b}
\end{align}
where a given component $F^{\eqg}_{ii}$ depends on the range parameters in all channels. In practice, the scattering amplitude at the threshold is complex except for the lowest energy channel, and hence Eq.\,\eqref{eq:requirement_b} provides $2N-1$ conditions in the $N$-channel problem for $N$ range parameters $b_{i}$. We determine the $\{b_i\}$ by minimizing the sum of the deviations, 
\begin{align}
\Delta F_g& \equiv \nonumber \\
 \sum_{i}& \left| F_{ii}(\rts=m_i+M_i)-F^{\eqg}_{ii}(E=\Delta M_i,\{b_i\}) \right|\,,
\label{eq:delF_thre_couple}
\end{align}
between the full chiral SU(3) amplitudes and those generated by the equivalent potential at the channel thresholds.

While Eq.\,\eqref{eq:requirement_b} guarantees that the requirement~\eqref{eq:matching} is satisfied near the thresholds, there can still be deviations distant from the thresholds, reflecting, for example, differences of the scattering equations used to calculate the respective amplitudes. To compensate for such deviations, we add an adjustment term, $\Delta V_{ij}$, to Eq.\,\eqref{eq:Vequiv_eff} at each energy as in Refs.\,\cite{Hyodo:2007jq,Miyahara:2015bya}:
\begin{align}
V^\delV_{ij}(\bm{r},E) =& \frac{e^{-r^2(1/2b_i^2+1/2b_j^2)}}{2(\pi b_ib_j)^{3/2}}\sqrt{\frac{M_iM_j}{s\mu_i\mu_j}} \notag \\
&\times\left[ V_{ij}(\rts) + \Delta V_{ij}(\rts) \right]\,.
\label{eq:Vequiv}
\end{align}
This adjustment permits to apply the equivalent potential over a wide energy range. The magnitude of $\Delta V_{ij}$ is expected to be small if the potential is properly constructed. The explicit $\Delta V_{ij}$ is chosen to reproduce the original amplitude at each energy $\rts = E + m_{\bar{K}} + M_N$. We then minimize the real quantity
\begin{align}
\Delta F(\rts) \equiv \sum_{i\leq j}\left|F_{ij}(\rts)-F_{ij}^\delV(E)\right|\,,
\label{eq:delF_delV}	
\end{align}
at each energy to determine $\Delta V_{ij}(\rts)$.

\subsection{Parametrization of the equivalent potential} \label{subsec:parameterization}

As shown in Eq.\,\eqref{eq:Vequiv}, the strength of the equivalent potential depends on the total energy $\sqrt{s}$. For practical convenience, we parametrize the energy dependence by a polynomial of the nonrelativistic energy $E=\sqrt{s}-M_N-m_{\bar{K}}$,
\begin{align}
V_{ij}^\eq(\bm{r},E) &= e^{-r^2(1/2b_{i}^2+1/2b_{j}^2)}
\sum_{\alpha=0}^{\alpha_{\rm max}} K_{\alpha,{ij}}\,
\left(\frac{E}{100 \text{ MeV}}\right)^\alpha\,, 
\label{eq:Vfit} 
\end{align}
where the degree of the polynomial, $\alpha_{\rm max}$, is to be determined as explained in the following. This parametrization also permits us to perform the analytic continuation of the amplitude into the complex energy plane, $\sqrt{s}\to z \in \mathbb{C}$, an important property in order to study the pole structure.\footnote{Note that an immediate analytic continuation of the potential in the form of Eq.\,\eqref{eq:Vequiv} is not possible because $\Delta V_{ij}(\rts)$ is not given as an analytic function of $\sqrt{s}$.}

As in Ref.\,\cite{Miyahara:2015bya}, the energy range for the parametrization of the potential strength in polynomial form is  optimized to reproduce $F_{\KN}$ in the complex energy plane. For a quantitative assessment of this optimization, we define the following dimensionless measure for the deviation of the amplitudes in the complex plane:
\begin{align}
\Delta f_{ij}(z) \equiv \left|\frac{F_{ij}(z)-F^{\rm equiv}_{ij}(z)}{F_{ij}(z)} \right|\,,
\label{eq:DelFz}
\end{align}
where $F^\text{equiv}_{ij}$ now denotes the scattering amplitudes calculated with the parametrized equivalent potentials\,\eqref{eq:Vfit}. We then define average deviations of the $\Delta f_{ij}(z)$ as follows:
\begin{align}
\Delta \bar{f}(z) &\equiv \frac{\Delta f_{\pS,\pS}(z)+\Delta f_{\pS,\KN}(z)+\Delta f_{\KN,\KN}(z)}{3} .
\end{align}
In Refs.\,\cite{Ikeda:2011pi,Ikeda:2012au}, theoretical uncertainties of the scattering amplitudes $F_{ij}(\rts)$ are estimated to be roughly 20\%. We take this uncertainty measure for guidance and regard the complex energy $z$ as being in an acceptable window if $\Delta \bar{f}(z) < 0.2$. The acceptable parameter range is then determined by maximizing the percentage measure,
\begin{align}
{\cal P} = \frac{\iint d(\re\,z)d(\im\,z)\,\Theta\left(0.2-\Delta \bar{f}(z)\right)}{\iint d(\re\,z)d(\im\,z)}\times 100~,
\label{eq:Pcomp}
\end{align}
where the integration region is set as
\begin{align}
1332\ {\rm MeV} \leq {\rm Re}\,z
\leq 1450\ {\rm MeV}~,  \\
-100\ {\rm MeV} \leq {\rm Im}\,z
\leq 50\ {\rm MeV}~, 
\end{align}
guided by Ref.\,\cite{Miyahara:2015bya}.\footnote{The lower boundary of $\re\,z$ is set at the $\pi\Sigma$ threshold. Although the coupled-channel potential is applicable beyond the $\pi\Sigma$ threshold, we use the same definition of ${\cal P}$ as in Ref.\,\cite{Miyahara:2015bya} in order to enable a direct comparison of the present results with the $\KN$ single-channel potential.}

\section{The $\KN$-$\pS$-$\pL$ local potential} \label{sec:results_const_pot}

In this section, we construct the strangeness $S = -1$ meson-baryon potential in the $\KN$-$\pS$-$\pi\Lambda$ coupled channels, following Sec.~\ref{sec:local_pot}. We start with the isospin $I=0$ potential in the coupled $\KN$-$\pS$ channels where the $\Lambda(1405)$ appears. The $I=1$ potential in the $\KN$-$\pS$-$\pL$ coupled channels is thereafter constructed in the same way. 

\subsection{$I=0$ potential }\label{subsec:I0}

\begin{figure}[tb]
\begin{center}
\centering
\subfigure{
\includegraphics[width=8 cm,bb=0 0 426 341]{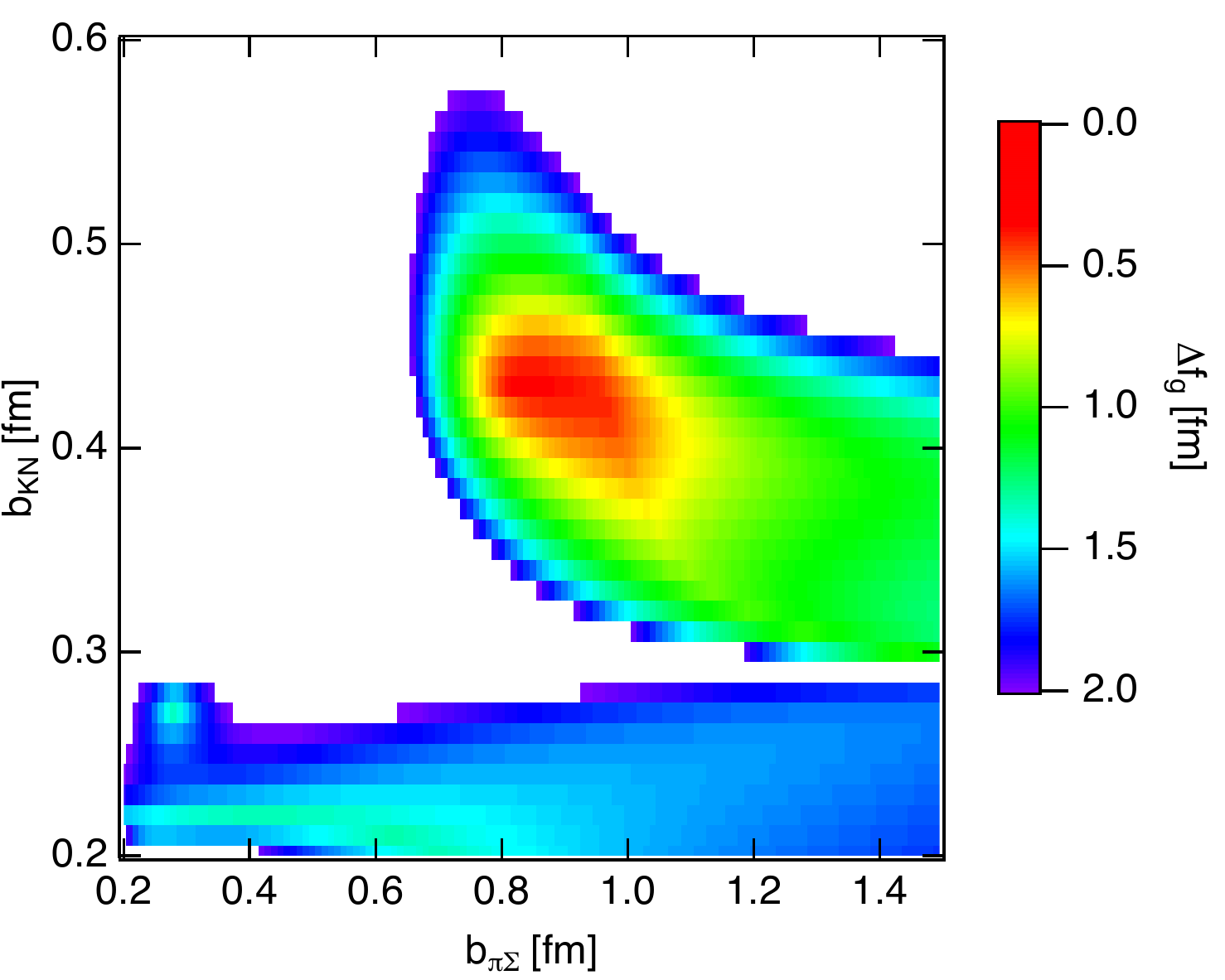}
}
\caption{Density plot of $\Delta F_g$ of Eq.~\eqref{eq:delF_thre_couple} on the plane of the range parameters $b_\pS$ and $b_\KN$ of the equivalent potential \eqref{eq:Vequiv_eff} in the isospin $I = 0$ channels. }
\label{fig:devF_IHW}  
\end{center}
\end{figure}

Consider now first the $I=0$ channel. As explained in Sec.\,\ref{subsec:equiv_pot}, the range parameters $b_i$ are determined by minimizing $\Delta F_g$ of Eq.\,\eqref{eq:delF_thre_couple} calculated with the potential $V^{\eq,g}_{ij}(\bm{r},E)$ of Eq.\,\eqref{eq:Vequiv_eff}. The behavior of $\Delta F_g$ under variations of the range parameters is shown in the density plot, Fig.\,\ref{fig:devF_IHW}. From this figure, it is seen that the range parameters can be uniquely determined within reasonably narrow margins. The optimized range parameters in the $I=0$ channels are found to be
\begin{align}
b_\pS^{I=0} = 0.80 \text{ fm},~~ \ b_\KN^{I=0} = 0.43 \text{ fm}.
\label{eq:bgau_I0}
\end{align}

%
%

\begin{figure*}[tb]
\subfigure{
\includegraphics[width=5.5cm,bb=0 0 846 594]{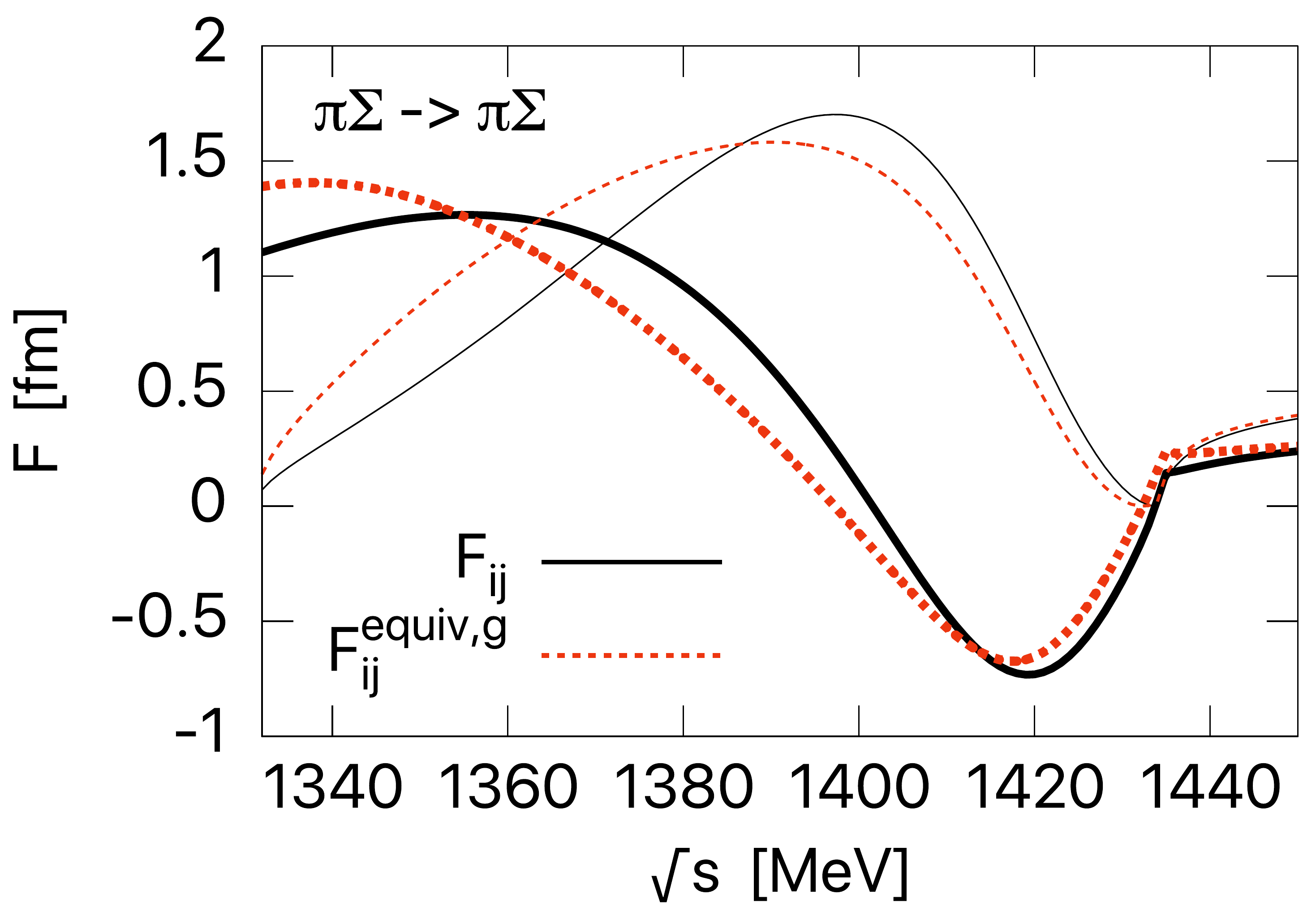}
}
\subfigure{
\includegraphics[width=5.5cm,bb=0 0 846 594]{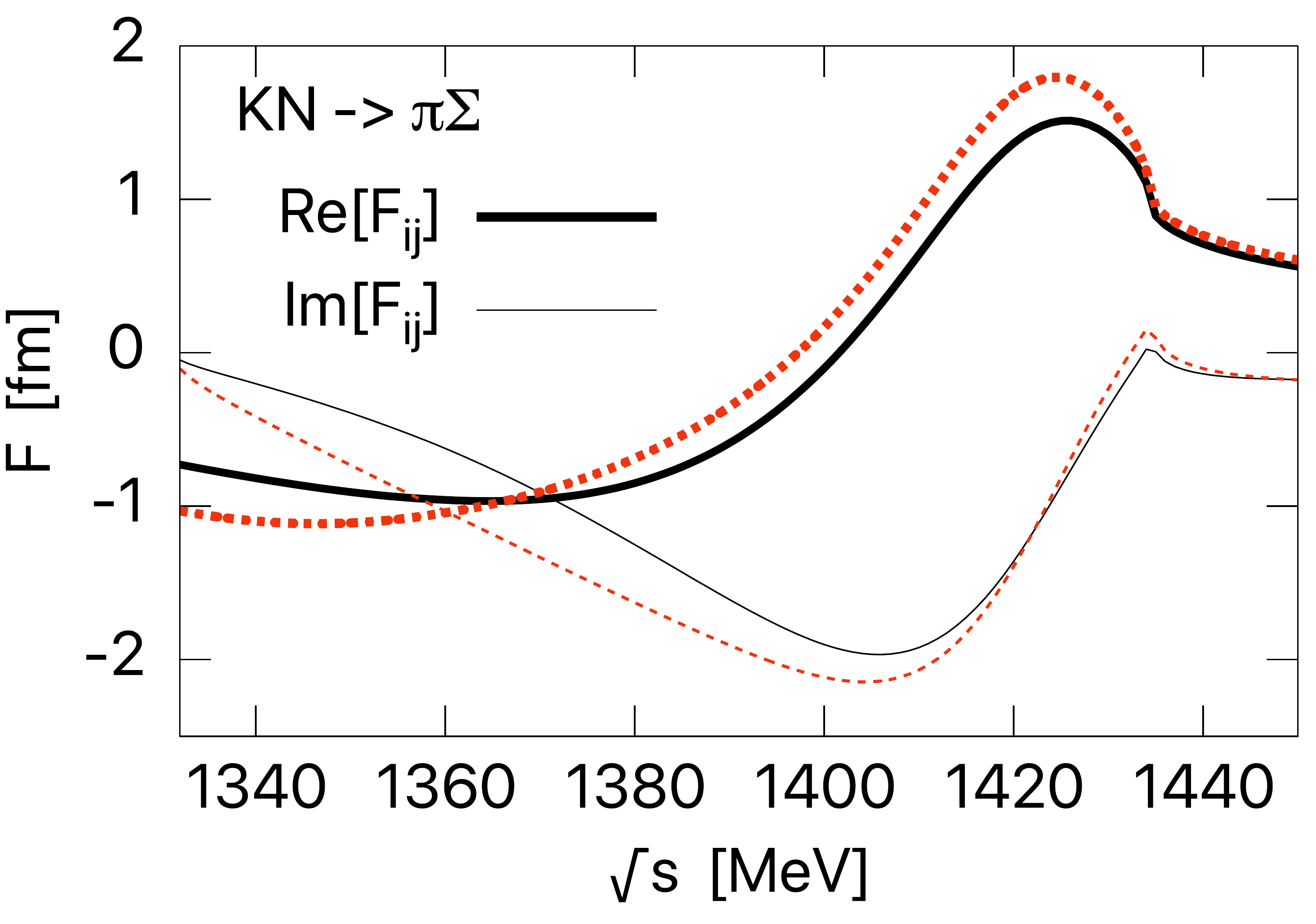}
}
\subfigure{
\includegraphics[width=5.5cm,bb=0 0 846 594]{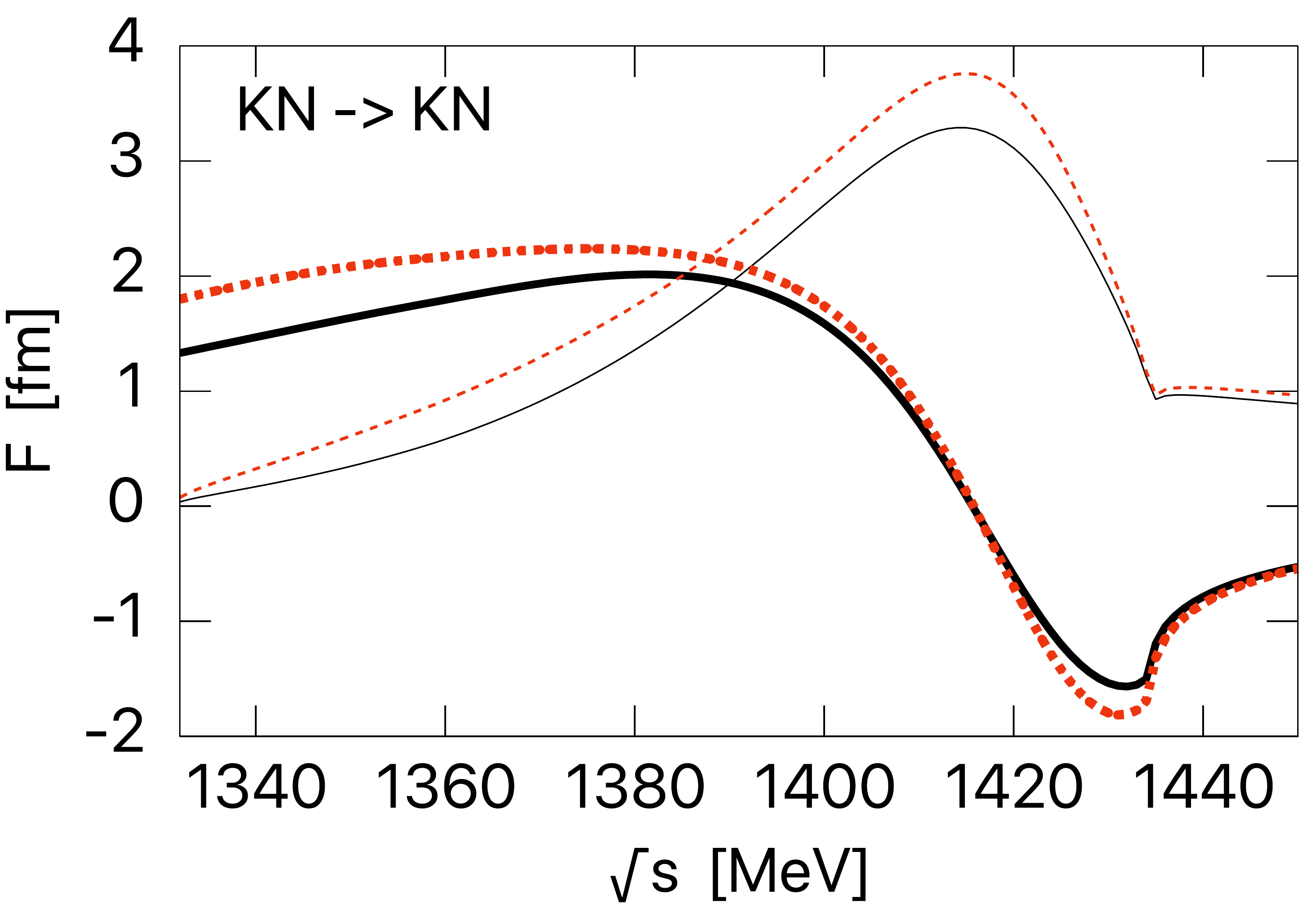}
}
\caption{Scattering amplitudes in the $I=0$ channel: $F_{ij}^{\eqg}$ (dotted lines) generated by the potential \eqref{eq:Vequiv_eff}, in comparison with the original amplitudes $F_{ij}$ (solid lines) from chiral SU(3) dynamics. The real (imaginary) parts are shown by the thick (thin) lines.}
\label{fig:F_Veff_IHW}  
\end{figure*}

In Fig.\,\ref{fig:F_Veff_IHW}, the scattering amplitude $F^{\eq,g}_{ij}(E)$ generated by the potential $V^{\eqg}_{ij}$ is compared with the original amplitude $F_{ij}(\rts)$. Even though the matching is done with minimal conditions, the potential $V^{\eqg}_{ij}$ reproduces the original amplitudes $F_{ij}$ remarkably well. This supports our prescription in Sec.\,\ref{subsec:equiv_pot} for constructing the potential. In Appendix\,\ref{app:range}, the validity and the physical interpretation of the range parameters $b_\pS$ and $b_\KN$ are discussed further. 

\begin{figure}[tbp]
\centering
\subfigure{
\includegraphics[width=8cm,bb=0 0 846 594]{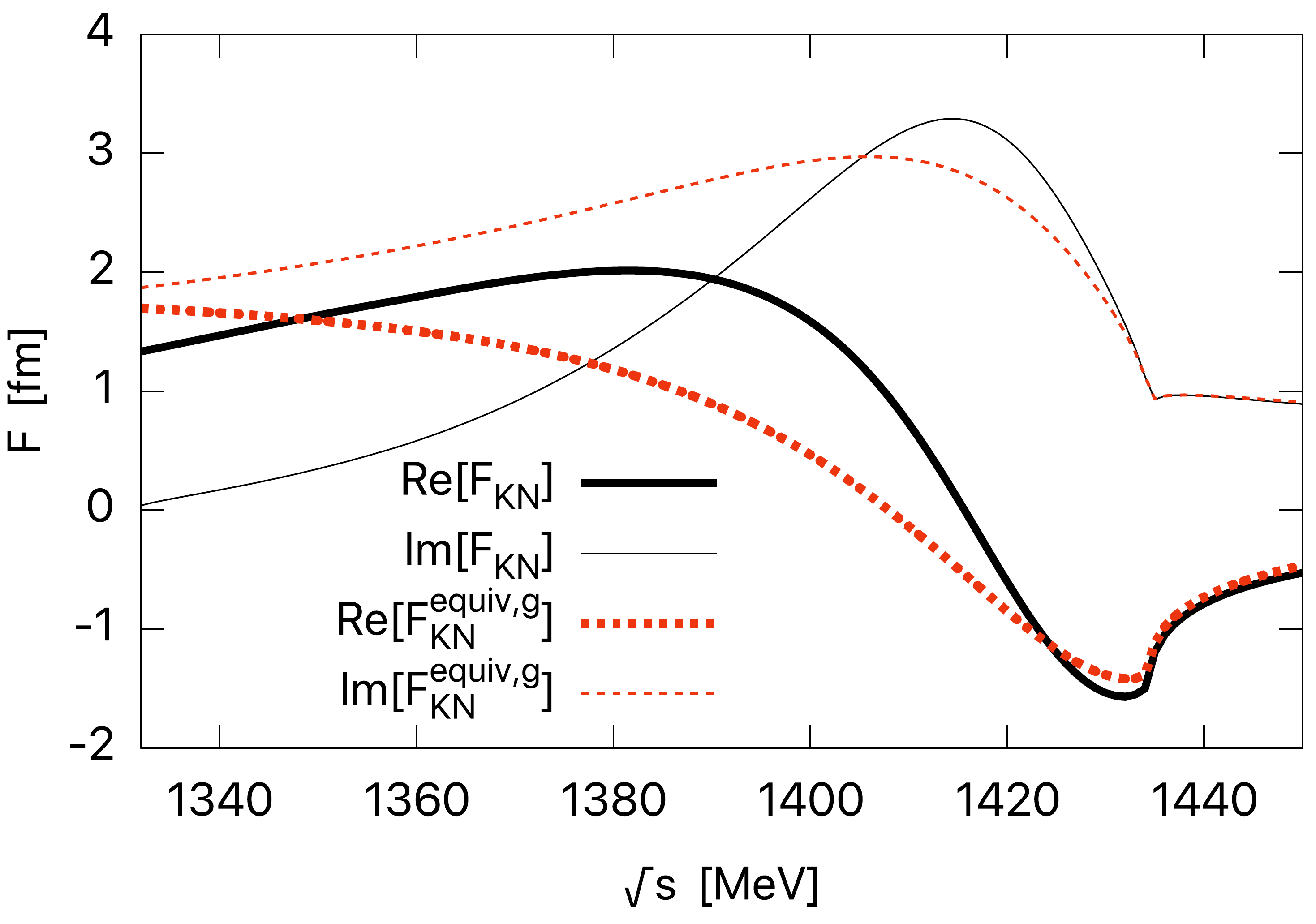}
}
\caption{Scattering amplitudes $F^{\eqg}_{\KN}$ produced by the single-channel $\KN$ potential (dotted lines) in comparison with the original chiral SU(3) dynamics amplitude $F_{\KN}$ (solid lines) in the $I=0$ channel. The real (imaginary) parts are shown by the thick (thin) lines.}
\label{fig:F_Veff_IHW_single}  
\end{figure}

It is instructive to compare the present result with the single-channel $\KN$ potential\,\cite{Miyahara:2015bya}, obtained by eliminating the $\pi\Sigma$ channel. In Fig.\,\ref{fig:F_Veff_IHW_single}, we show $F^{\eqg}$ in the $\KN$ single-channel case for comparison.\footnote{Here the normalization factor in Eq.~\eqref{eq:flux_factor} is adopted, in contrast to the previous work in Ref.\,\cite{Miyahara:2015bya}.}
While this single-channel potential is designed to reproduce $F_\KN$ near and above the $\KN$ threshold, the deviation becomes larger at lower energies, as seen by comparison with Fig.\,\ref{fig:F_Veff_IHW}. This indicates the importance of treating the $\pS$ channel explicitly in this lower energy region as one moves closer to the $\pS$ threshold. 

%
%
%
%

The analytic continuation of the scattering amplitude produced by the coupled-channel potential reveals two poles in the relevant energy region. The pole positions of $F^{\eqg}_{ij}$ are listed in Table~\ref{tab:pole_Fequiv_couple}. In comparison to the original chiral SU(3) amplitude, it turns out that $V^{\eqg}$ reproduces the position of the high-mass pole within a few MeV, while the position of the low-mass pole deviates from the original one beyond the theoretical uncertainties reported in Ref.\,\cite{Ikeda:2012au}. The ``accuracy measure" ${\cal P}$ in Eq.\,\eqref{eq:Pcomp} is relatively low, indicating that the amplitude in the complex plane is not reproduced very well. As the pole positions are essential for the detailed analysis of the $\Lambda(1405)$ and possible $\bar{K}$ nuclei, the potential needs to be further improved, and this is accomplished by adding the adjustment term $\Delta V_{ij}$.\footnote{We recall that the low-mass pole is not generated by the single-channel $\KN$ potential of Ref.\,\cite{Miyahara:2015bya} unless this adjustment term is added. The occurrence of the low-mass pole in the absence of the adjustment term $\Delta V_{ij}$ (although its position is not determined accurately) points once again to the importance of treating the $\pS$ channel explicitly in the coupled-channel potential.}

%
\begin{table*}[bt]
\begin{center}
\caption{
Results of computations using the equivalent coupled-channel potentials, $V^{\eqg}_{ij}$ of Eq.\,\eqref{eq:Vequiv_eff} and $V^\eq_{ij}$ of Eq.~\eqref{eq:Vfit}. Shown are, in this sequence, the polynomial order of $V^\eq_{ij}$, the energy range used for parameter fixing, the ``accuracy measure" given by the percentage ${\cal P}$, and the pole positions in the $I = 0$ scattering amplitude. The theoretical uncertainties of the original chiral SU(3) dynamics pole positions are taken from Ref.\,\cite{Ikeda:2012au}.
}
\begin{ruledtabular}
\begin{tabular}{lcccc}
Potential (polynomial in $E$) & 
Energy range [MeV] & ${\cal P}$ & High-mass pole [MeV] & Low-mass pole [MeV] \\ \hline
$V^{\eqg}$ &  & 32 &$1425-23i$ & $1336-69i$ \\
$V^\eq$ (first order) & 1403--1440 & 84 & $1423-26i$ & $1378-80i$ \\
$V^\eq$ (second order) & 1362--1511 & 99 & $1424-27i$ & $1380-81i$ 
\\ 
Original poles\,\cite{Ikeda:2012au} &  &  & $1424^{+3}_{-23}-26^{+3}_{-14}i$ & $1381^{+18}_{-6}-81^{+19}_{-8}i$
\end{tabular}
\label{tab:pole_Fequiv_couple}
\end{ruledtabular}
\end{center}
\end{table*}%
%

\begin{figure}[tb]
\centering
\subfigure{
\includegraphics[width=8cm,bb=0 0 846 594]{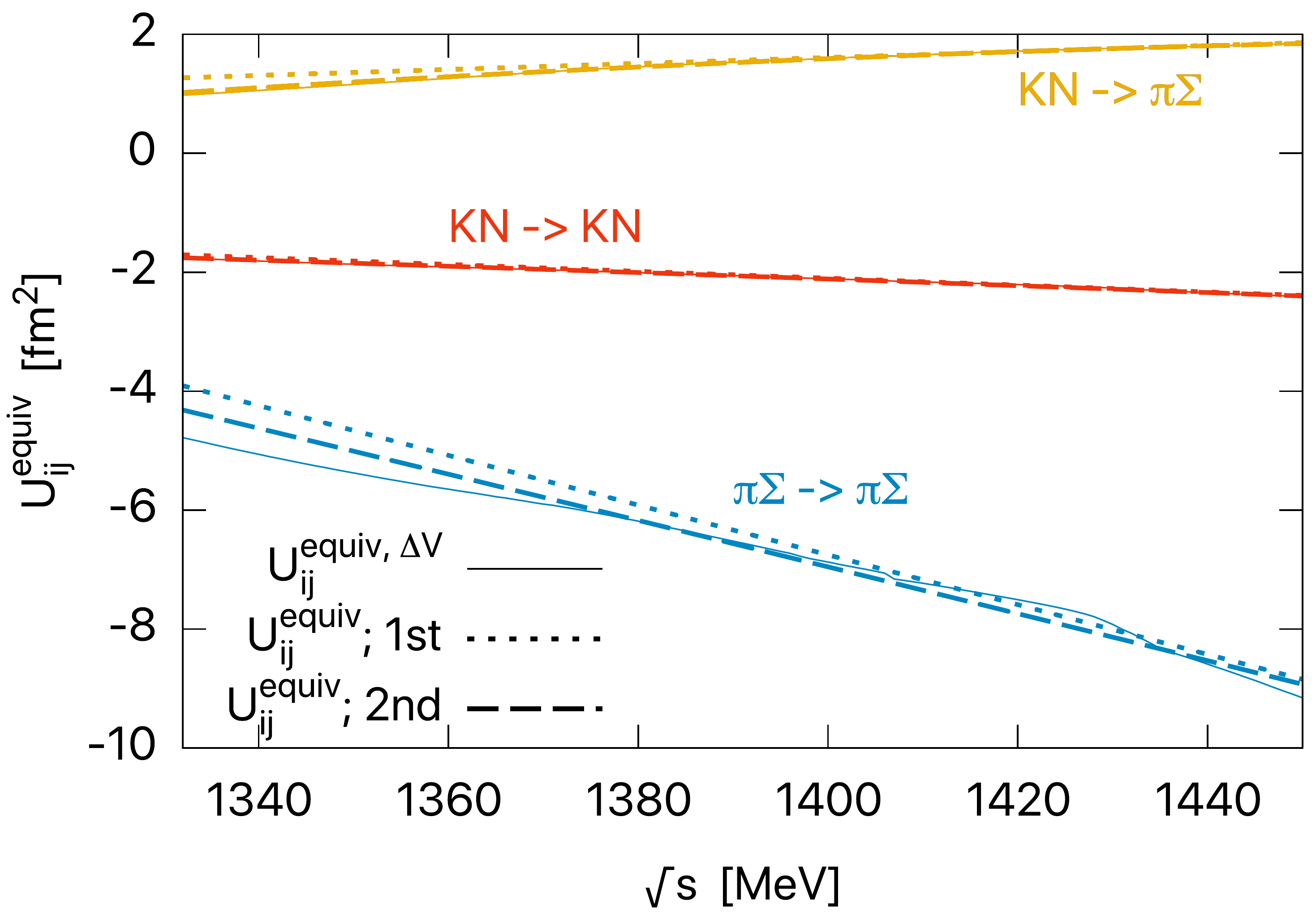}
}
\caption{Solid lines: volume integrals of equivalent potentials including the adjustment term $\Delta V_{ij}(\rts)$, $U^\delV_{ij}(\rts)$ of Eq.\,\eqref{eq:volint}, in the isospin $I=0$ channels ($\bar{K}N\rightarrow \bar{K}N, \pi\Sigma\rightarrow \pi\Sigma$, and $\bar{K}N\rightarrow\pi\Sigma$). Shown for comparison are the parametrizations $U^\eq_{ij}(\rts)$ according to Eq.\,\eqref{eq:Vfit} with first- and second-order polynomial expansions (dotted and dashed lines, respectively). The energy range for fitting the first-order (second-order) polynomial representations of $V^\eq_{ij}$ is 1403--1440 MeV (1362--1511 MeV).
}
\label{fig:V_fit_IHW}  
\end{figure}

As discussed in Sec.~\ref{subsec:equiv_pot}, we determine the adjustment term $\Delta V_{ij}(\rts)$
by minimizing $\Delta F$ of Eq.~\eqref{eq:delF_delV}.
A useful quantity for further demonstration is the volume integral of the potential in the diagonal and nondiagonal channels, 
\begin{align}
   U^{\delV}_{ij}(\rts)=\int d^{3}r\,V^{\delV}_{ij}(\bm{r},E) ~,
\label{eq:volint}
\end{align} 
shown in Fig.\,\ref{fig:V_fit_IHW} by solid lines. It is seen that the energy dependence of the $U_{ij}$ is almost linear in the region of interest. This energy dependence is primarily generated by the leading-order Tomozawa-Weinberg term in the chiral Lagrangian, plus contributions from the next-to-leading-order terms and from the elimination of channels with higher energy thresholds.\footnote{Note that actually the energy dependence is approximately linear in the nonrelativistic energy $E$. The denominator proportional to $1/\rts$ in Eq.~\eqref{eq:Vequiv} can be expanded as $(M_{N}+m_{\bar{K}})^{-1}\{1+\mathcal{O}[E/(M_{N}+m_{\bar{K}})]\}$ in the relevant energy region.} 
As an additional bonus, the nonanalytic behavior at the $\pi\Sigma$ threshold found in the single $\bar{K}N$ potentials\,\cite{Hyodo:2007jq,Miyahara:2015bya} does not appear when the $\pi\Sigma$ channel is treated explicitly, and so the potential is applicable in this entire energy region. 

Not surprisingly, the potential strengths seen in Fig.\,\ref{fig:V_fit_IHW} reflect qualitatively the trends already expected from the leading-order (LO, Tomozawa-Weinberg) terms of the chiral SU(3) meson-baryon Lagrangian. For example, the LO $I=0$ $\bar{K}N$ diagonal potential at threshold, when integrated over volume, gives $U_{\KN\rightarrow\KN} \simeq-3/(4f^2)\simeq -3.4$ fm$^2$, with the pseudoscalar meson decay constant $f\simeq 92$ MeV. The corresponding LO  $I=0$ $\pi\Sigma$ diagonal potential is slightly stronger and gives $U_{\pS\rightarrow\pS} \simeq - 1/f^2\simeq -4.5$ fm$^2$. Next-to-leading-order terms are important, of course, and contribute to the more detailed quantitative behavior of the $U_{ij}$.

The smooth energy dependence of $U^{\delV}_{ij}(\rts)$ in Fig.\,\ref{fig:V_fit_IHW} justifies terminating the polynomial expansion \eqref{eq:Vfit} of the parametrized potential $V^{\eq}_{ij}$ at low orders (i.e., first or second order, $\alpha_{\rm max}=1,2$). The energy range of validity for this parametrization is determined by maximizing ${\cal P}$ as discussed in Sec.\,\ref{subsec:equiv_pot}. The lower boundary of this energy window is varied in steps of one MeV upward from 1200 MeV, while the upper boundary is chosen below 1660 MeV in order to avoid the nonanalytic behavior at the threshold of the (eliminated) $\eta\Lambda$ channel. By this procedure, the energy window of optimized fitting is determined as 1403--1440 MeV (1362--1511 MeV) for the first-order (second-order) polynomial. The resulting polynomial coefficients, $K_{\alpha,ij}$, are summarized in Table\,\ref{tab:K_couple_I0}. They display excellent convergence in the following sense: The $K_{2}$ coefficients are an order of magnitude smaller than $K_{0}$ and $K_{1}$. The latter do not change significantly when including the $K_{2}$ terms. This indicates the dominance of the linear energy dependence and justifies the truncation of the expansion at the second order.  The volume integral $U^{\eq}_{ij}(\sqrt{s})$ is shown in Fig.\,\ref{fig:V_fit_IHW} by dashed (first-order parametrization) and dotted (second-order parametrization) lines.

%
%
%
\begin{table*}[bt]
\begin{center}
\caption{Polynomial coefficients $K_{\alpha,ij}$ in Eq.~\eqref{eq:Vfit} of the equivalent coupled-channel potentials in the $I=0$ channels. Results of the coefficients of the first- and second-order polynomials in the energy $E= \rts - m_K - M_N$ are summarized. In both cases, the range parameters are $b_\pS=0.80$ fm and $b_\KN=0.43$ fm.}
\begin{ruledtabular}
\begin{tabular}{lllll}
Polynomial type & Channel & $K_0$ [MeV]  & $K_1$ [MeV]  &  $K_2$ [MeV]  \\  \hline
First order & ${\pS,\pS}$ & $-5.67\times10^2$ & $-2.90\times10^2$  & \\
& ${\pS,\KN}$ & $\phantom{-}4.11\times10^2$ & $\phantom{-}1.16\times10^2$ &   \\
& ${\KN,\KN}$ & $-1.03\times10^3$ & $-2.58\times10^2$ &  \\ 
Second order & ${\pS,\pS}$ & $-5.76\times10^2$ & $-2.74\times10^2$ & $-3.93\times10^{0}$   \\
& ${\pS,\KN}$ & $\phantom{-}4.11\times10^2$ & $\phantom{-}1.05\times10^2$ & $-6.42\times10^{1}$  \\
&  ${\KN,\KN}$ & $-1.03\times10^3$ & $-2.59\times10^2$ & $-1.86\times10^{1}$
\end{tabular} 
\label{tab:K_couple_I0} 
\end{ruledtabular}
\end{center}
\end{table*}%
%

\begin{figure*}[tb]
\centering
\subfigure{
\includegraphics[width=5.5cm,bb=0 0 846 594]{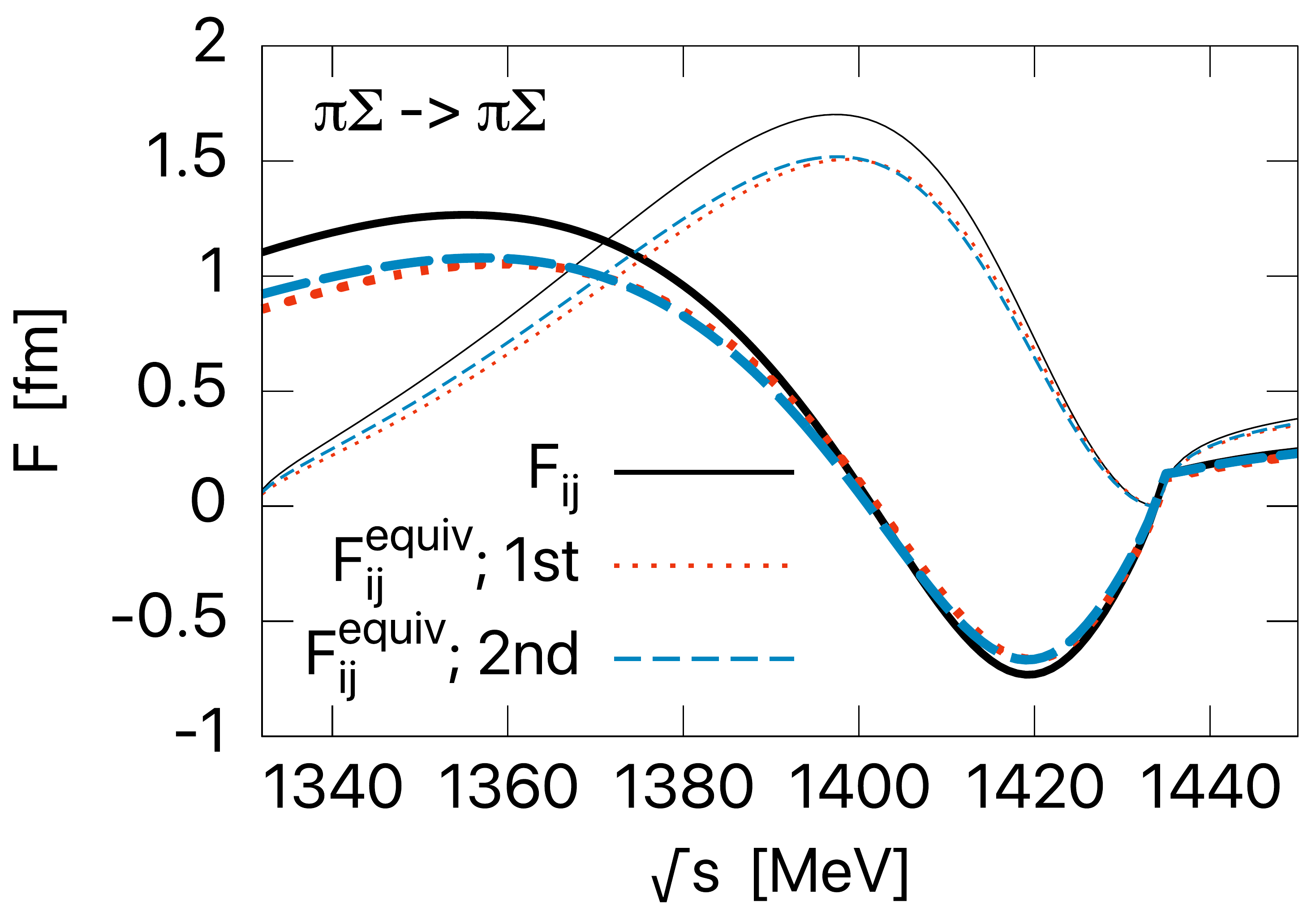}
}
\subfigure{
\includegraphics[width=5.5cm,bb=0 0 846 594]{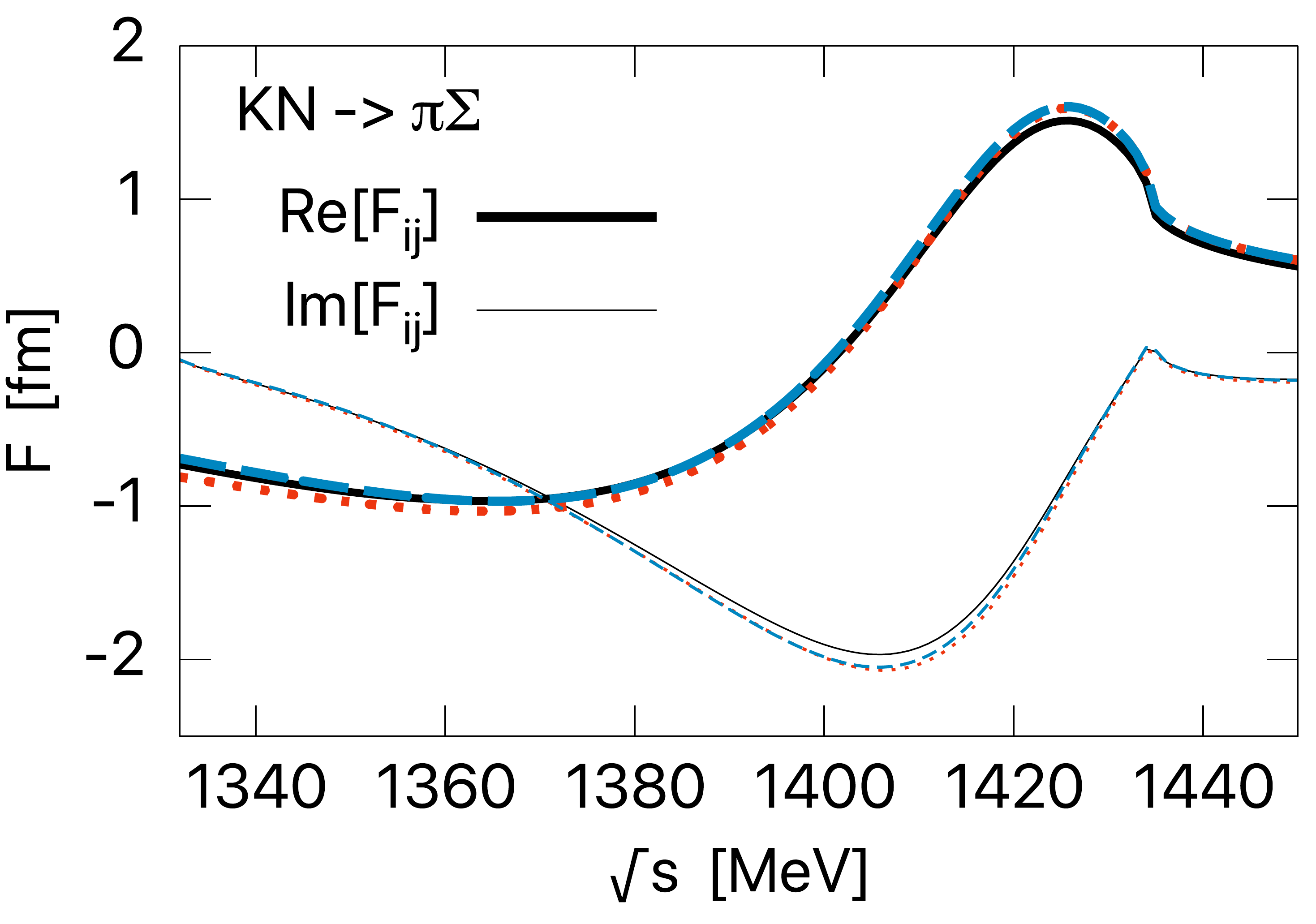}
} 
\subfigure{
\includegraphics[width=5.5cm,bb=0 0 846 594]{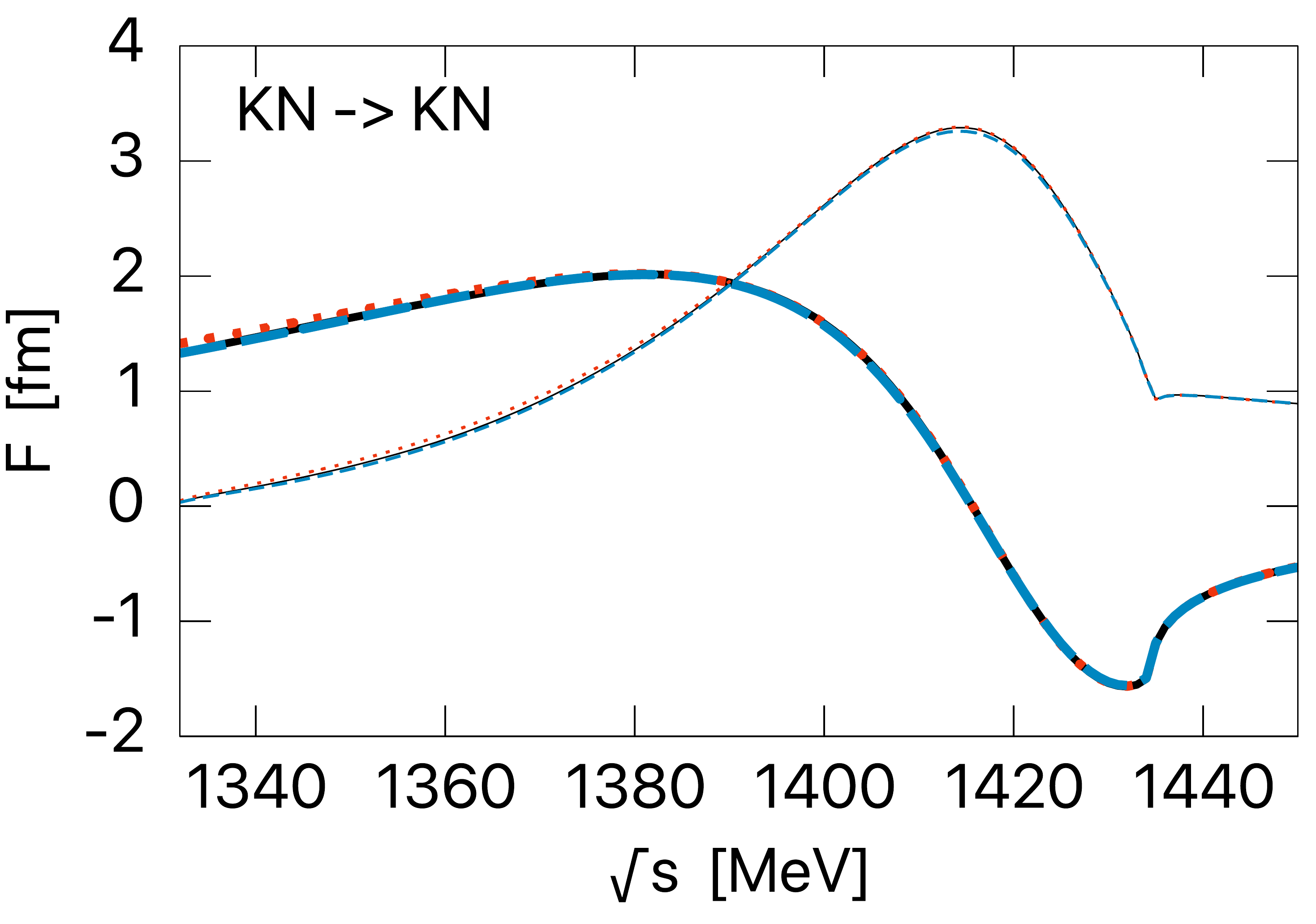}
}
\caption{Scattering amplitudes $F^\eq_{ij}$ calculated using the potential in Eq.~\eqref{eq:Vfit} with first-order (dotted lines) and second-order (dashed lines) polynomials in comparison with the original chiral SU(3) dynamics amplitudes $F_{ij}$ (solid lines) in the $I=0$ channel. The real (imaginary) parts are shown by the thick (thin) lines. }
\label{fig:F_fit_IHW}  
\end{figure*}

The scattering amplitudes calculated using the optimized potential $V^{\rm equiv}_{ij}$ of Eq.\,\eqref{eq:Vfit}, with first- and second-order polynomials, are compared with the original chiral SU(3) dynamics amplitudes in Fig.\,\ref{fig:F_fit_IHW}. The results of both the first- and second-order parametrizations are now significantly improved from those of $F^{\eqg}$ in Fig.\,\ref{fig:F_Veff_IHW}, thanks to the added adjustment term. It is worth noting that the potential with the first-order polynomial properly extrapolates the amplitude down to the region near the $\pi\Sigma$ threshold even though the lower boundary of the  energy range for parameter adjustment is around 1400 MeV, far above the $\pS$ threshold at $\sim$1330 MeV. This can be understood by the almost linear energy dependence of the potential strength seen in Fig.\,\ref{fig:V_fit_IHW}. 

%
%

\begin{figure*}[tb]
\centering
\subfigure{
\includegraphics[width=5.2cm,bb=0 0 373 341]{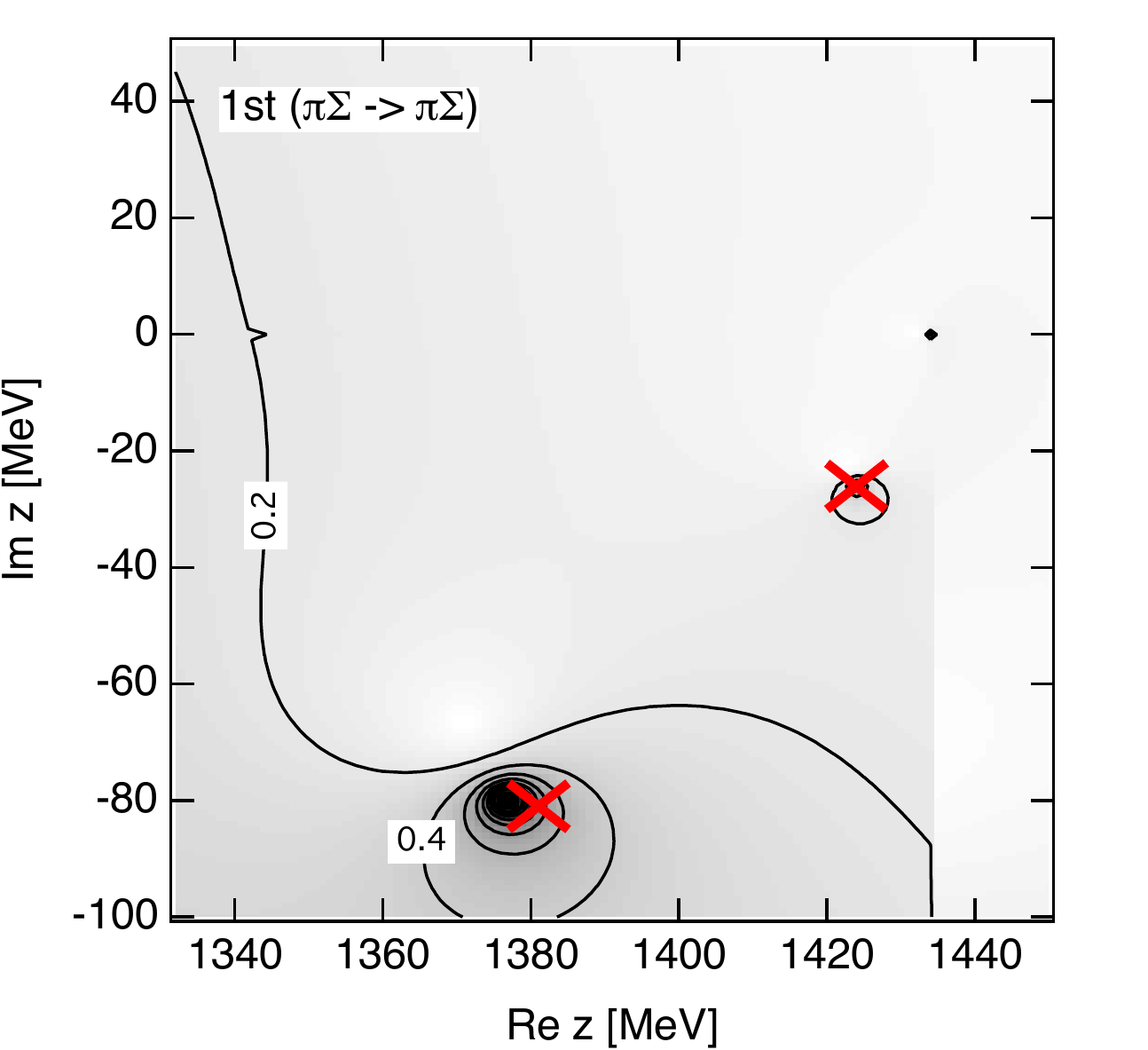}
}
\subfigure{
\includegraphics[width=5.2cm,bb=0 0 373 341]{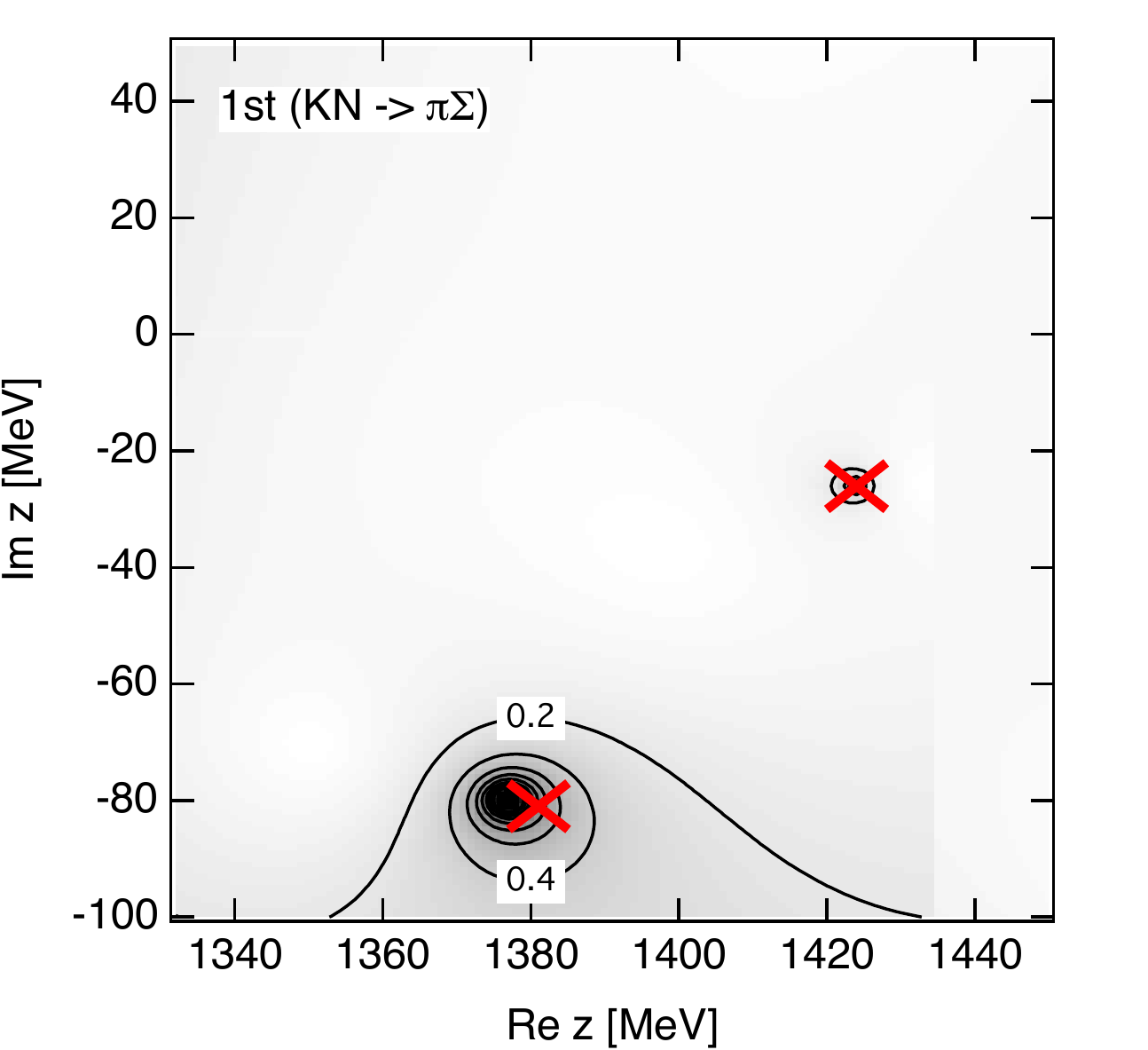}
}
\subfigure{
\includegraphics[width=6.33cm,bb=0 0 454 341]{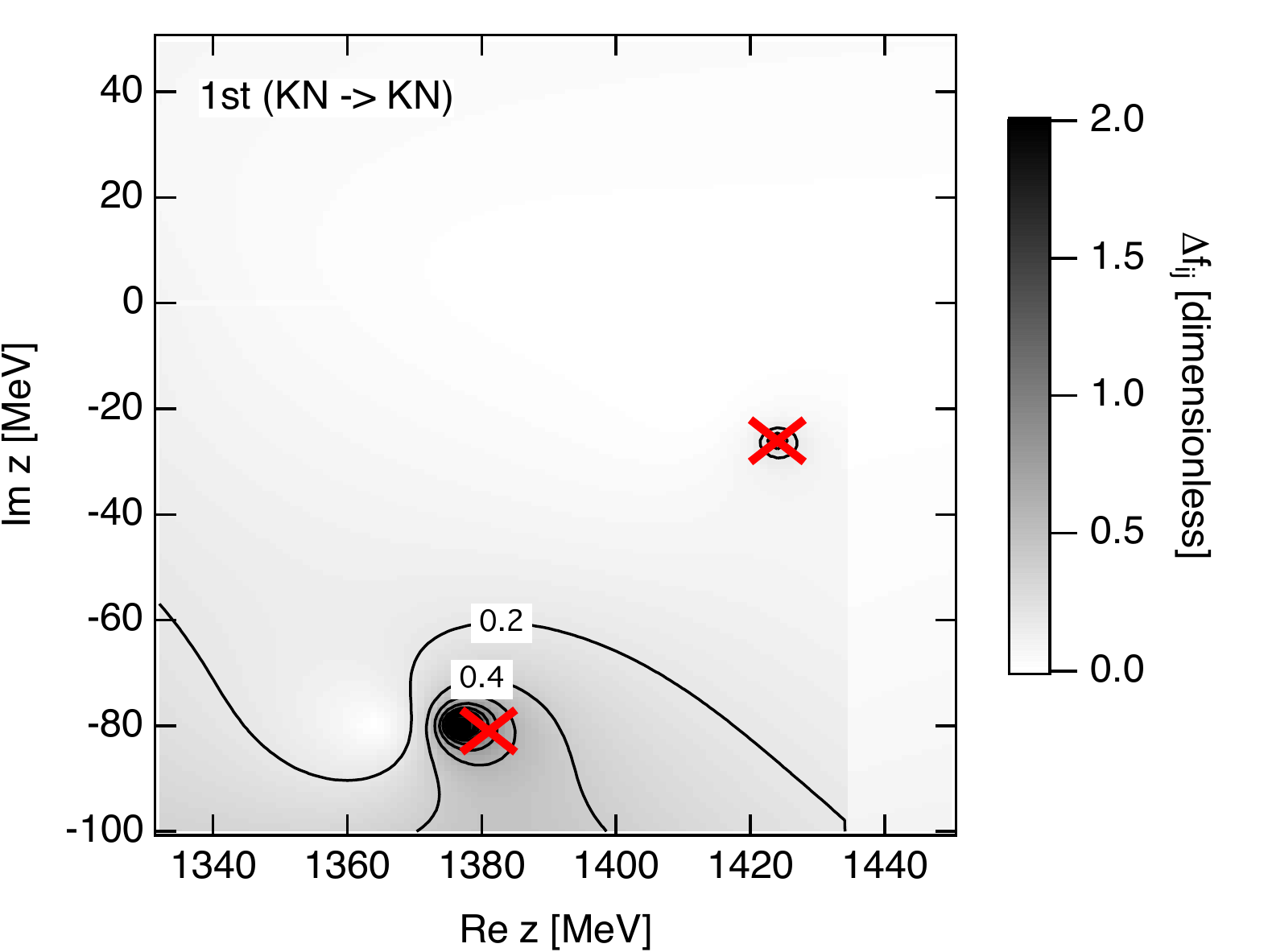}
}
\subfigure{
\includegraphics[width=5.2cm,bb=0 0 373 341]{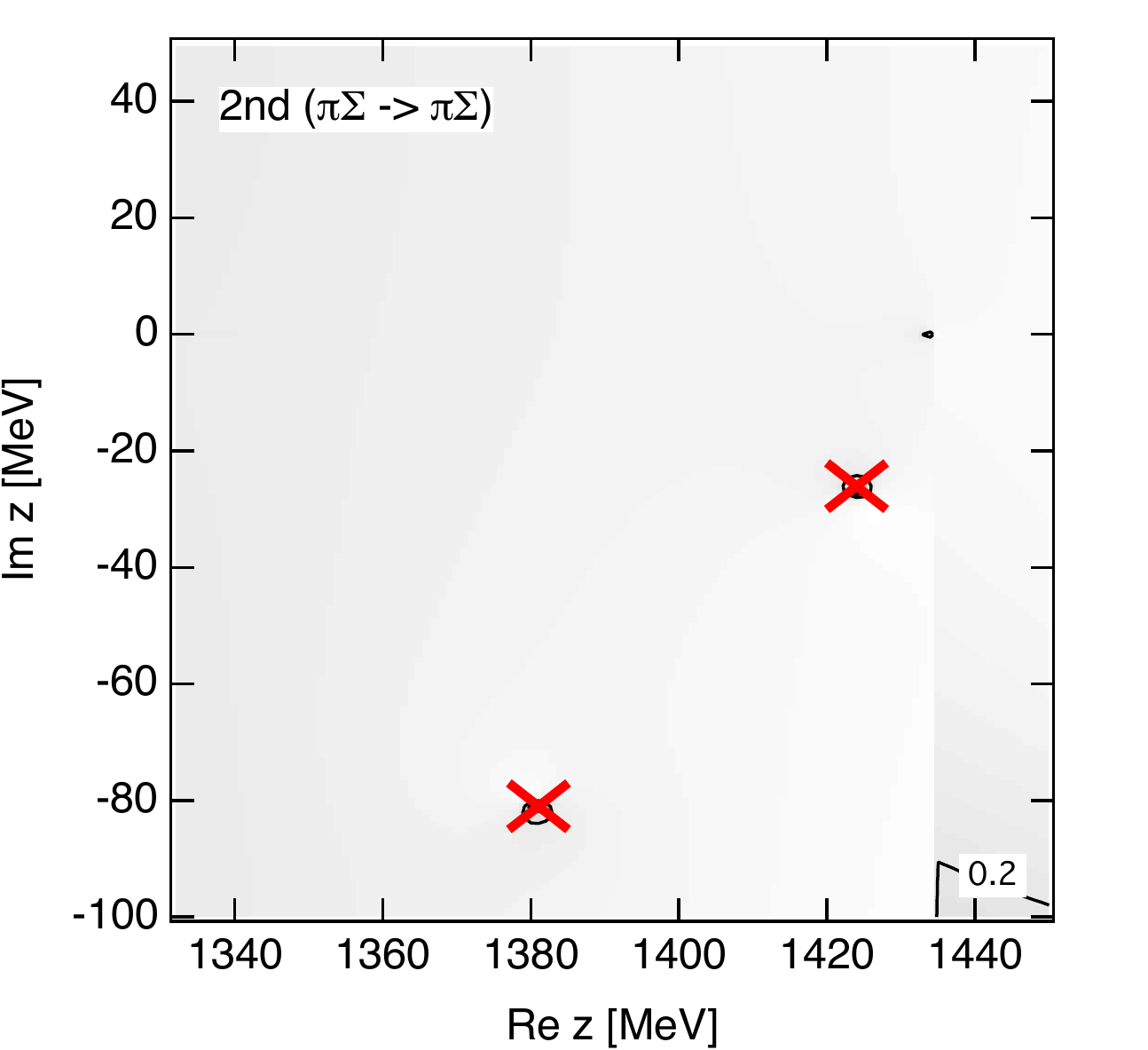}
}
\subfigure{
\includegraphics[width=5.2cm,bb=0 0 373 341]{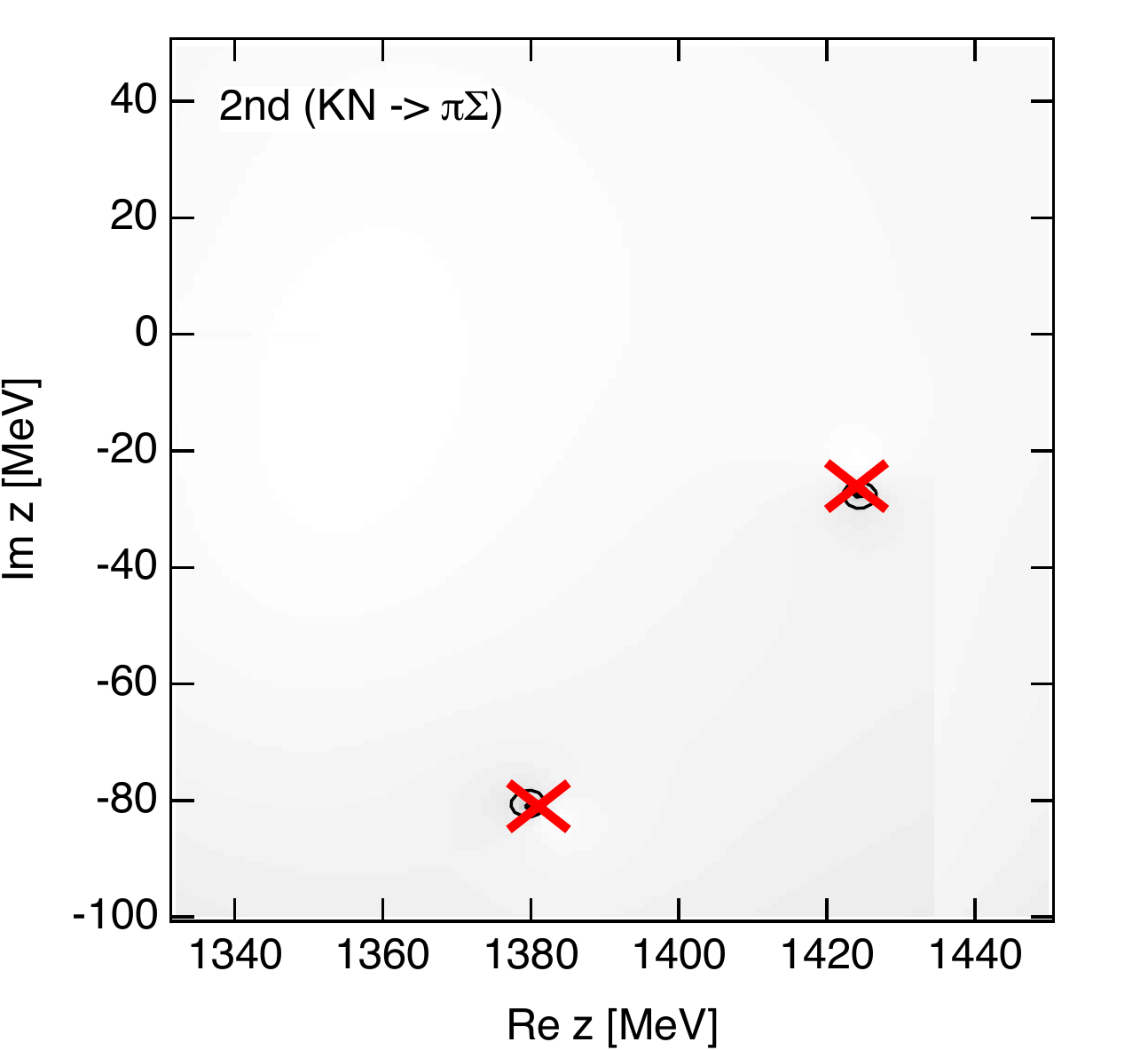}
}
\subfigure{
\includegraphics[width=6.33cm,bb=0 0 454 341]{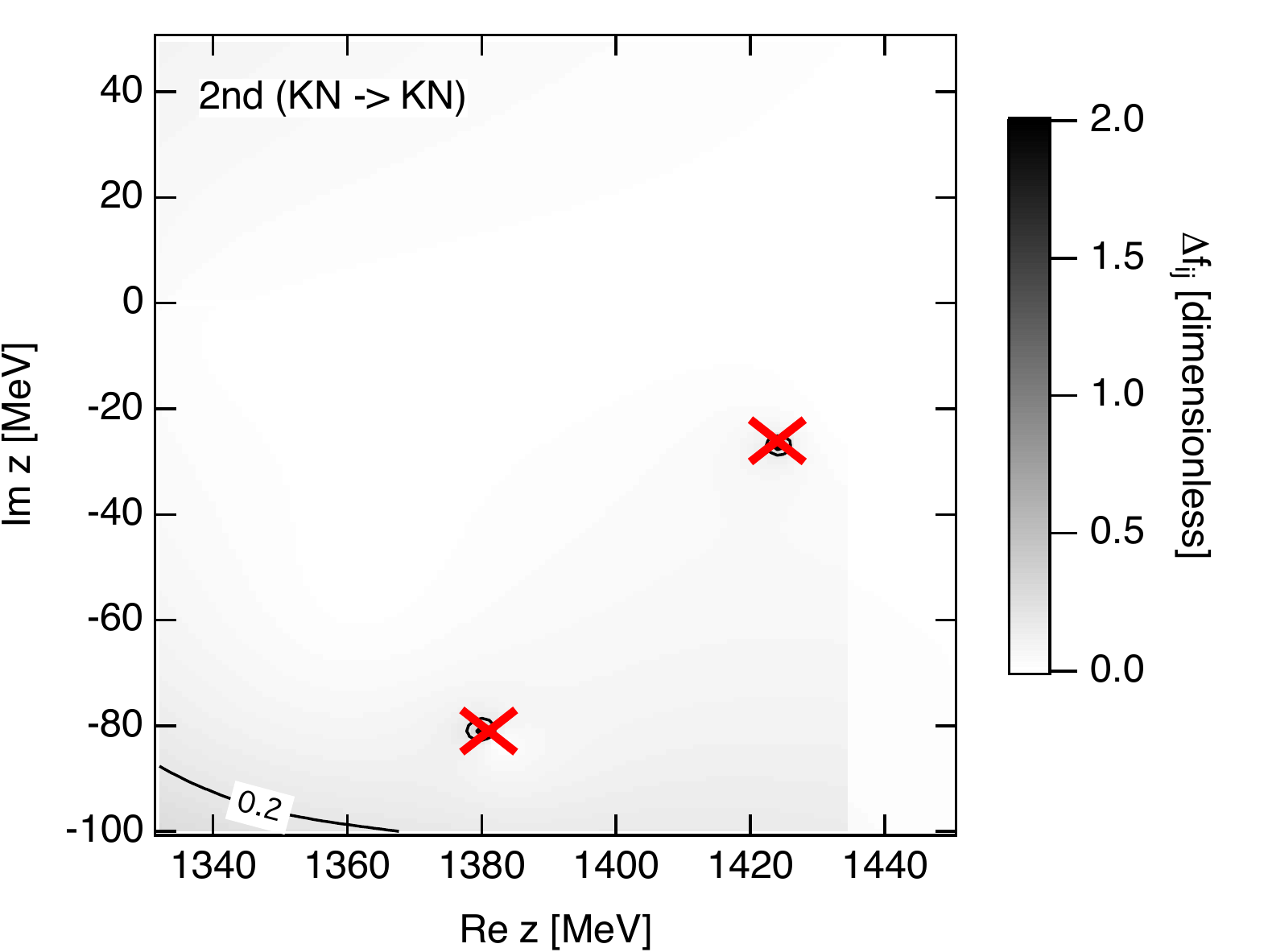}
}
\caption{Deviations $\Delta f_{ij}(z)$ [see Eq.\,\eqref{eq:DelFz}] of the  $I=0$ amplitudes in the complex energy plane relative to the original chiral SU(3) amplitudes, visualized as contours. Upper and lower figures represent the results for $\Delta f_{ij}$ computed with first- and second-order polynomial parametrizations of the potential strengths, respectively. From the left, each figure displays $\Delta f_{\pS,\pS}$, $\Delta f_{\pS,\KN}$, and $\Delta f_{\KN,\KN}$. Crosses denote positions of the two poles of the original amplitude in the complex plane. The sequence of contour lines are given in steps of 0.2.
}
\label{fig:delF_IHW}  
\end{figure*}

In order to investigate the pole structure of the $\Lambda(1405)$, the scattering amplitudes are analytically continued into the region of complex energies. In Fig.\,\ref{fig:delF_IHW}, we plot the deviations of the amplitudes, $\Delta f_{ij}(z)$ of Eq.\,\eqref{eq:DelFz}, in the complex energy plane. With both the first- and second-order polynomial potentials, each component of the original chiral SU(3) amplitude matrix is reproduced with 20\% accuracy, including the energy region of the high-mass ($\bar{K}N$-dominated) pole of the $\Lambda(1405)$. The low-mass pole can likewise be covered when the second-order polynomial is used. For a more quantitative assessment, the pole positions and the accuracy measure ${\cal P}$ defined in Eq.\,\eqref{eq:Pcomp} are summarized in Table\,\ref{tab:pole_Fequiv_couple}. The first-order polynomial potential reproduces the pole positions within the theoretical uncertainties given in Ref.\,\cite{Ikeda:2012au}. The second-order polynomial version of the potential further improves these pole positions, which are then reproduced to an accuracy of 1 MeV.  The value of ${\cal P}$ is as high as 84 (99) with the first-order (second-order) potential. This result is comparable with or better than that of the single-channel $\KN$ potential in Ref.\,\cite{Miyahara:2015bya}, which gives ${\cal P}=96$. Recalling that the complete set of available experimental data for $K^{-}$p scattering and reactions is reproduced accurately by the original amplitude of chiral SU(3) dynamics, the equivalent potential in its second-order polynomial representation and with its explicit treatment of coupled channels can justifiably be called a {\it realistic} $\KN$-$\pS$ potential.

It is remarkable that the energy-dependent strengths of the coupled-channel potential can be parametrized very well by minimal polynomial orders. This is in strong contrast to the single-channel $\KN$ effective potential for which the parametrization of the energy dependence requires a tenth-order polynomial\,\cite{Miyahara:2015bya}. This important difference can be traced to the explicit treatment of the $\pS$ channel. In Refs.\,\cite{Hyodo:2007jq,Kamiya:2017pcq}, it is shown by switching off the $\pS\leftrightarrow\KN$ channel coupling that the low-mass and high-mass poles are dynamically generated, respectively, by the attractive single-channel $\pS$ and $\KN$ interactions in chiral SU(3) dynamics. In the single-channel $\KN$ potential\,\cite{Miyahara:2015bya}, a nontrivial strong energy dependence necessarily emerges through the condition to incorporate the low-mass pole that appears in the eliminated $\pS$ channel. Using the coupled-channel potential, this low-mass pole is now generated dynamically in the explicitly included $\pS$ channel. 

%
%

A point of practical importance is the observation that the coupled-channel potential represented by a first-order polynomial in the energy works already very well in a reasonably broad energy interval, including extensions to the complex energy plane. An application of this potential to few-body $\bar{K}$-nuclear calculations would be of some interest. A linear $E$ dependence of the potential can be renormalized into an equivalent nonlocality (see, e.g., Ref.\,\cite{Suzuki:2007wi}). This provides a way to avoid ambiguities related to the energy dependence of the potential, which are a prime source of theoretical uncertainties in computations of few-body $\bar{K}$ nuclei\,\cite{Ohnishi:2017uni}.

\subsection{$I=1$ potential}\label{subsec:I1}

The $\KN$-$\pS$-$\pL$ local coupled-channel potential in the $I=1$ channel is constructed in the same manner as the potential in the $I=0$ channel. The range parameters are determined to minimize the deviation $\Delta F_g$ in Eq.\,\eqref{eq:delF_thre_couple} for the $\pi\Lambda$, $\pS$, and $\KN$ channels. The results for the range parameters in the $I=1$ channel are
\begin{align}
b_\pL^{I=1}=0.43\text{ fm},\ b^{I=1}_\pS=0.51\text{ fm},\ b_\KN^{I=1}=0.35\text{ fm}.
\label{eq:range_I1}
\end{align}
An assessment of these parameters and their validity is again given in Appendix~\ref{app:range}. Figure\,\ref{fig:F_Veff_IHW_I1} summarizes the scattering amplitudes $F^{\eq,g}_{ij}$ resulting from the potential \eqref{eq:Vequiv_eff} (without $\Delta V_{ij}$), in comparison with the original amplitudes from chiral SU(3) dynamics.%
\footnote{The nonanalytic behavior around 1360 MeV in Fig.~\ref{fig:F_Veff_IHW_I1} is related to the treatment of unphysical subthreshold cuts of the $u$-channel Born term in the on-shell formalism\,\cite{Borasoy:2005ie,Nissler:2007}.}
As in the $I=0$ case, the qualitative agreement is already quite acceptable at this stage. 

%
\begin{figure*}[tb]
\subfigure{
\includegraphics[width=5.5cm,bb=0 0 846 594]{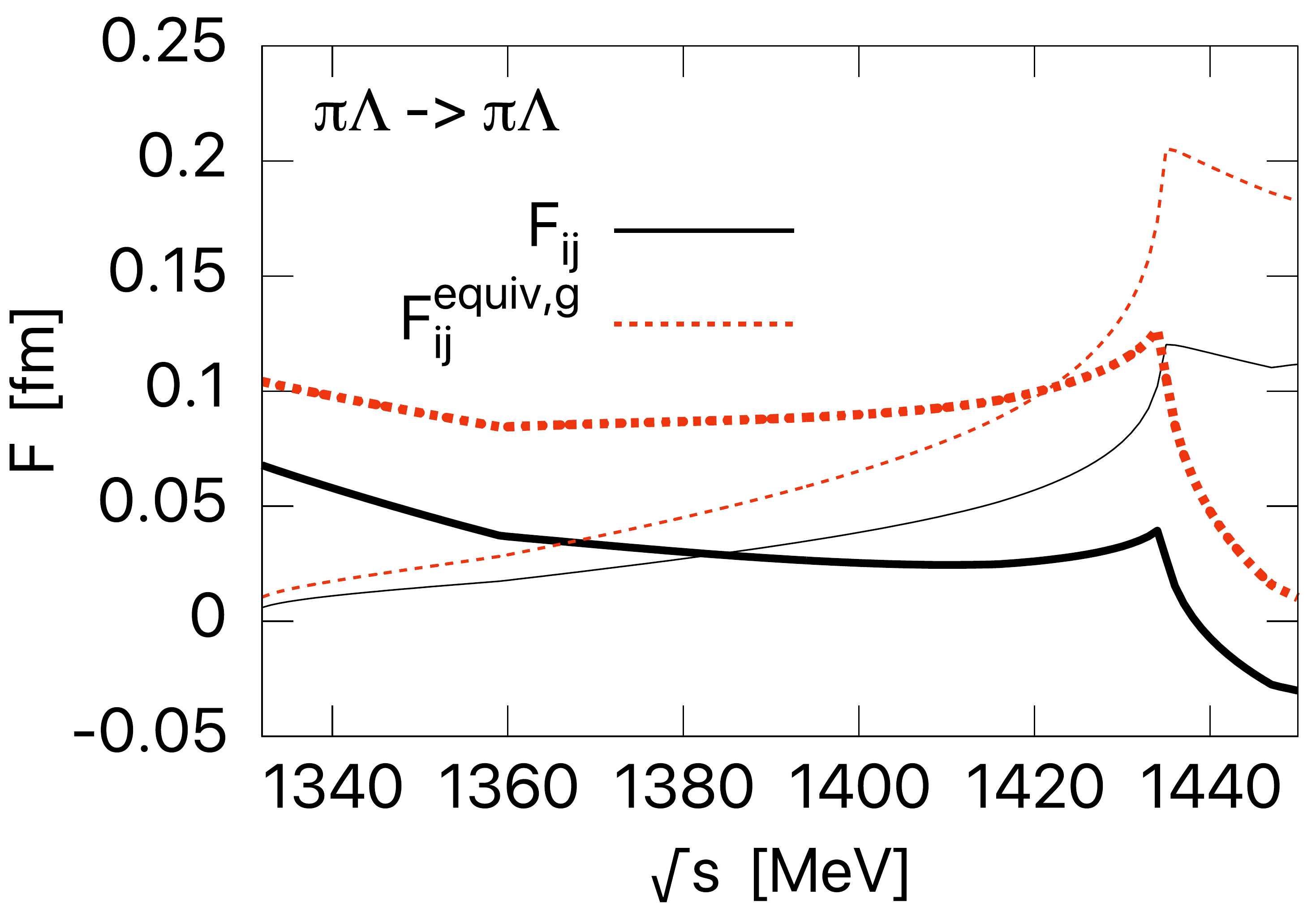}
}
\subfigure{
\includegraphics[width=5.5cm,bb=0 0 846 594]{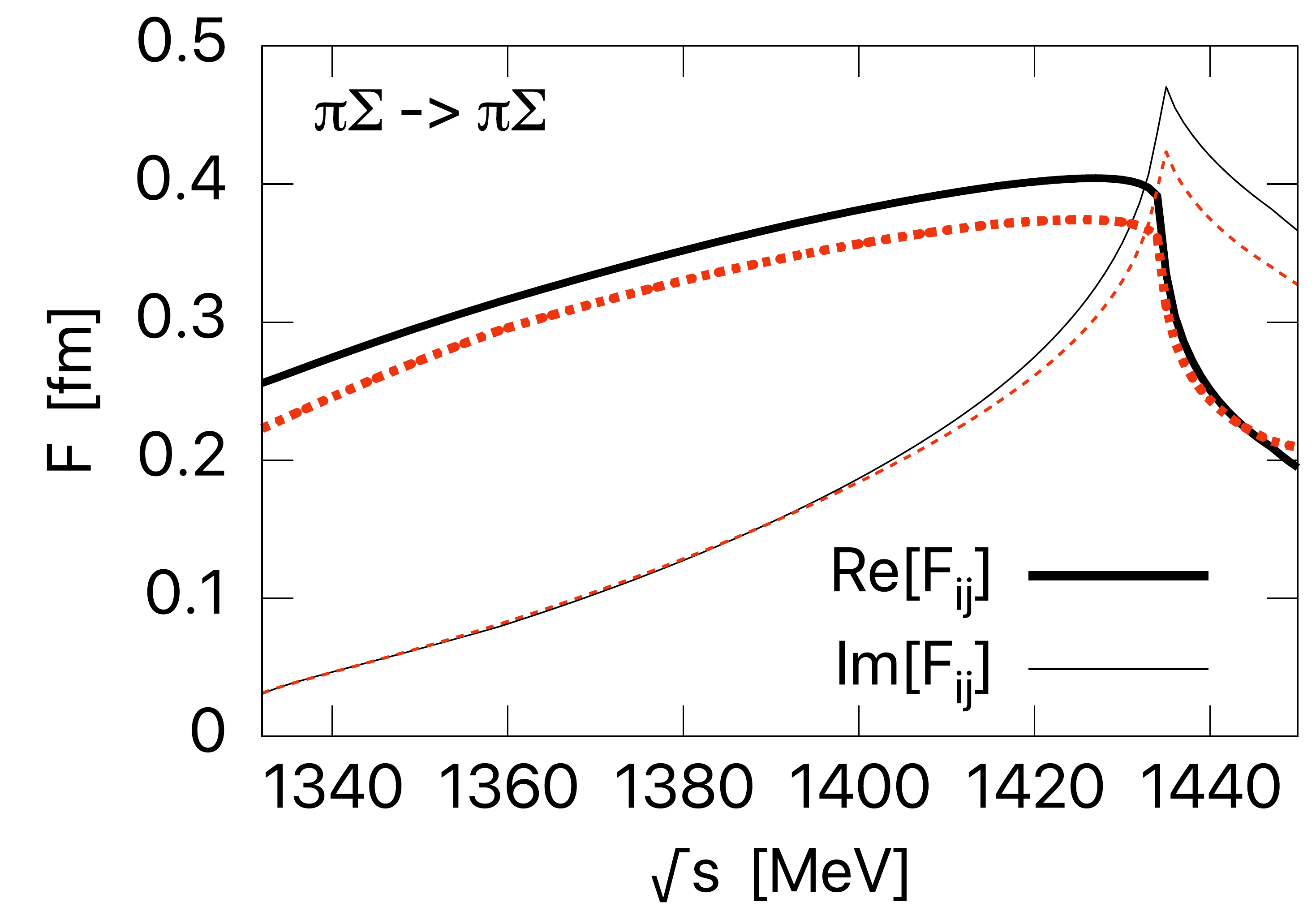}
}
\subfigure{
\includegraphics[width=5.5cm,bb=0 0 846 594]{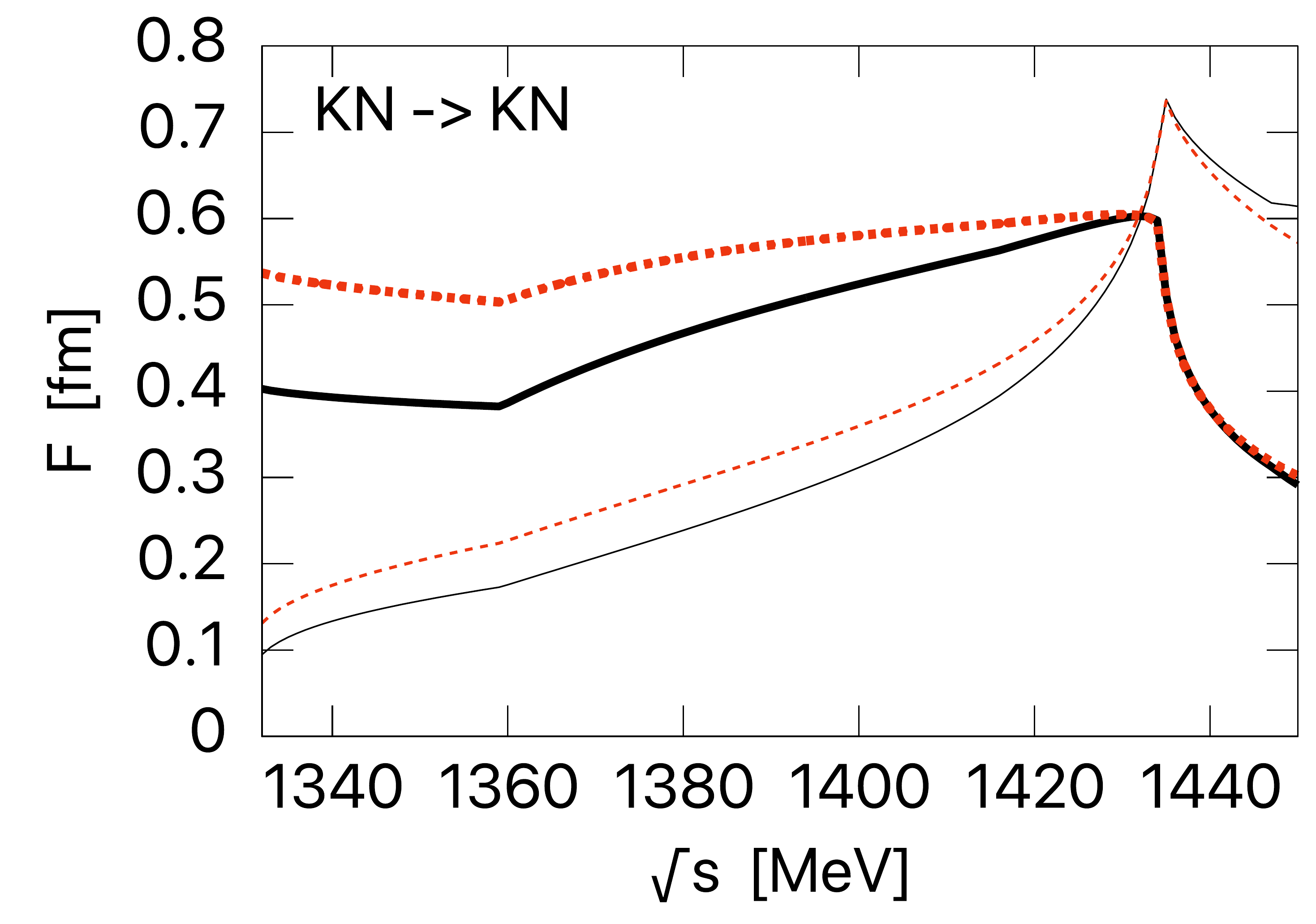}
}
\subfigure{
\includegraphics[width=5.5cm,bb=0 0 846 594]{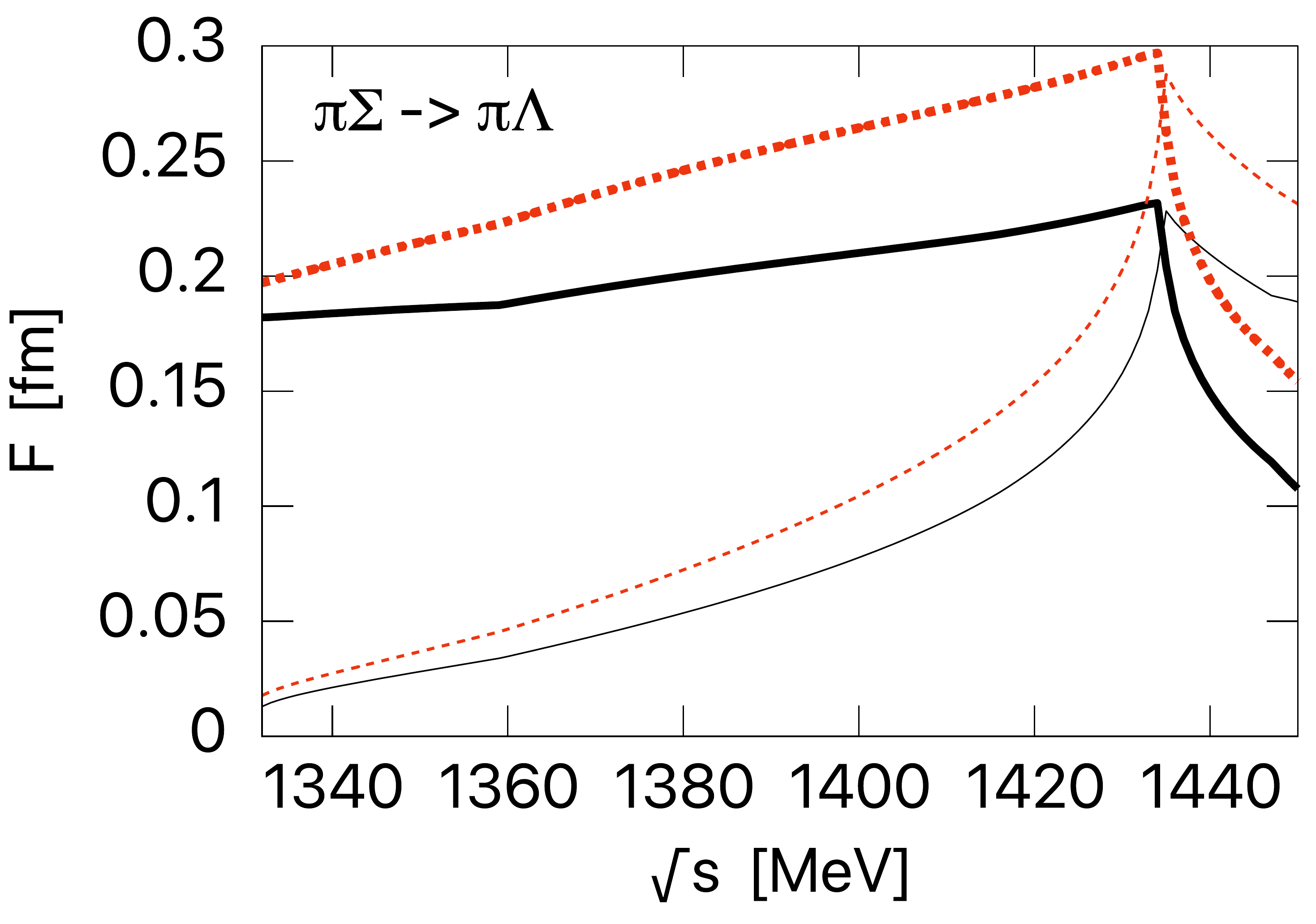}
}
\subfigure{
\includegraphics[width=5.5cm,bb=0 0 846 594]{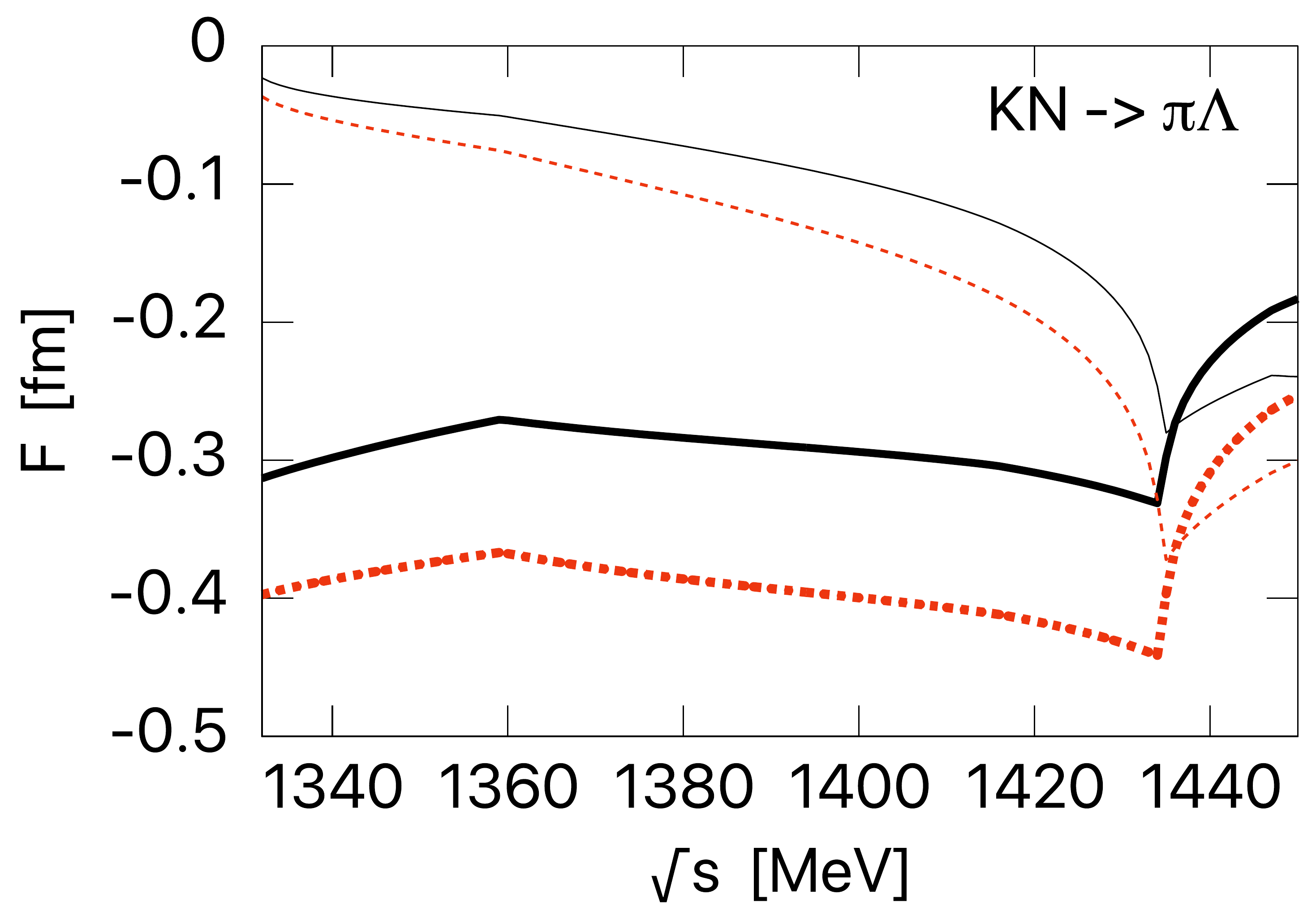}
}
\subfigure{
\includegraphics[width=5.5cm,bb=0 0 846 594]{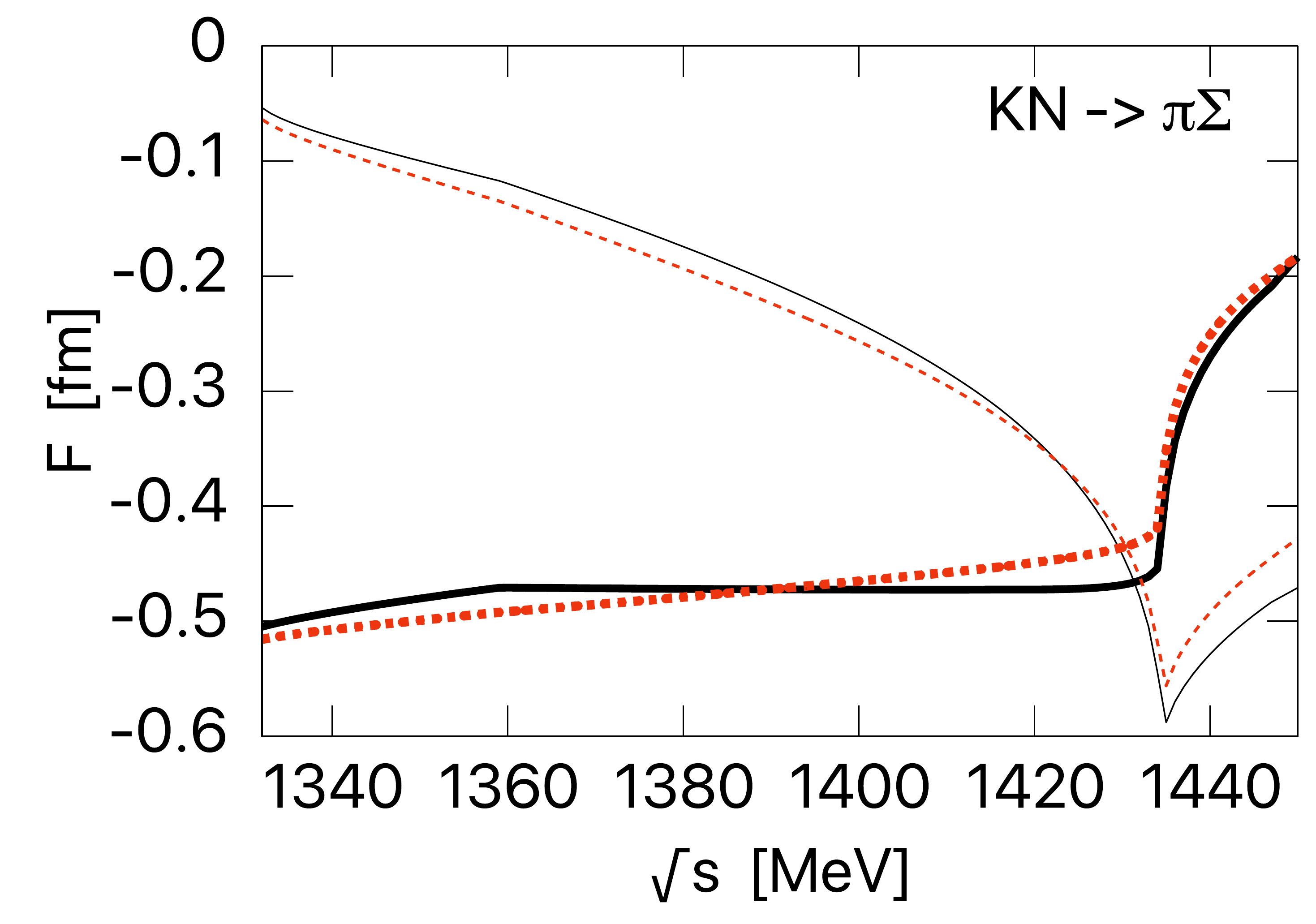}
}
\caption{Scattering amplitudes $F_{ij}^{\eqg}$ (dotted lines) resulting from the equivalent potential \eqref{eq:Vequiv_eff} in comparison with the original chiral SU(3) dynamics amplitudes (denoted by $F$, sold lines) in the $I=1$ channel. The real (imaginary) parts are shown by the thick (thin) lines.}
\label{fig:F_Veff_IHW_I1}  
\end{figure*}%
%

\begin{figure*}[tb]
\subfigure{
\includegraphics[width=5.5cm,bb=0 0 846 594]{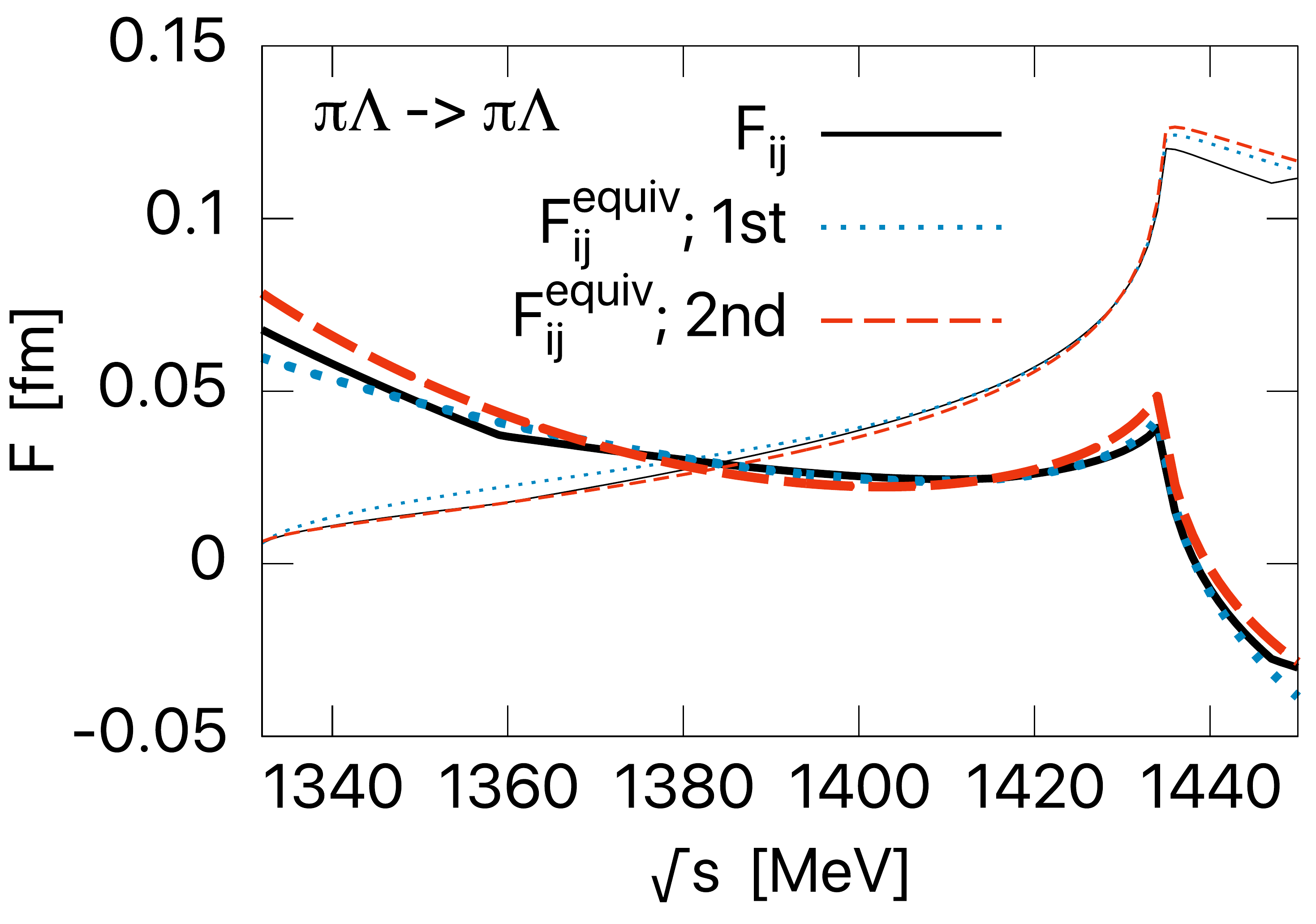}
}
\subfigure{
\includegraphics[width=5.5cm,bb=0 0 846 594]{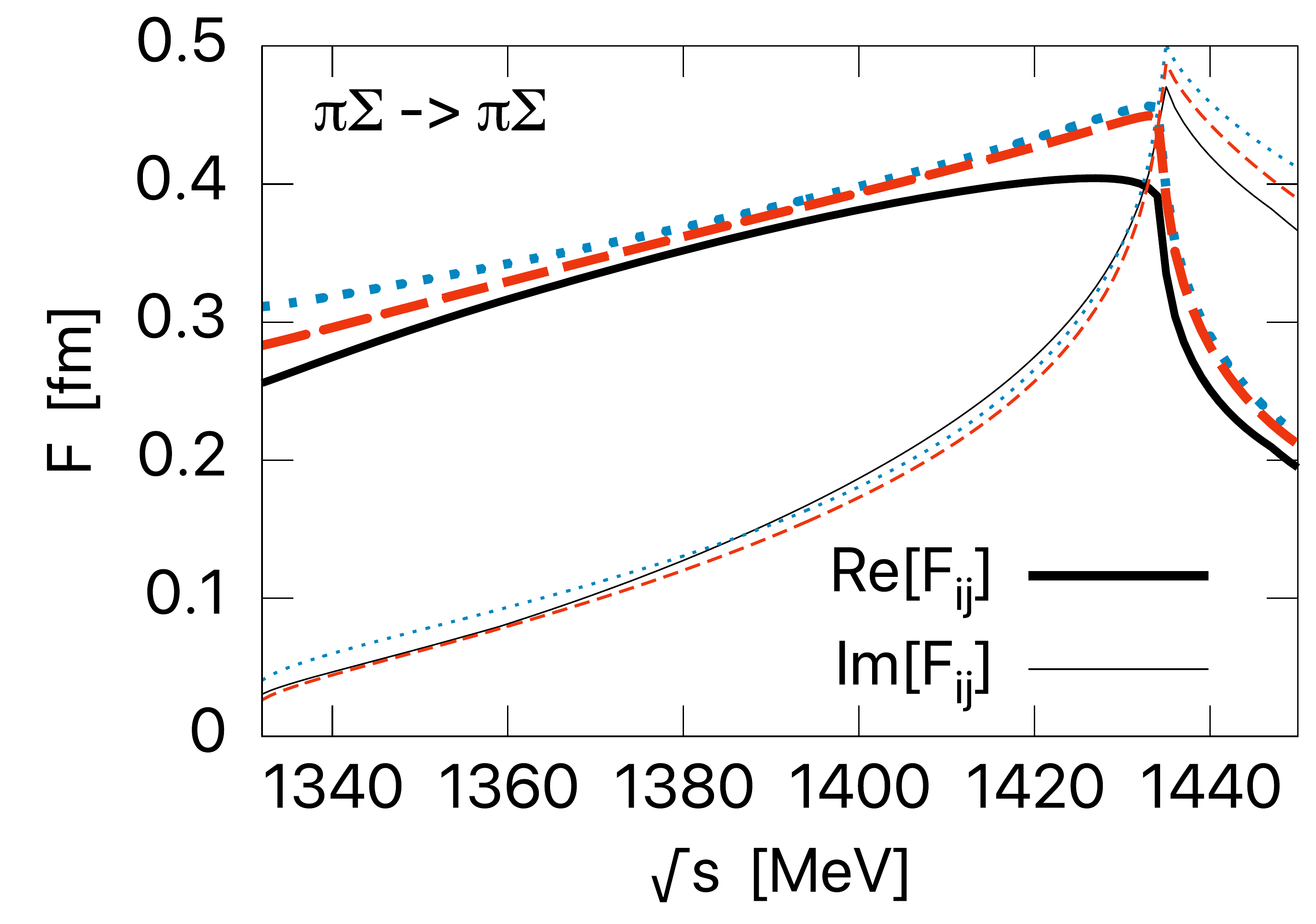}
}
\subfigure{
\includegraphics[width=5.5cm,bb=0 0 846 594]{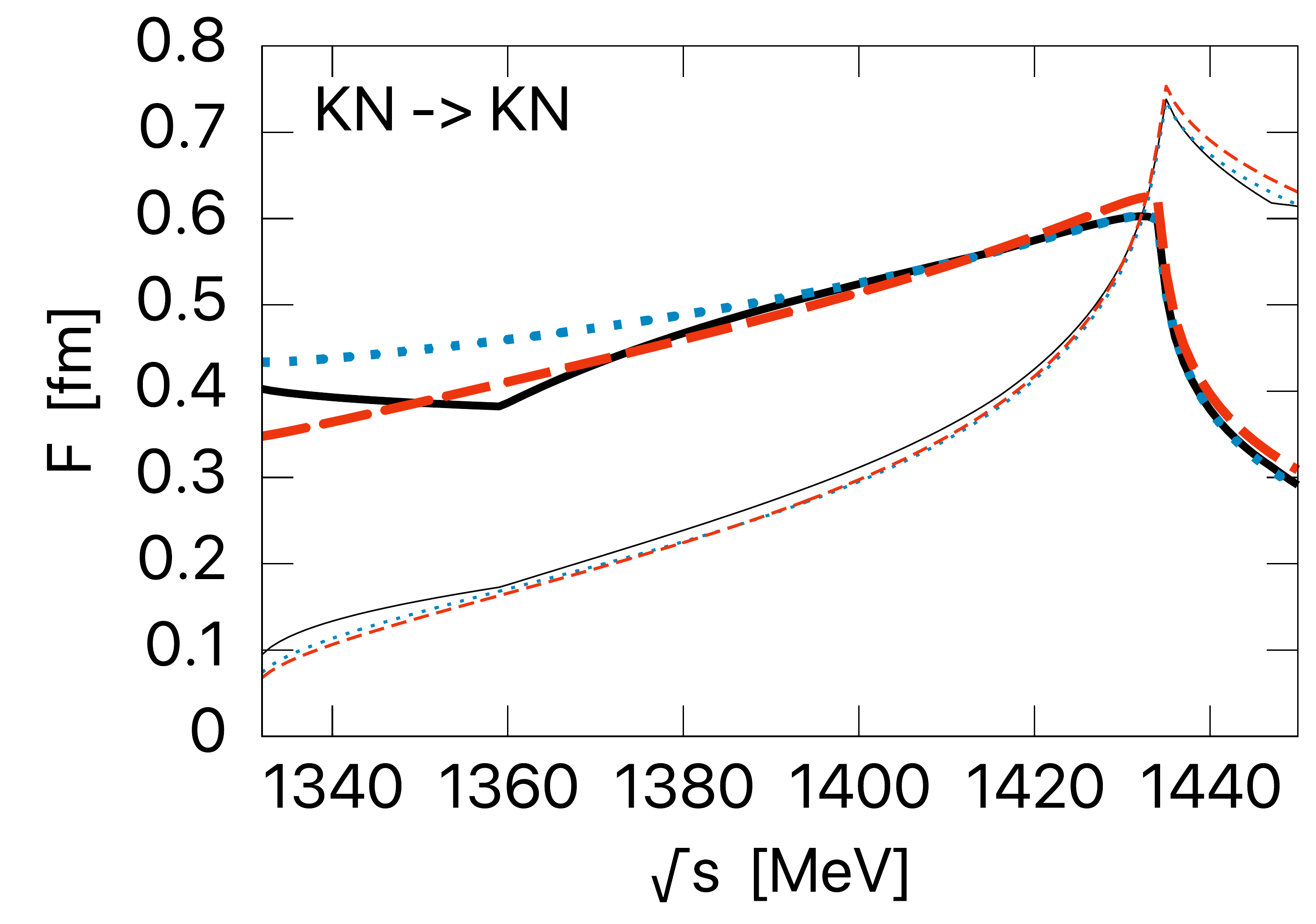}
}
\subfigure{
\includegraphics[width=5.5cm,bb=0 0 846 594]{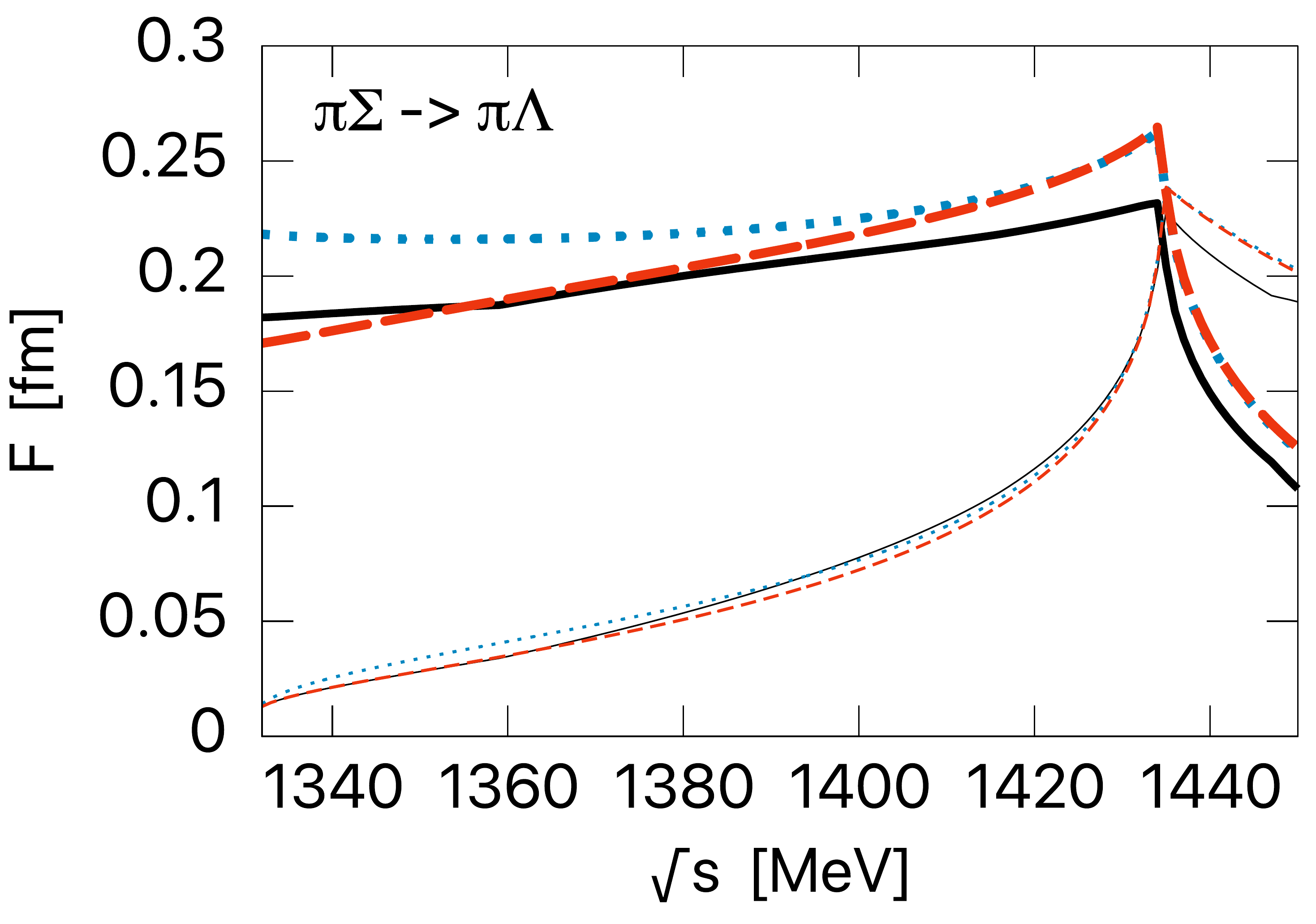}
}
\subfigure{
\includegraphics[width=5.5cm,bb=0 0 846 594]{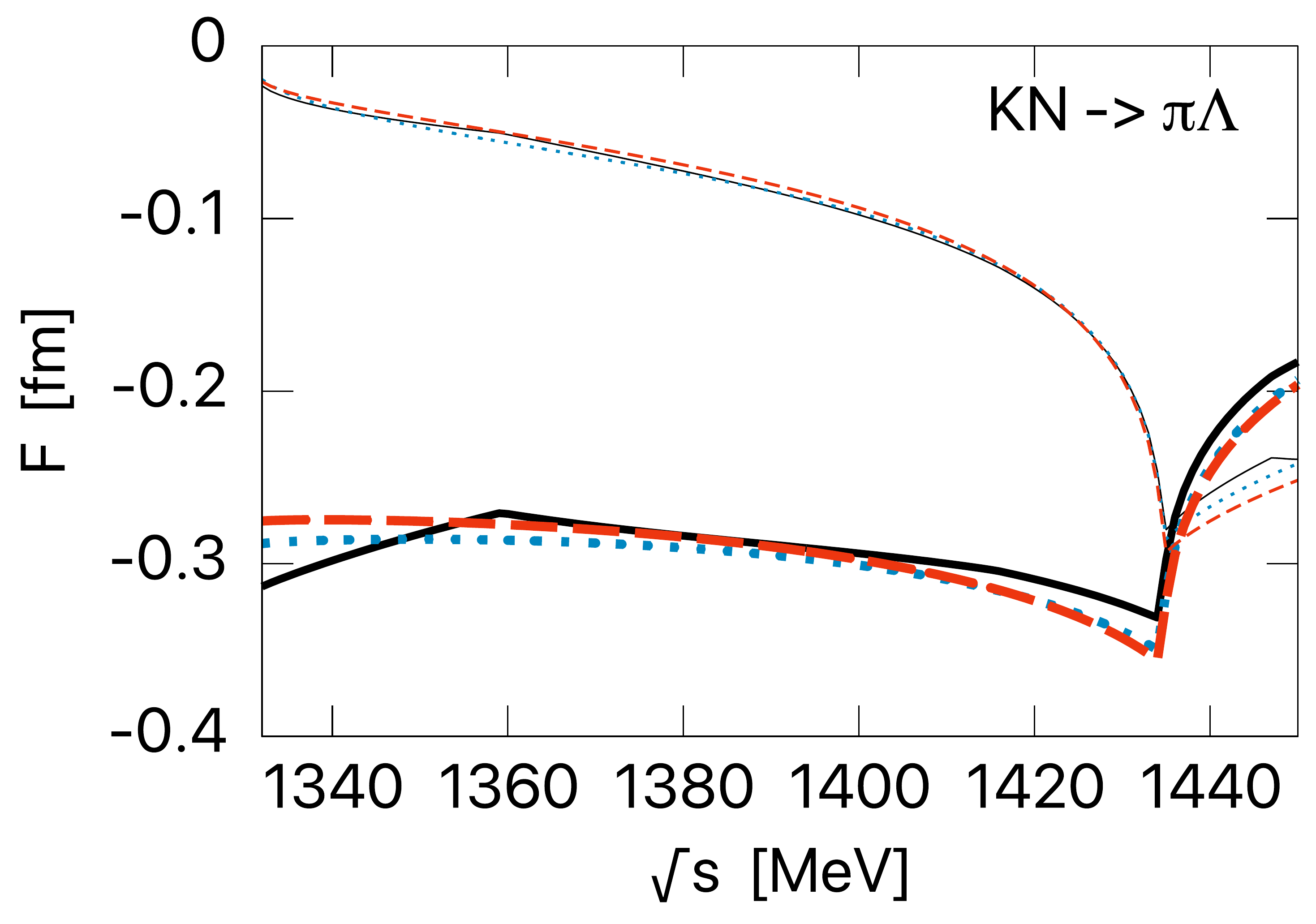}
}
\subfigure{
\includegraphics[width=5.5cm,bb=0 0 846 594]{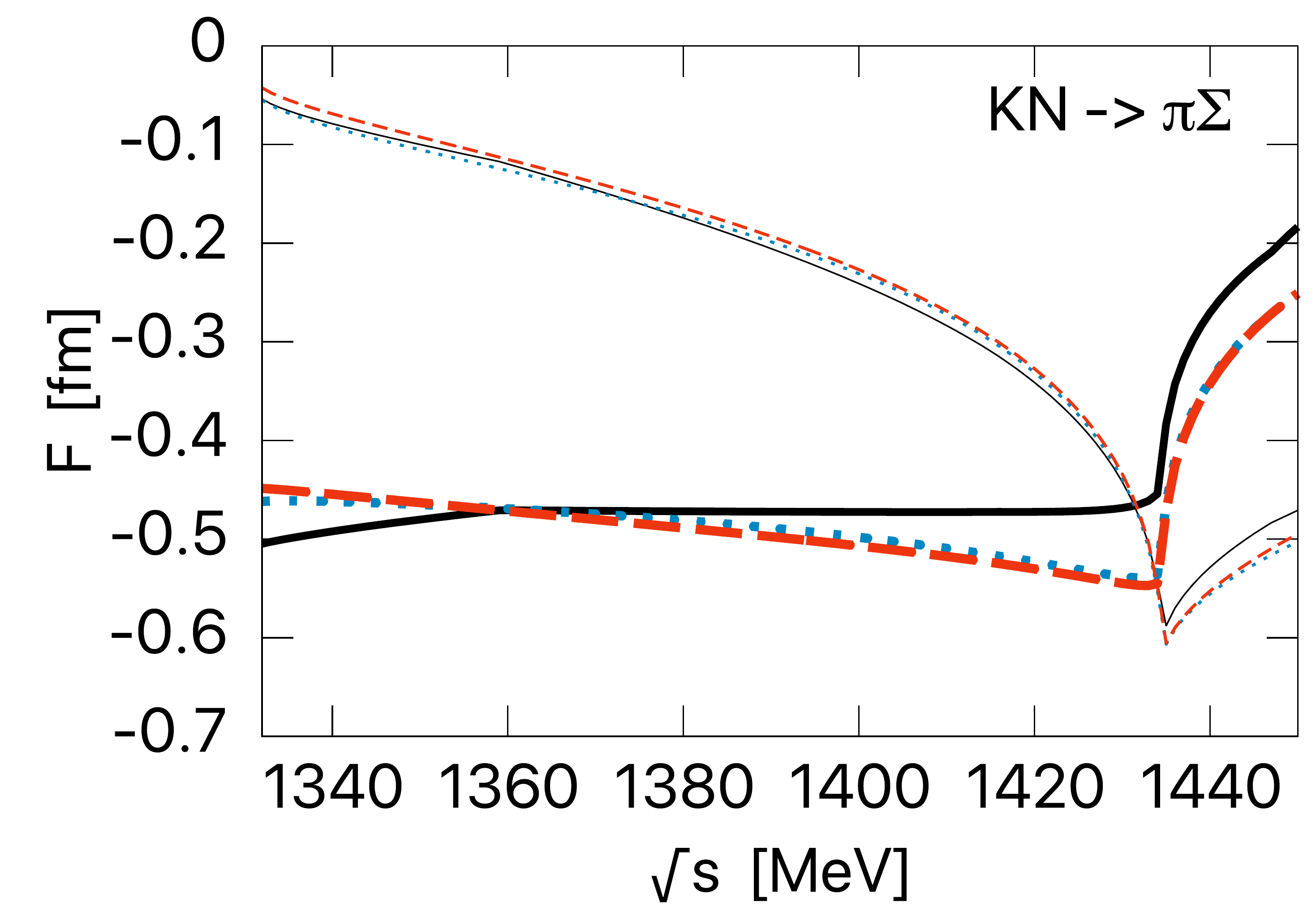}
}
\caption{Scattering amplitudes $F^\eq_{ij}$ resulting from the potential \eqref{eq:Vfit} with first-order (dotted lines) and second-order (dashed lines) polynomials, in comparison with the original chiral SU(3) amplitudes (denoted by $F_{ij}$, solid lines) in the $I=1$ channel. The real (imaginary) parts are shown by the thick (thin) lines. }
\label{fig:F_par_IHW_I1}  
\end{figure*}

Next we add the adjustment term $\Delta V_{ij}$ as in Eq.\,\eqref{eq:Vequiv}. The optimal $\Delta V_{ij}$ is determined to minimize the deviation $\Delta F(\rts)$ in Eq.\,\eqref{eq:delF_delV} at each energy. We find again that the energy dependence of the volume integrals of the potentials is almost linear in each $I=1$ channel. The strengths of these optimized equivalent potentials are then parametrized by first- and second-order polynomials as in the $I=0$ case. We use the same energy ranges for parametrization as before, namely 1403--1440 MeV and 1362--1511 MeV for the first- and second-order polynomial expansions, respectively. Once again, low-order polynomials turn out to be sufficient for the present purpose. Table\,\ref{tab:K_couple_I1} lists the polynomial coefficients of the parametrized potentials. 
The amplitudes $F^{\eq}_{ij}$ resulting from the optimized equivalent potential are shown in Fig.\,\ref{fig:F_par_IHW_I1}. The original chiral SU(3) amplitudes are now quantitatively well reproduced. (We note that, in the $I=1$ channel, theoretical uncertainties of the $F_{ij}$ from chiral SU(3) dynamics are larger than those in the $I=0$ channel\,\cite{Ikeda:2012au,Kamiya:2016jqc}.)

%
\begin{table*}[bt]
\begin{center}
\caption{
Coefficients $K_{\alpha,ij}$ in Eq.~\eqref{eq:Vfit} of the strength of the equivalent potentials in the $I=1$ channel. The results of the first- and second-order polynomials are summarized. In both cases, the range parameters
are $b_{\pi\Lambda}=0.43$ fm, $b_\pS=0.51$ fm, and $b_\KN=0.35$ fm.}
\begin{ruledtabular}
\begin{tabular}{lclll}
Polynomial type & Channel & $K_0$ [MeV]  & $K_1$ [MeV]  &  $K_2$ [MeV]  \\  \hline
First order & ${\pi\Lambda,\pi\Lambda}$ & $\phantom{-}4.73\times10^2$ & $\phantom{-}3.58\times10^2$  &  \\
& ${\pS,\pS}$ & $-4.87\times10^2$ & $-1.77\times10^2$ &  \\
& ${\KN,\KN}$ & $-5.68\times10^2$ & $-2.69\times10^2$ &  \\
& ${\pi\Lambda,\pS}$ & $-3.25\times10^2$ & $\phantom{-}4.11\times10^{1}$ &   \\
& ${\pi\Lambda,\KN}$ & $\phantom{-}6.05\times10^2$ & $\phantom{-}6.30\times10^{1}$ &   \\
& ${\pS,\KN}$ & $\phantom{-}6.37\times10^2$ & $-2.62\times10^{1}$ &   \\ 
Second order & ${\pi\Lambda,\pi\Lambda}$ & $\phantom{-}4.35\times10^2$ & $\phantom{-}1.63\times10^2$  & $-2.99\times10^{2}$ \\
& ${\pS,\pS}$ & $-4.65\times10^2$ & $-1.74\times10^2$  & $\phantom{-}1.59\times10^{0}$ \\
& ${\KN,\KN}$ & $-5.83\times10^2$ & $-3.78\times10^2$ & $\phantom{-}9.13\times10^{1}$ \\
& ${\pi\Lambda,\pS}$ & $-3.27\times10^2$ & $-2.83\times10^{0}$ & $\phantom{-}9.09\times10^{1}$  \\
& ${\pi\Lambda,\KN}$ & $\phantom{-}5.95\times10^2$ & $\phantom{-}2.89\times10^{1}$ & $\phantom{-}2.20\times10^{1}$  \\
& ${\pS,\KN}$ & $\phantom{-}6.35\times10^2$ & $-1.37\times10^2$ & $-2.87\times10^{1}$  \\
\end{tabular} 
\end{ruledtabular}
\label{tab:K_couple_I1} 
\end{center}
\end{table*}%
%
%
%

\section{Analysis of $\Lambda(1405)$ composition and structure}  \label{sec:L1405}  

\subsection{Normalization and wave function with energy-dependent potential}\label{subsec:composite_Edep}  

Our coupled-channel local potential is energy dependent as seen from Eq.\,\eqref{eq:Vequiv}. In order to analyze the structure of the $\Lambda(1405)$ in this context, one must first examine the proper normalization condition for its wave function. A system with an energy-dependent and real potential requires a modification of the normalization scheme and orthogonality condition for eigenstates in order to satisfy fundamental rules of quantum mechanics\,\cite{Lepage:1997cs,Lepage:1977gd,Caswell:1978mt,Sazdjian:1986qn,Formanek2004,2003quant.ph.12148Z,Benchikha:2013mba,Miyahara:2015bya}. The generalization to a non-Hermitian system with a single-channel energy-dependent potential has been performed in Ref.\,\cite{Miyahara:2015bya}, referring to the treatment of a resonance eigenstate with a complex potential\,\cite{Hokkyo:1965,Berggren:1968zz,non_hermitian}. 

The normalization condition for the coordinate space wave function $\psi(\bm{r})$ of a discrete eigenstate is derived,  starting from the continuity equation, as
\begin{align}
1 &= \int d^3r\ \psi(\bm{r}) \left[ 1 - \frac{\del V}{\del E}(\bm{r},E) \right] \psi(\bm{r}), 
\end{align}
where $V$ is the energy-dependent potential. This condition can be directly extended to a coupled-channels system:
\begin{align}
1 &= \int d^3r\  \sum_{i,j} \psi_{i}(\bm{r}) \left[ \delta_{ij} - \frac{\del V_{ij}}{\del E}(\bm{r},E) \right] \psi_{j}(\bm{r})\,,
\label{eq:norm_Gamow}
\end{align}
as explained in Appendix~\ref{app:E_dep}. However, in this derivation, the physical interpretation of the $\del V_{ij}/\del E$ term needs to be clarified. In the present work we give an interpretation of the modified norm by deriving Eq.\,\eqref{eq:norm_Gamow} in an alternative way using the Feshbach projection method\,\cite{Feshbach:1958nx,Feshbach:1962ut}. 

The energy-dependent potential is defined in a certain model space (such as $\bar{K}N$-$\pi\Sigma$), called ``$P$ space'' for later convenience. We consider the $P$ space as a subspace of the ``full space'' in which the Hamiltonian is energy independent. In other words, we assume that the energy dependence of the potential emerges from the elimination of implicit channels in the full space. In our case of meson-baryon interactions, the implicit channels can be, for example, a one-body discrete state representing a ``bare" $\Lambda^{*}$ as a three-quark state, higher energy meson-baron channels, meson-meson-baryon channels, and so on. Thus, we first prepare the state vector in the full space $|\psi\rangle$ and the corresponding Gamow state $|\psi^\dag\rangle$,

\begin{align}
|\psi\rangle =
\left(
\begin{array}{c}
|\psi_1\rangle \\
|\psi_2\rangle \\
\vdots
\end{array}
\right),\ \ \ 
|\psi^\dag\rangle =
\left(
\begin{array}{c}
|\psi_1^\dag\rangle \\
|\psi_2^\dag\rangle \\
\vdots
\end{array}
\right),
\end{align}
conceptually including all possible channels (that is, the set of $|\psi_i\rangle$  represents all one-body to many-body systems of any relevant degrees of freedom acting in the full space). The Gamow vector $|\psi^\dag\rangle$ is introduced to properly normalize resonance wave functions. 

These state vectors satisfy the Schr\"odinger equations,\footnote{In Ref.\,\cite{Hokkyo:1965}, the Gamow vector is shown to satisfy the Schr\"odinger equation, $\hH^\dag|\psi^\dag\rangle=E^*|\psi^\dag\rangle$, which is equivalent to Eq.\,\eqref{eq:sch_fesh_Gamow}. The Hermitian conjugate of $\hH$ is defined with proper boundary conditions for both $\psi$ and $\psi^\dag$ (see Ref.\,\cite{Hokkyo:1965} for more details).}
\begin{align}
\hat{H}|\psi\rangle = (\hat{H}_0 + \hat{V}) |\psi\rangle = E|\psi\rangle,
\label{eq:sch_fesh} \\
\langle \psi^\dag|\hat{H} = \langle \psi^\dag| (\hat{H}_0 + \hat{V}) = \langle \psi^\dag| E,
\label{eq:sch_fesh_Gamow}
\end{align}
with an energy-independent interaction $\hat{V}$ and a free Hamiltonian $\hat{H}_0$ which is diagonal for each channel. A resonance wave function can be normalized as $\bra{\psi^{\dag}}\kket{\psi}=1$ employing the Gamow state vector. 

Suppose now that the full space is reduced to a model space ($P$ space). The effective interaction acting on $P$ space will then be energy dependent. The reduction of channels can be performed by the Feshbach projection method\,\cite{Feshbach:1958nx,Feshbach:1962ut}. Let $\hP$ be the projection operator onto $P$ space. The projection operator to the eliminated channels is denoted by $\hQ$. These operators meet the usual relations for general projection operators, $\hP + \hQ =1$, $\hP\hQ = 0$, $\hP^2 = \hP$, $\hQ^2 = \hQ$. We introduce the quantities $X_i$ and $Z$ as the norm of channel $i$ in the $P$ space and the sum of the norms of the channels in $Q$ space, respectively:
\begin{align}
X_i &= \langle \psi^\dag_i| \psi_i\rangle \ \ \ (i\in P)\,, \quad
Z = \langle \psi^\dag| \hQ|\psi\rangle.
\end{align}
Inserting $\hP + \hQ =1$ in the normalization of the state vector, $\langle \psi^\dag|\psi\rangle=1$, implies the following sum rule for $X_i$ and $Z$:
\begin{align}
\langle \psi^\dag| \hP+\hQ|\psi\rangle &= \langle \psi^\dag| \hP\hP |\psi\rangle + \langle \psi^\dag|\hQ\hQ|\psi\rangle \notag \\
&=\sum_{i\in P}X_i + Z = 1\,\,.
\label{eq:sum_rule_XZ}
\end{align}
When the $P$ space consists of only two-body states [such as the $\bar{K}N$-$\pi\Sigma$ coupled-channel system of the $\Lambda (1405)$], $X_i$ and $Z$, respectively, correspond to the compositeness and the ``elementarity" of the states $|\psi_i\rangle\in P$\,\cite{Hyodo:2011qc,Aceti:2012dd,Hyodo:2013nka,Sekihara:2014kya,Kamiya:2015aea,Sekihara:2015gvw,Kamiya:2016oao,Sekihara:2016xnq}. Historically, $Z$ has been introduced in quantum field theory as the renormalization constant of a bare field\,\cite{Weinberg:1965zz}. Its interpretation as elementarity has later been extended to stand for the contribution from the implicit channels including continuum states\,\cite{Sekihara:2014kya}. The Feshbach projection formalism provides a foundation for this interpretation, with $Z$ including the contributions from all channels in $Q$ space. Operating with $\hP$ or $\hQ$ from the left (right) to Eqs.\,\eqref{eq:sch_fesh} and Eq.\,\eqref{eq:sch_fesh_Gamow}, the state vectors in $P$ space and $Q$ space are related as
\begin{align}
\hQ |\psi\rangle 
&=\frac{1}{E-\hQ\hH\hQ}(\hQ\hV\hP) \hP|\psi\rangle\, ,
\label{eq:Qwf} \\
\langle \psi^\dag | \hQ 
&= \langle \psi^\dag |\hP (\hP\hV\hQ) \frac{1}{E-\hQ\hH\hQ}\, .
\label{eq:Qwf_G}
\end{align}
The reduced Schr\"odinger equation for the $P$-space channels, $\hP|\psi\rangle$, becomes
\begin{align}
(\hP\hH^0\hP) \hP|\psi\rangle + \hV^\eff(E) \hP|\psi\rangle = E \hP|\psi\rangle\,,
\end{align}
with the effective potential
\begin{align}
\hV^\eff(E) = \hP\hV\hP + (\hP\hV\hQ)\frac{1}{E-\hQ\hH\hQ}(\hQ\hV\hP) \,.
\label{eq:Veff_Fesh}
\end{align}
The second term in Eq.\,\eqref{eq:Veff_Fesh} introduces the energy dependence of this effective potential. Acting on $\hP|\psi\rangle$, $\hat{V}^\eff(E)$ is constructed such that it exactly reproduces the wave functions $|\psi\rangle$ of the full Schr\"odinger equation for the channels within the restricted model space.

From Eqs.\,\eqref{eq:Qwf} and \eqref{eq:Qwf_G}, the norm $Z$ of the eliminated $Q$-space channels can be expressed in terms of $P$-space quantities as follows:
\begin{align}
Z &= \langle \psi^\dag| \hQ \hQ|\psi\rangle \notag \\
&=\langle \psi^\dag| \hP (\hP\hV\hQ)\left(\frac{1}{E-\hQ\hH\hQ}\right)^2(\hQ\hV\hP)\hP|\psi\rangle \notag \\
&=\langle \psi^\dag | \hP \left(-\frac{\del \hV^\eff}{\del E}{(E)}\right) \hP|\psi\rangle.
\label{eq:ele_delVdelE}
\end{align}
This is the general expression for the elementarity in operator form. It is related to the additional term appearing in Eq.\,\eqref{eq:norm_Gamow}. Denoting the wave function in $P$ space as $P\ket{\psi}\equiv\ket{\psi_{P}}$, the general form of the normalization condition for $\ket{\psi_{P}}$ with the energy-dependent effective $P$-space potential is
\begin{align}
1 &=
\bra{\psi^{\dag}_{P}}
\left(I-
 \frac{\del \hV^\eff}{\del E}{(E)}
\right)|\psi_{P}\rangle .
\label{eq:norm_full}
\end{align}

In the present case of the $\Lambda(1405)$, the $P$ space consists of the two-body system of coupled $\KN$ and $\pS$ channels. All other channels such as $\eta\Lambda$, $K\Xi$, and bare $\Lambda^{*}$ are eliminated and included in $Q$ space. The completeness relation is written as
\begin{align}
1 &= \hP + \hQ 
=\sum_{i\in P} \int d^3r_i\ |\bm{r}_i\rangle \langle \bm{r}_i| + \hQ\,,
\end{align}
where $\bm{r}_{i}$ denotes the relative coordinates in the $P$-space two-body channels. Inserting the completeness relation into Eq.\,\eqref{eq:ele_delVdelE} gives
\begin{align}
Z &=\sum_{i,j \in P} \iint d^3 r_i d^3r_j \,\psi_i(\bm{r}_i)\left(-\frac{\del V_{ij}^\eff}{\del E}(\bm{r}_i,\bm{r}_j;E) \right) \psi_j(\bm{r}_j)\,,
\end{align}
where $\bra{\bm{r}_{i}}\kket{\psi}=\bra{\psi^{\dag}}\kket{\bm{r}_{i}}\equiv\psi_{i}(\bm{r}_{i})$ is the  $P$ space wave function. If the effective interaction is local as in Sec.\,\ref{sec:results_const_pot}, with $V^\eff\propto\delta^{(3)}(\bm{r}_i-\bm{r}_j)$, this relation reduces to
\begin{align}
Z &= \sum_{i,j\in P} \int d^3r\ \psi_i(\bm{r}) \left( -\frac{\del V^\eff_{ij}}{\del E}(\bm{r};E) \right) \psi_j(\bm{r})\,.
\label{eq:Z_Fesh}
\end{align}
This is exactly the same as the second term in Eq.\,\eqref{eq:norm_Gamow}.
The compositeness $X_i$ for channel $i$ in $P$ space can simply be written as
\begin{align}
X_i &= \langle \psi^\dag_i| \psi_i\rangle  
=\int d^3r \,\psi_i^2(\bm{r})\ \ \ (i\in P).
\label{eq:Xidef}
\end{align}
The normalization of the full wave function $|\psi\rangle$, or equivalently, the sum rule \eqref{eq:sum_rule_XZ}, leads to the normalization condition of the $P$-space wave function in Eq.\,\eqref{eq:norm_Gamow}.
In this way, using the Feshbach projection method, we derive an appropriate normalization condition of the wave function for non-Hermitian systems with energy-dependent potentials. At the same time, this formulation substantiates the ``elementarity" interpretation of the energy-derivative term, previously discussed in Ref.\,\cite{Sekihara:2016xnq}.

We comment briefly on the relation between the energy dependence of the potential and positivity aspects in $Q$ space. For a stable bound state, both compositeness and elementarity are given by absolute values squared and hence should be non-negative\,\cite{Hyodo:2013nka}. In this case, Eq.\,\eqref{eq:Z_Fesh} implies that the energy derivative of the potential should be negative. However, the $Q$ space is not necessarily a physical space. In the present context, it is introduced as an auxiliary means to interpret the energy dependence of the potential. In such a case, negative norm states are not unusual as an effective description (see Refs.\,\cite{Kaplan:1996nv,Braaten:2007nq}).

Next, consider the expectation value of an operator $\hat{O}$ in the full $P+Q$ space: 
\begin{align}
\langle \hat{O} \rangle 
&= \langle \psi^\dag| \hP \hat{O} \hP |\psi\rangle + \langle \psi^\dag| \hQ \hat{O} \hQ |\psi\rangle \notag \\
\quad &+ \langle \psi^\dag| \hP \hat{O} \hQ |\psi\rangle + \langle \psi^\dag| \hQ \hat{O} \hP |\psi\rangle\,.
\label{eq:exval}
\end{align}
Like the normalization condition \eqref{eq:norm_full}, we wish to express $\langle \hat{O} \rangle $ within $P$ space only. If $\hat{O}$ is diagonal with respect to the channels, the last two terms in Eq.\,\eqref{eq:exval} vanish. The first term represents the expectation value in $P$ space and can be straightforwardly calculated. The second term stands for the contribution from $Q$ space. One might naively expect an expression analogous to the normalization condition, namely $\langle \psi^\dag|\hP \hat{O} \left( -\frac{\partial \hV^\eff}{\partial E} \right) \hP|\psi\rangle$, in terms of $P$-space quantities. However, with Eqs.\,\eqref{eq:Qwf} and \eqref{eq:Qwf_G}, the correct expression becomes
\begin{align}
&\langle \psi^\dag| \hQ \hat{O} \hQ |\psi\rangle \notag \\
&= \langle \psi^\dag | \hP (\hP\hV\hQ)\frac{1}{E-\hQ\hH\hQ} \hat{O} \frac{1}{E-\hQ\hH\hQ}(\hQ\hV\hP)\hP|\psi\rangle\,.
\end{align}
This form must be maintained unless $\hat{O}$ commutes with all other operators. Therefore, in contrast to the normalization condition, the calculation of the full $\langle\hat{O}\rangle$ can generally not be reduced to $P$ space only. The limited information that can be extracted is the channel expectation value of the $i$th component, $\langle\psi^\dagger_i| \hat{O}|\psi_i \rangle \equiv \langle\hat{O}\rangle_i$ in $P$ space. For example, the mean-squared distance of a two-body system in channel $i$ is written as 
\begin{align}
\langle \hat{\bm{r}}^2 \rangle_i &= \int d^3r\, r^2 \psi_i^2(\bm{r})\ \ \ (i\in P)\,.
\end{align}
Note that for a resonance, with its normalization involving the Gamow state vector, this quantity will in general be complex, reflecting the instability of that resonant state. 

\subsection{Application to $\Lambda(1405)$}\label{subsec:analy_Lamb}  
%

We are now prepared to calculate the norms of the $\KN$ and $\pS$ components of the $\Lambda(1405)$ as a composite two-body object, together with its ``mean distance," using the realistic $\KN$-$\pS$ potentials in the $I=0$ channel constructed in Sec.\,\ref{sec:results_const_pot}. We recall that the detailed properties of the $\Lambda(1405)$ are strongly influenced by the energy dependence of the (real) coupled-channel potentials $V_{ij}^\eq(\bm{r},E)$ of Eq.\,\eqref{eq:Vfit} which we now identify with $V^\eff_{ij}$ of Eq.\,\eqref{eq:Veff_Fesh}. In the present context, the energy dependence can be thought of as coming from two sources. First, there is the primary energy dependence of the chiral interaction which has its origin in the ``integrating out" of high-energy degrees of freedom when constructing the low-energy chiral EFT.\footnote{As an example, consider a linear $\sigma$ model in which pseudoscalar and scalar fields interact with Fermions through (energy-independent) Yukawa couplings. In the low-energy limit with spontaneously broken chiral symmetry, eliminating the (heavy) scalar field implies pseudovector derivative couplings of the (pseudoscalar) Nambu-Goldstone bosons in the resulting nonlinear $\sigma$ model, with energy dependence generated by time derivatives.} Second, restricting the active degrees of freedom to the $\KN$ and $\pS$ channels as elements of $P$ space means relegating other channels with higher mass thresholds to $Q$ space, which generates additional energy dependence in $V_{ij}^\eq$. The complete $E$ dependence of the potential is then determined by reproducing empirical data and parametrized in the polynomial form \eqref{eq:Vfit}.

 In general, the $\KN$-$\pS$ two-component wave functions at an energy $E$ in Eq.\,\eqref{eq:Sch_psi} are subject to boundary conditions for the incident and outgoing states. At the energy corresponding to a pole of the scattering amplitude, the wave function behaves like a discrete eigenstate, satisfying an outgoing-wave boundary condition. Such wave functions are then calculated at the energies of the high-mass and low-mass poles of the $\Lambda(1405)$.
Both these poles are located on the second Riemann sheet in the $\pS$ channel. In particular, the $\pS$ component of the wave function, $\psi_\pS(\bm{r})$, diverges at $r\to\infty$. To calculate matrix elements, we regularize the wave function using the complex scaling method\,\cite{Aguilar:1971ve,Balslev:1971vb,Myo:2014ypa}. The relative coordinate $\bm{r}$ and the wave function $\psi_i$ are transformed as
\begin{align}
\bm{r} &\to \bm{r}e^{i\theta}, \\
\psi_i(\bm{r}) &\to e^{i\frac{3\theta}{2}}\psi_i(\bm{r}e^{i\theta}),
\end{align}
with a real parameter $\theta$. It is known that expectation values with respect to discrete eigenstates remain unchanged under this transformation.
Hence, the compositeness $X_i$, the elementarity $Z$, and the expectation value of $\hat{\bm{r}}^{2}$ can be calculated as
\begin{align}
X_i &= {1\over {\cal N}}\,\int d^3r\, \psi_{i}^2(\bm{r}e^{i\theta})\,,  \\
Z &= {1\over {\cal N}}\displaystyle{\sum_{i,j}}\int d^3r\,\psi_{i}(\bm{r}e^{i\theta})\nonumber \\
& \times \left[-\frac{\del V^\eq_{ij}(\bm{r}e^{i\theta},E)}{\del E} \right] \psi_{j}(\bm{r}e^{i\theta})\, , \\
\langle r^{2} \rangle_i &=  {1\over {\cal N}}\int d^3r\,r^{2}e^{2i\theta} \psi_{i}^2(\bm{r}e^{i\theta}) \,,
\label{eq:ith_exp_Gamow}
\end{align}
with
\begin{align}
{\cal N} = 
\displaystyle{\sum_{i,j}}&\int d^3r\, \psi_{i}(\bm{r}e^{i\theta}) \left[ \delta_{ij}-\frac{\del V^\eq_{ij}(\bm{r}e^{i\theta},E)}{\del E} \right] \psi_{j}(\bm{r}e^{i\theta})\,, 
\end{align}
where the sums over $i,j$ refer to the $P$-space channels, $\KN$ and $\pS$. We note that $X_i, Z$, and $\langle r^{2} \rangle_i$ involving regularized integrals are independent of the parameter $\theta$.
These quantities are computed for both poles of the $\Lambda(1405)$, using the realistic coupled-channel potentials in Sec.\,\ref{sec:results_const_pot}, with strengths parametrized by first- or second-order polynomials. 

Results of the compositeness $X_i$ in channel $i$ and the ``elementarity" $Z$ are summarized in Table\,\ref{tab:norm_couple}. The unstable nature of the resonances and their description in terms of Gamow states has a consequence that the $X_i$ and $Z$ emerge as complex numbers. While the imaginary parts add up to zero in the sum rule $\sum_i X_i + Z = 1$, their physical interpretation in the individual terms is not straightforward. A natural criterion is proposed in view of the similarity of the resonance wave function with that of a stable bound state\,\cite{Sekihara:2014kya,Kamiya:2016oao}: If the compositeness of a channel $i$ is close to unity with small imaginary part, then this channel dominates the structure of the resonance. With this criterion, we conclude that the high-mass pole is indeed dominated by the $\KN$ channel. In fact, this upper pole moves to the real axis and becomes a $\KN$ bound state when the coupling to the $\pS$ channel is turned off. With this coupling activated, the $\Lambda(1405)$ figures as a $\KN$ quasibound state embedded in the $\pS$ continuum. 

The low-mass pole, on the other hand, is characterized by a large imaginary part; i.e., the pole position is far removed from the real axis. In this case, following the discussion in Refs.\,\cite{Kamiya:2015aea,Sekihara:2015gvw,Kamiya:2016oao,Sekihara:2016xnq}, a definite interpretation concerning the physical composition and detailed structure associated with this pole is not possible.

From Table\,\ref{tab:norm_couple}, one finds that deviations between results calculated with different parametrizations of $V_{ij}^\eq(\bm{r},E)$ are less than 0.1 for the high-mass pole and about 0.2 for the low-mass pole. The larger deviations in the latter can be understood by differences in the position of the low-mass pole as shown in Table\,\ref{tab:pole_Fequiv_couple}. 

Alternatively, one can make use of the complex numbers in Table\,\ref{tab:norm_couple} and introduce real quantities,
\begin{align}
\tilde{X}_i = &{|X_i|\over \sum_j|X_j| + |Z|}~, ~~\tilde{Z} = {|Z|\over \sum_j|X_j| + |Z|}~,\nonumber\\
&\sum_i\tilde{X_i} + \tilde{Z} = 1~,
\end{align}
which permit a probabilistic interpretation\,\cite{Kamiya:2015aea,Sekihara:2015gvw,Kamiya:2016oao,Sekihara:2016xnq}. For the realistic $V_{ij}^\eq$ in its second-order polynomial version, this yields the following values at the $\KN$-dominated high-mass pole,
\begin{align}
\tilde{X}_{\KN} = 0.62~,~~\tilde{X}_{\pS} = 0.15~,~~\tilde{Z} = 0.23~,
\end{align}
whereas for the low-mass pole one finds
\begin{align}
\tilde{X}_{\KN} = 0.11~,~~\tilde{X}_{\pS} = 0.35~,~~\tilde{Z} = 0.54~.
\label{eq:Xtildelow}
\end{align}
This confirms the dominance of the $\bar{K}N$ component in the high-mass pole. For the low-mass pole, the results, Eq.\,\eqref{eq:Xtildelow}, do not offer a straightforward interpretation because of the large imaginary parts of $X_{\pS}$ and $Z$.

%
\begin{table*}[tb]
\caption{
Compositeness $X_i$ and elementarity $Z$ for each pole of the $\Lambda(1405)$ coupled-channel system, calculated using the $\KN$-$\pS$ potential of Eq.\,\eqref{eq:Vfit}. Results from the $\KN$ single-channel potential in Ref.\,\cite{Miyahara:2015bya} are also shown for comparison, together with residues of the poles of the scattering amplitudes evaluated in Ref.\,\cite{Sekihara:2014kya}.
}
\begin{center}
\begin{ruledtabular}
\begin{tabular}{llccc}
State & Method  & $X_{\pS}$ & $X_{\KN}$ & $Z$ \\ \hline
High-mass pole & Coupled-channel $\KN$-$\pS$ potential (first order) & $-0.05-0.23i$ & $0.96-0.15i$ & $0.09+0.38i$  \\ 
& Coupled-channel $\KN$-$\pS$ potential (second order)  & $-0.02-0.25i$ & $1.01-0.13i$ & $0.01+0.37i$ \\ & Single-channel $\KN$ potential\,\cite{Miyahara:2015bya} &  & $1.01-0.07i$ &  \\ 
&Residue of the pole\,\cite{Sekihara:2014kya}  & $-0.19-0.22i$ & $1.14+0.01i$ & $0.05+0.21i$  \\ 
Low-mass pole & Coupled-channel $\KN$-$\pS$ potential (first order) & $0.31+0.86i$ & $-0.22+0.03i$  & $0.91-0.90i$  \\
& Coupled-channel $\KN$-$\pS$ potential (second order) & $0.18+0.97i$ & $-0.30+0.07i$ & $1.12-1.04i$ \\ 
& Single-channel $\KN$ potential\,\cite{Miyahara:2015bya} &  & $-0.33-0.03i$ &  \\ 
&Residue of the pole\,\cite{Sekihara:2014kya}  & $0.66+0.52i$ & $-0.39-0.07i$ & $0.73-0.45i$ 
\end{tabular}
\end{ruledtabular}
\end{center}
\label{tab:norm_couple}
\end{table*}%
%

It is instructive to compare the present results with those of other evaluations based on the same scattering amplitudes in Refs.\,\cite{Ikeda:2011pi,Ikeda:2012au}. First, using the single-channel $\bar{K}N$ potential constructed in Ref.\,\cite{Miyahara:2015bya}, we evaluate the compositeness of the $\bar{K}N$ channel, also shown in Table\,\ref{tab:norm_couple}. Remarkably, $X_{\bar{K}N}$ is quantitatively close to the corresponding quantity, resulting from the second-order coupled-channel potential, for both high-mass and low-mass poles. This confirms that the $\bar{K}N$ component of the wave function can be properly determined even with the single-channel potential as starting point, once the normalization condition \eqref{eq:norm_Gamow} is applied. 

The compositeness can also be looked at by studying the residues at the poles of the on-shell scattering amplitudes in Ref.\,\cite{Sekihara:2014kya}. We show the results of Ref.\,\cite{Sekihara:2014kya} by rewriting $Z+X_{\eta\Lambda}+X_{K\Xi}\to Z$ in order to be consistent with the present model space. While these numbers display a similar tendency compared with the results obtained from the coupled-channel potential, there are nonetheless sizable deviations. Compositeness and elementarity are in general model-dependent quantities except for near-threshold states\,\cite{Hyodo:2013nka,Sekihara:2016xnq,Kamiya:2015aea}. The norm \eqref{eq:Xidef} of the wave function depends on the off-shell behavior of the amplitude. In the present calculations, off-shell behavior is reflected in the spatial distribution of the potential, while it is implicitly determined by dimensional regularization in the formulation of Ref.\,\cite{Sekihara:2014kya}. Hence, we may regard the difference of those results as a measure of model dependence related to off-shell behavior. Nevertheless, the $\bar{K}N$ dominance of the high-mass pole is a robust conclusion in all of these studies, as also indicated by an approach based on a generalized weak-binding relation\,\cite{Kamiya:2015aea,Kamiya:2016oao}.

Results for root mean-squared distances, $\sqrt{\langle r^2\rangle_i}$, are summarized in Table\,\ref{tab:rmsr_couple}, together with those obtained using the single-channel $\bar{K}N$ potential of Ref.\,\cite{Miyahara:2015bya}. Small deviations between values of $\sqrt{\langle r^2\rangle_i}$ calculated with different potentials show a tendency seen before in the compositeness: The difference between first- and second-order polynomial parametrizations of the potential is larger in the low-mass pole results, reflecting the difference of the pole positions in those cases. 
On the other hand, $\sqrt{\langle r^2 \rangle_\KN}$ calculated using the coupled-channel potential in second-order polynomial form differs from the value found with the single-channel $\KN$ potential by less than 0.1 fm. Both these potentials are based on the same scattering amplitude, so the properly constructed equivalent potentials give consistent spatial distributions of the wave functions. 

The interpretation of the complex $\sqrt{\langle r^2\rangle_i}$, likewise a consequence of the unstable nature of the resonance states, is again not straightforward. In Ref.\,\cite{Miyahara:2015bya}, the real-valued spatial size associated with the high-mass pole is estimated to be 1.44 fm from the behavior of the wave function at large distance. This indicates that the size of the $\Lambda(1405)$ is larger than that of ordinary hadrons. A similar tendency is seen for the magnitudes of $\sqrt{\langle r^2\rangle_{\KN}}$. The wave function resulting from the coupled-channel potential displays an unusually large distance scale in the diagonal $\KN$ matrix element of $\bm{r}^2$.  We thus conclude that the large spatial extension of the $\Lambda(1405)$ is confirmed by the present calculations using the $\KN$-$\pS$ equivalent potential.

\begin{table*}[tb]
\caption{Root-mean-squared distance $\sqrt{\langle r^2\rangle}_i$ in each channel, calculated using Eq.\,\eqref{eq:ith_exp_Gamow}. The $\KN$-$\pS$ coupled-channel potentials with first- and second-order polynomial representations are employed. Results obtained using the $\KN$ single-channel potential in previous work\,\cite{Miyahara:2015bya} are shown for comparison.}
\begin{center}
\begin{ruledtabular}
\begin{tabular}{llcc}
State & Method & $\sqrt{\langle r^2 \rangle_\pS}$ [fm] & $\sqrt{\langle r^2 \rangle_\KN}$ [fm]   \\ \hline
High-mass pole & Coupled-channel $\KN$-$\pS$ potential (first order) & $0.46+0.17i$ & $1.03-0.60i$    \\
& Coupled-channel $\KN$-$\pS$ potential (second order) & $0.45+0.21i$ & $1.05-0.62i$    \\ 
& Single-channel $\KN$ potential\,\cite{Miyahara:2015bya} &  & $1.04-0.61i$   \\ 
Low-mass pole & Coupled-channel $\KN$-$\pS$ potential (first order)  & $0.38-0.56i$ & $0.12+0.36i$    \\
& Coupled-channel $\KN$-$\pS$ potential (second order)  & $0.42-0.57i$ & $0.17+0.42i$    \\ 
& Single-channel $\KN$ potential\,\cite{Miyahara:2015bya} &  & $0.13+0.41i$ 
\end{tabular}
\end{ruledtabular}
\end{center}
\label{tab:rmsr_couple}
\end{table*}%
\section{Summary}
\label{sec:summary}

In the present work, we have constructed a quantitatively reliable $\KN$-$\pS$-$\pL$ coupled-channel local potential. This potential accurately reproduces the subthreshold amplitudes based on chiral SU(3) dynamics, with stringent threshold constraints from the SIDDHARTA kaonic hydrogen data. This novel potential is suitable for systematic and detailed computations using few-body equations in theoretical studies of the $\Lambda(1405)$ and of $\bar{K}$-nuclear systems, relevant for the analysis and interpretation of current and future experiments.

The determination of the energy-dependent potential strengths is systematically performed by imposing matching conditions for the scattering amplitudes. In the practical application to the $\KN$-$\pS$-$\pL$ system of coupled channels, Gaussian spatial distributions are adopted, with range parameters uniquely determined to reproduce the scattering amplitudes near thresholds. In comparison with the previously developed effective single-channel $\bar{K}N$ potential\,\cite{Miyahara:2015bya}, it is noteworthy that the explicit treatment of the $\pi\Sigma$ channel naturally generates the low-mass pole within the two-pole structure of the $\Lambda(1405)$. Furthermore, a second-order polynomial turns out to be sufficient as a quantitatively successful parametrization of the energy dependence of the coupled-channel potential. With this representation, the results are comparable to those using the single-channel $\bar{K}N$ potential\,\cite{Miyahara:2015bya}, which, however, required a tenth-order polynomial to achieve a similar level of accuracy.

Using the wave functions derived from the so-constructed equivalent coupled-channel potential, the detailed structure and composition of the $\Lambda(1405)$ has been analyzed. For this purpose, it is necessary to establish a proper normalization condition of the resonance wave functions generated by the energy-dependent potential. This energy dependence introduces a specific additional term in the normalization condition. Its derivation is demonstrated using the Feshbach projection method. This scheme offers a well-posed interpretation for each part of the normalization condition in terms of the notions of compositeness and elementarity that have recently been used to investigate the structure of hadrons. When applied to the calculation of properties of the $\Lambda(1405)$, it is found that the $\KN$ component of the properly normalized wave function of the coupled-channel system is consistent with the one obtained using the single-channel effective $\KN$ potential. We demonstrate that the high-mass pole of the $\Lambda(1405)$ is dominated by the $\KN$ component which features a characteristic spatial distance scale significantly larger than that of ordinary hadrons, and supporting the picture of the $\Lambda(1405)$ as a quasibound $\KN$ molecular state embedded in the $\pS$ continuum.

\section*{Acknowledgments}

We thank Wataru Horiuchi for useful discussions.
This work is supported in part by JSPS KAKENHI Grants No. JP16K17694 and No. JP17J11386, by JSPS Japan-France Joint Research Project, and by the Yukawa International Program for Quark-Hadron Sciences (YIPQS). One of the authors (W. W.) acknowledges partial support by the DFG Cluster of Excellence ``Origin and Structure of the Universe."


\appendix
\section{Scattering solutions of coupled-channel Schr\"odinger equation} \label{app:wf}

In this Appendix, we present practical procedures to obtain scattering solutions of the $N$ coupled-channel Schr\"odinger equation~\eqref{eq:Sch_psi}. For a spherically symmetric potential $V^\eq_{ij}(r,E)$, the $s$-wave radial Schr\"odinger equation reads
\begin{align}
\left[ \left(-\frac{1}{2\mu_i}\frac{d^{2}}{dr^{2}} + \Delta M_i\right) \delta_{ij} + V^\eq_{ij}(r,E) \right]u_{j}(r) = E u_{i}(r)~.
\label{eq:Sch_psi_App}
\end{align}
where $u_{i}(r)=r\psi_{i}^{l=0}(r)$. To extract the coupled-channel $S$ matrix, the wave function must satisfy the boundary condition in Eq.\,\eqref{eq:asymptotic}, namely,
\begin{align}
\left.u_{i,j}(r)\right|_{r\to \infty}
& \propto
e^{-ik_{i}r}\delta_{ij}
-\sqrt{\frac{\mu_{i}k_{j}}{\mu_{j}k_{i}}}\,
S_{ij}(E)\,
e^{ik_{i}r}  \\
& \equiv
e^{-ik_{i}r}
\delta_{ij}
-e^{ik_{i}r}\tilde{S}_{ij}(E) , \label{eq:asymptotic_App} 
\end{align}
with $k_{i} =\sqrt{2\mu_{i}(E-\Delta M_{i})}$. This wave function represents the scattering solution with an incoming wave in channel $j$ which is then scattered into channel $i$ with the weight determined by $\tilde{S}_{ij}$. 

Equation\,\eqref{eq:Sch_psi_App} is a second-order differential equation. Its general solution for a scattering state is specified by the boundary condition at $r=0$, as in the single-channel problem. For a given energy $E$, by setting 
\begin{equation}
u_{i}(r=0)=0
\end{equation} 
and choosing a value for the derivative,
\begin{equation}
\frac{du_{i}(r=0)}{dr}\equiv u_{i}^{\prime}(r=0),
\end{equation}
a particular solution is obtained, which we denote $\bar{u}^{(1)}_i(r)$. At sufficiently large $r=R$, where the potential vanishes, the wave function behaves as a superposition of the incoming and outgoing waves. In general, it contains incoming waves in all channels. It can therefore be expressed as a linear combination of solutions \eqref{eq:asymptotic_App} as
\begin{align}
\bar{u}^{(1)}_i(R)
= \sum_{j}[e^{-ik_{i}R}\delta_{ij}-e^{ik_{i}R}\tilde{S}_{ij}(E)]\,A^{(1)}_j~\label{eq:linearcomb_app} ,
\end{align}
with weight factors $A^{(1)}_j$ ($j=1,\dots, N$).

To obtain the solution with the proper boundary condition at large $r$, we now prepare a set of $N$ solutions $\bar{u}^{(1)}_i, \dots, \bar{u}^{(N)}_i$ by choosing different $u_{i}^{\prime}(r=0)$. Because the weight factors in Eq.\,\eqref{eq:linearcomb_app} depend on the choice of $u_{i}^{\prime}(r=0)$, we can construct $N$ linearly independent solutions. These are again written as linear combinations of Eq.\,\eqref{eq:asymptotic_App}:
\begin{align}
\bar{u}^{(\alpha)}_i(R)
&= \sum_{j}[e^{-ik_{i}R}\delta_{ij}-e^{ik_{i}R}\tilde{S}_{ij}(E)]\,A^{(\alpha)}_j~, \label{eq:ubar_app}
\\ 
\alpha &=1,\dots,N .
\end{align}
Their derivatives with respect to $r$ become
\begin{align}
\bar{u}^{(\alpha){\prime}}_i(R)
&= \sum_{j}ik_{i}[-e^{-ik_{i}R}\delta_{ij}-e^{ik_{i}R}\tilde{S}_{ij}(E)]\,A^{(\alpha)}_j~ .
\label{eq:ubarprime_app}
\end{align}

The weight matrix $A^{(\alpha)}_j$ is then determined from Eqs.\,\eqref{eq:ubar_app} and \eqref{eq:ubarprime_app}:
\begin{align}
A^{(\alpha)}_j
&= e^{ik_{j}R}\frac{ik_{j}\bar{u}^{(\alpha)}_j(R)-\bar{u}^{(\alpha){\prime}}_j(R)}{2ik_{j}} ~.
\end{align}
Constructing the inverse of the matrix $A^{(\alpha)}_j$, the $S$ matrix can be calculated as
\begin{align}
\tilde{S}_{ij}(E)&=\sqrt{\frac{\mu_{i}k_{j}}{\mu_{j}k_{i}}}\,S_{ij}(E)\\
&= e^{-2ik_{i}R}\delta_{ij}
-e^{-ik_{i}R}\sum_{\alpha}\bar{u}^{(\alpha)}_i{(A^{-1})}^{(\alpha)}_ j ~.
\end{align}

\section{Range parameters of the potential} \label{app:range}

\begin{figure*}[tb]
\subfigure{
\includegraphics[width=5.5cm,bb=0 0 846 594]{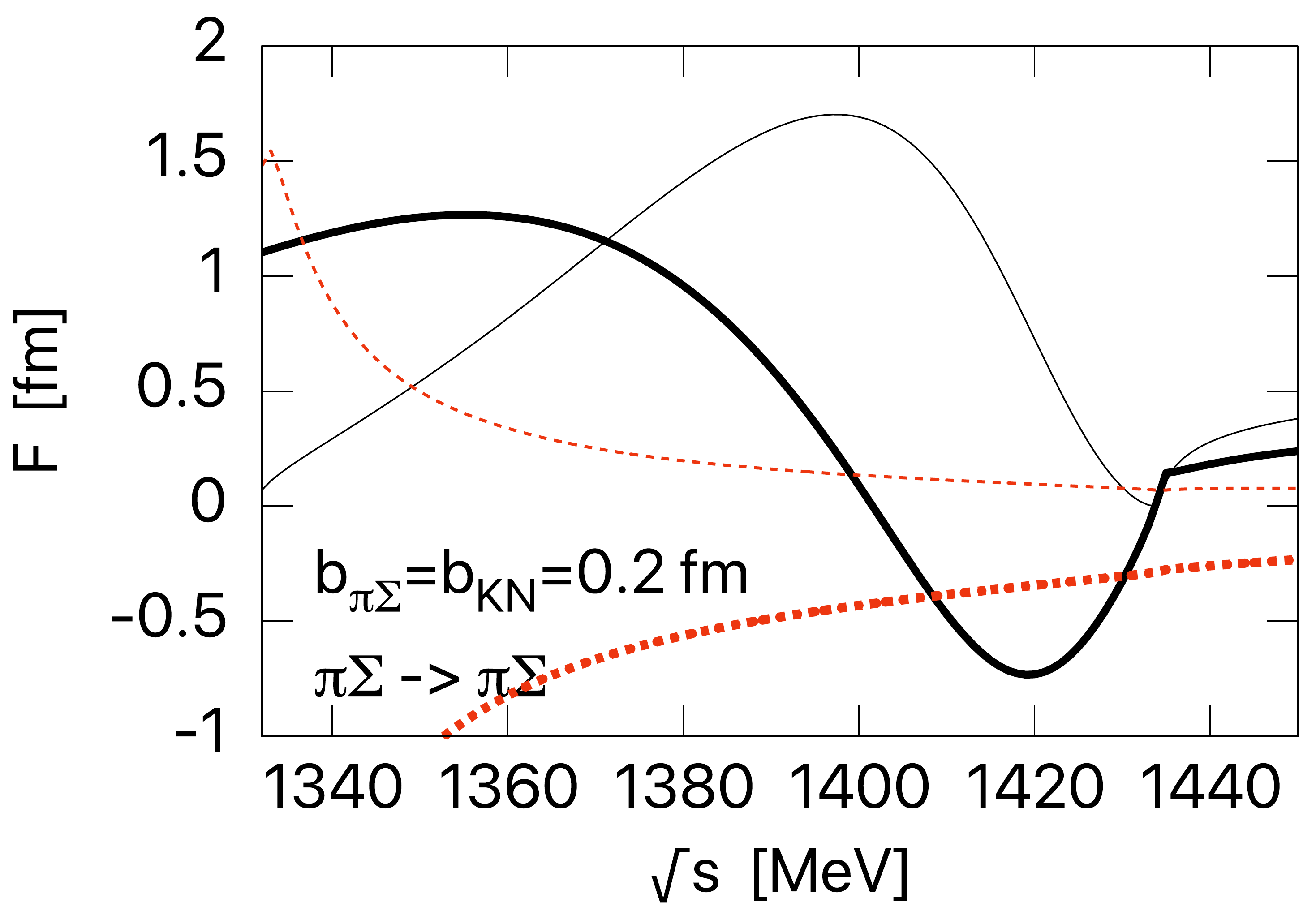}
}
\subfigure{
\includegraphics[width=5.5cm,bb=0 0 846 594]{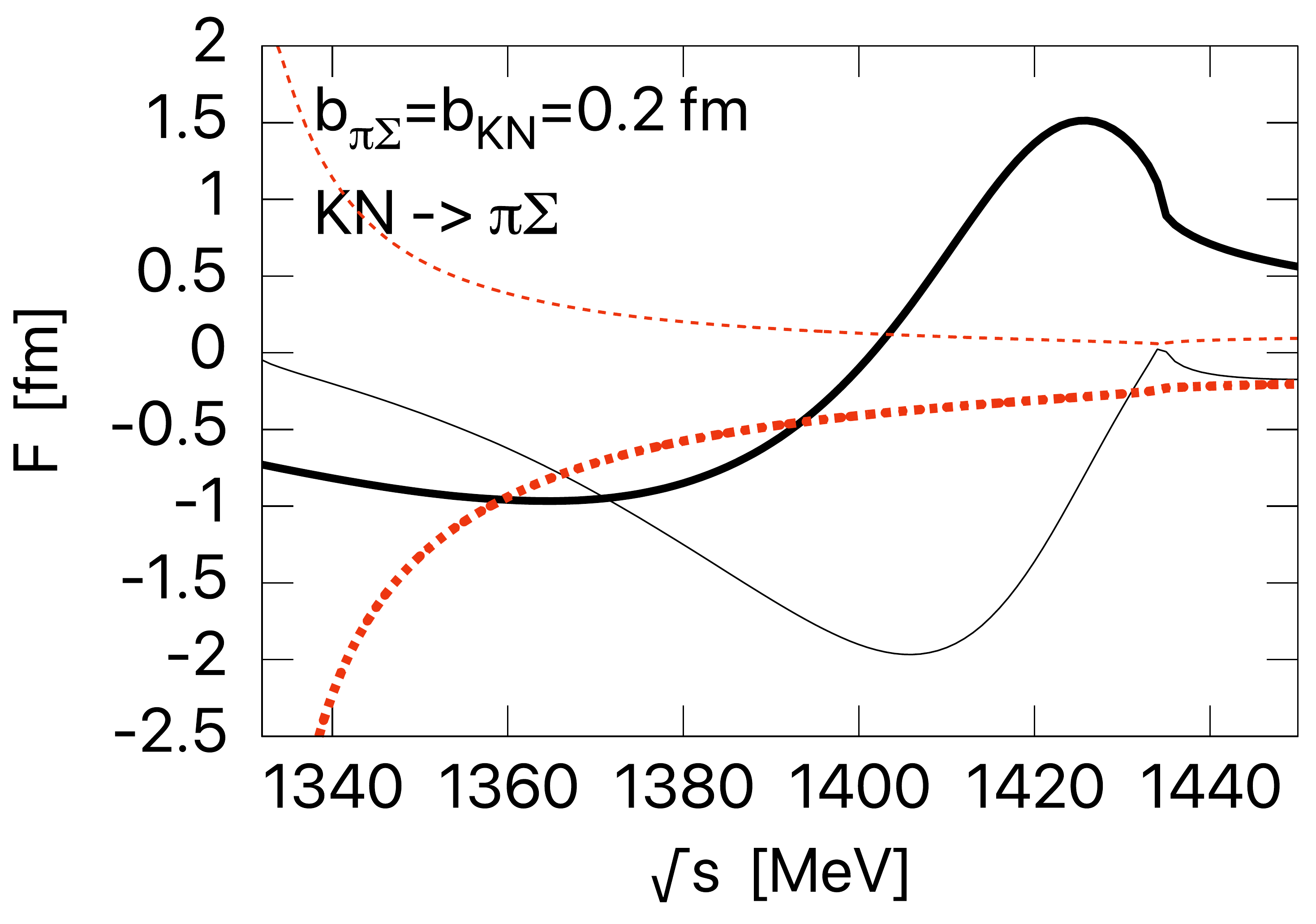}
}
\subfigure{
\includegraphics[width=5.5cm,bb=0 0 846 594]{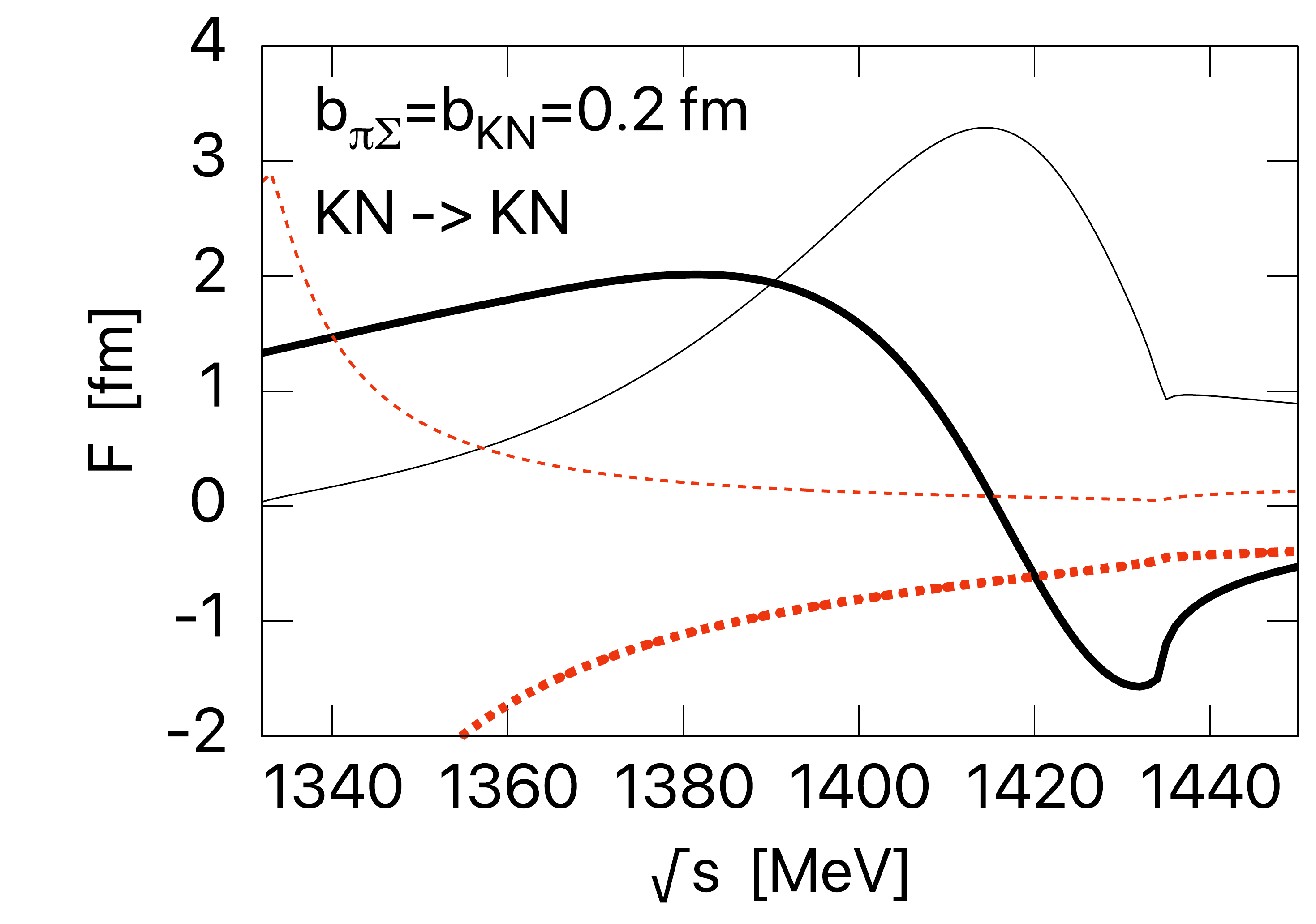}
}

\subfigure{
\includegraphics[width=5.5cm,bb=0 0 846 594]{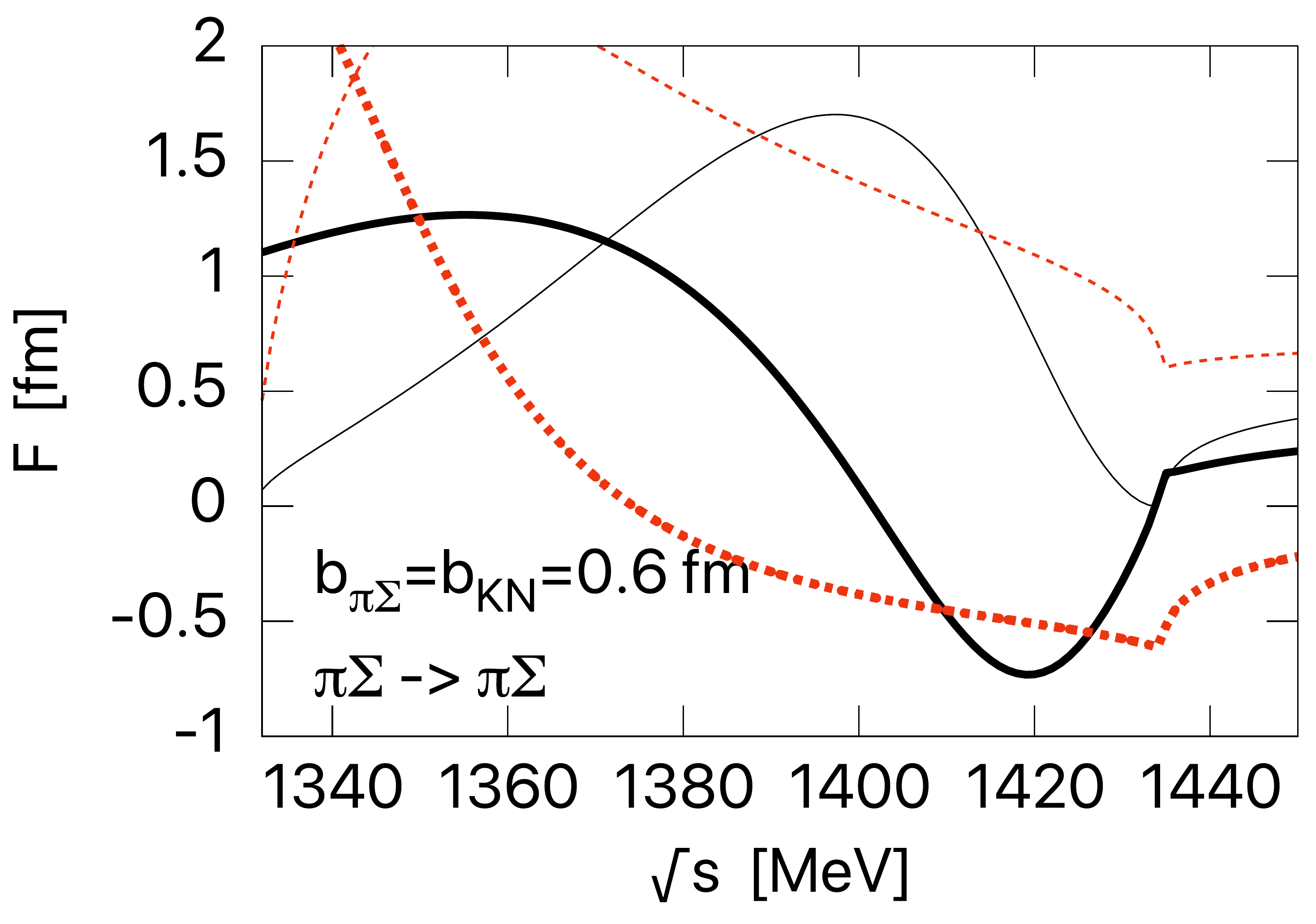}
}
\subfigure{
\includegraphics[width=5.5cm,bb=0 0 846 594]{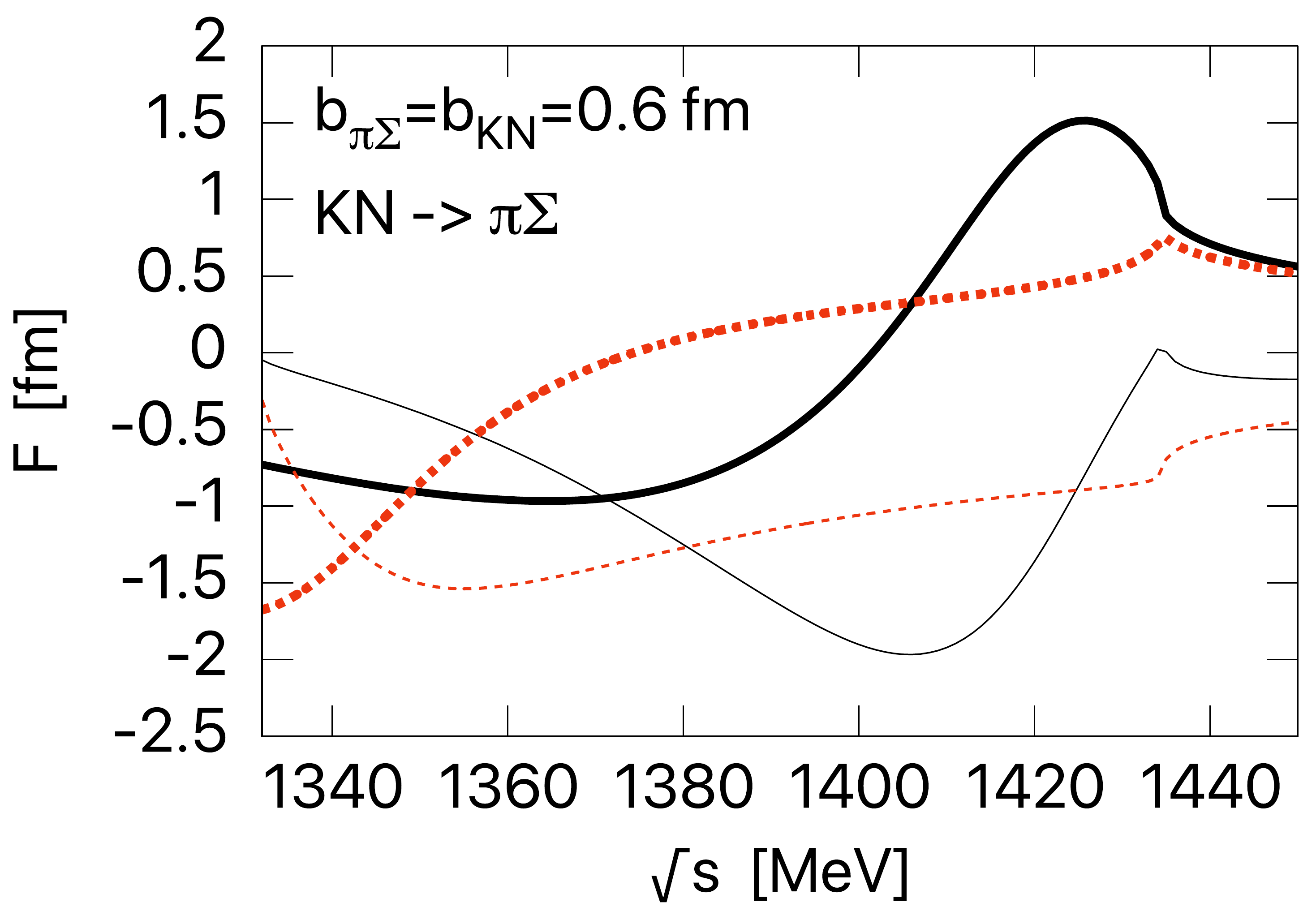}
}
\subfigure{
\includegraphics[width=5.5cm,bb=0 0 846 594]{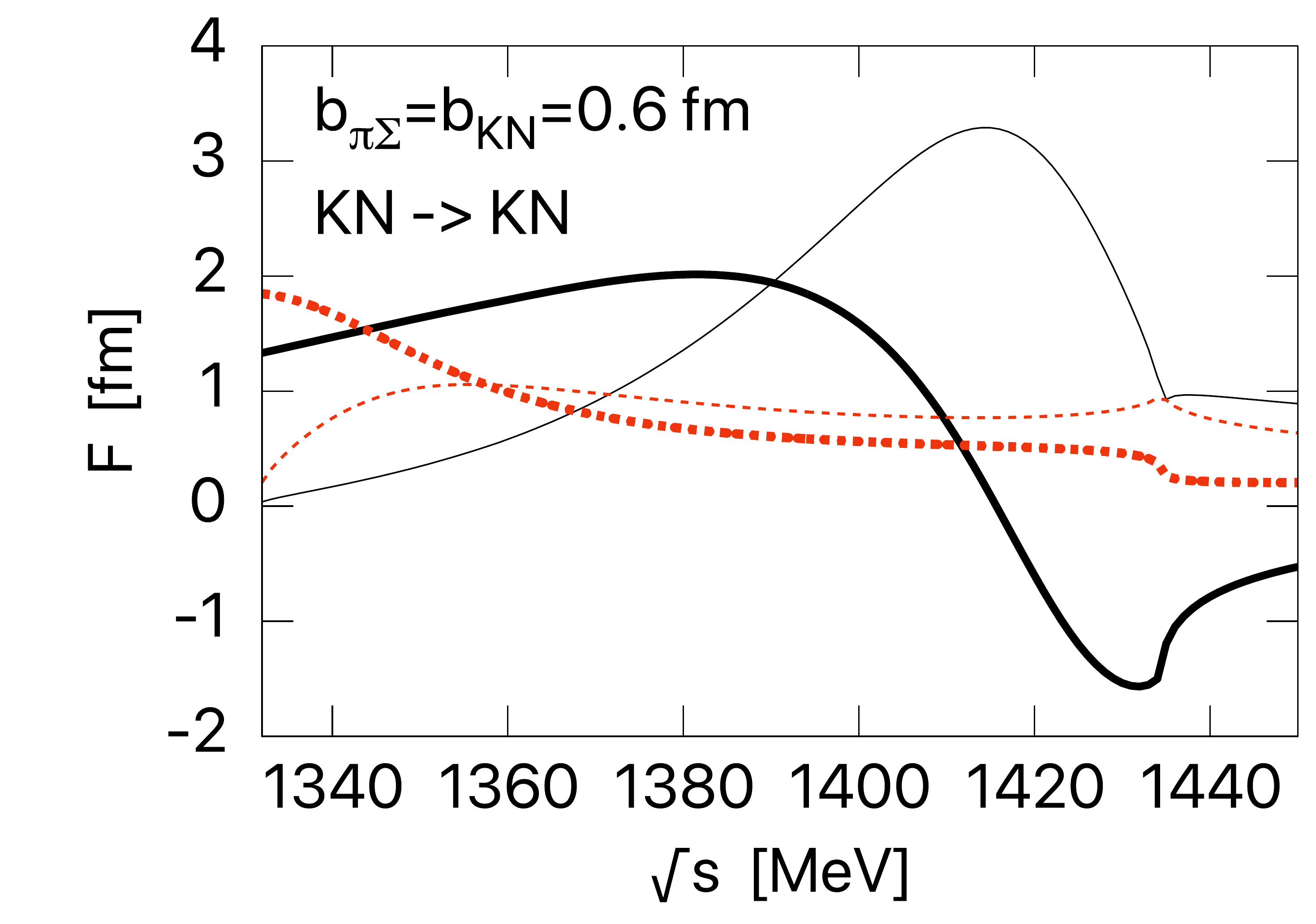}
}

\subfigure{
\includegraphics[width=5.5cm,bb=0 0 846 594]{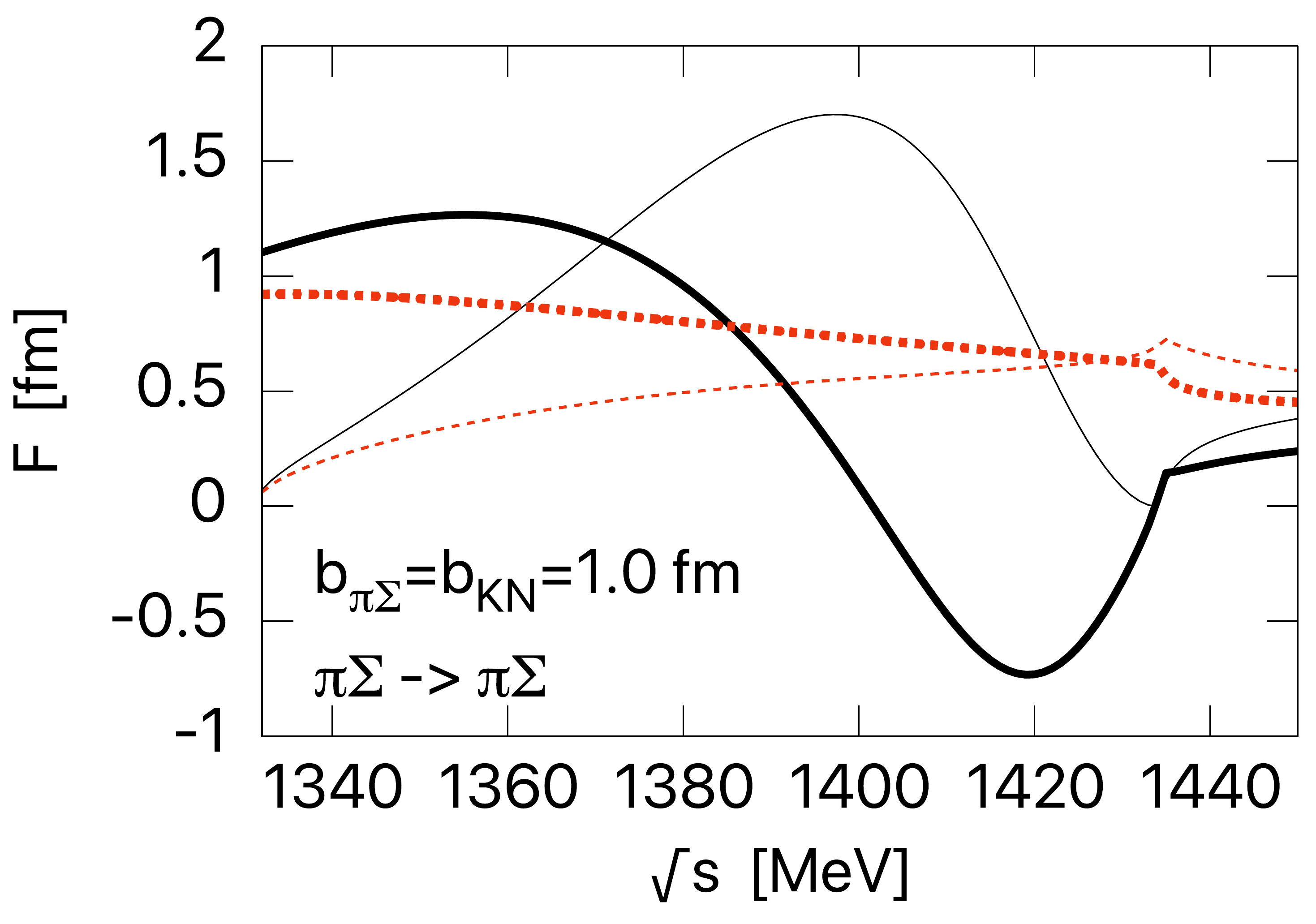}
}
\subfigure{
\includegraphics[width=5.5cm,bb=0 0 846 594]{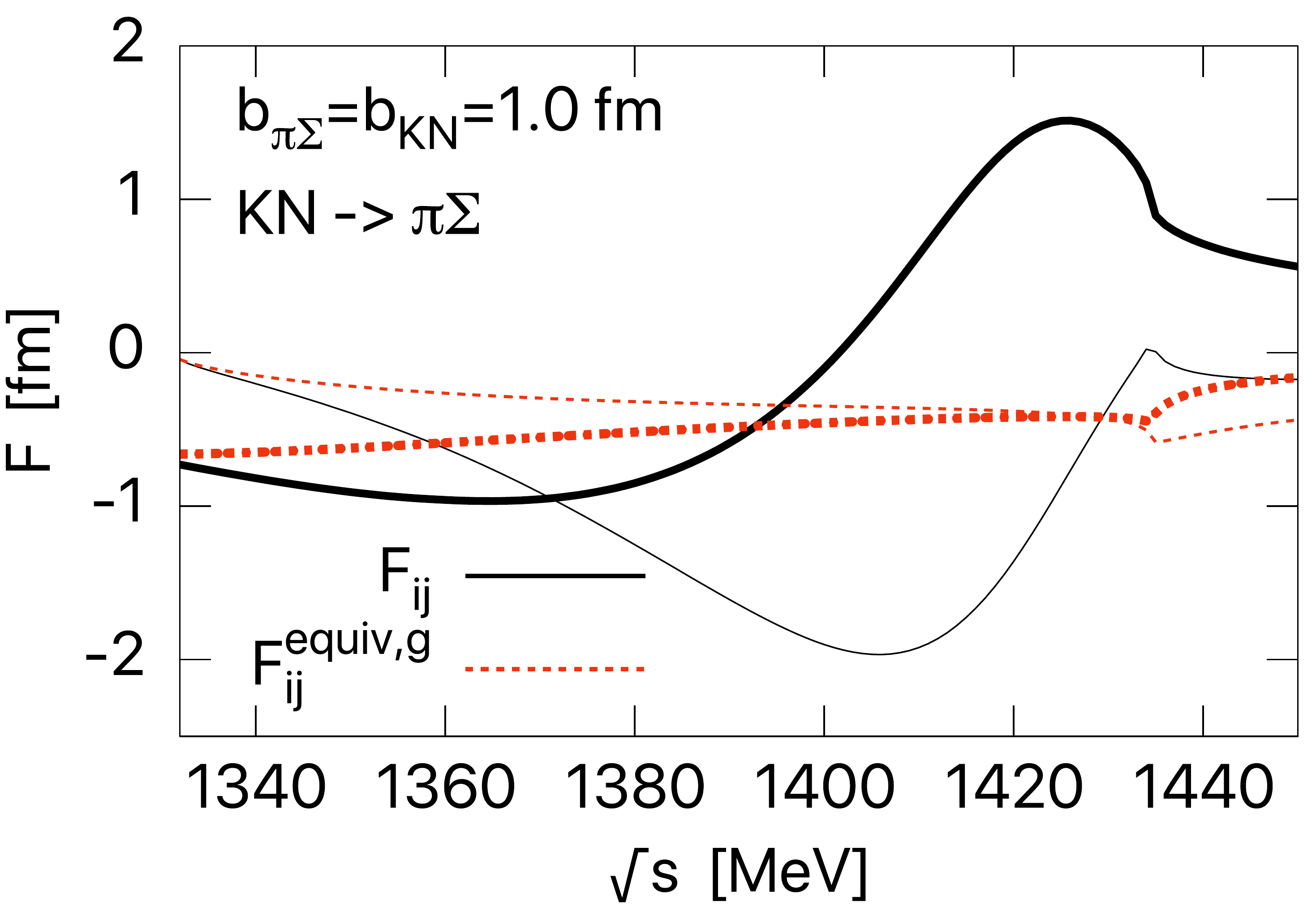}
}
\subfigure{
\includegraphics[width=5.5cm,bb=0 0 846 594]{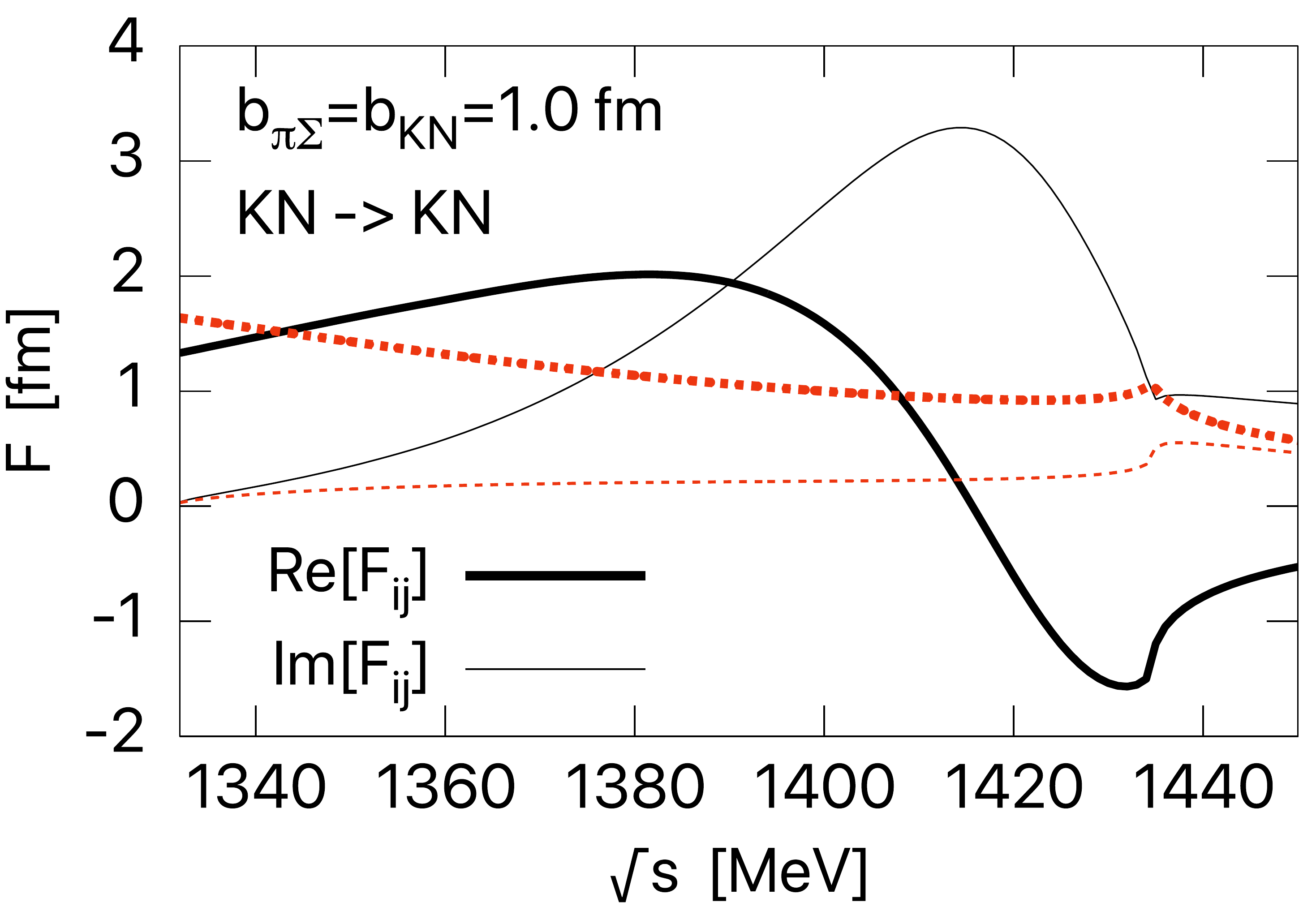}
}
\caption{
Scattering amplitudes $F_{ij}^{\eqg}$ (dotted lines) produced by the potential of Eq.\,\eqref{eq:Vequiv_eff} in comparison with the original amplitudes, $F$, from chiral SU(3) dynamics (solid lines) in the $I=0$ channel. 
In the series of figures from top to bottom, the range parameters are chosen $b_\pS=b_\KN=$ 0.2, 0.6, and 1.0 fm, respectively. Real (imaginary) parts of the amplitudes are displayed as thick (thin) lines.
}
\label{fig:Feff_diff_bgau}  
\end{figure*}

Here we discuss the validity and interpretation of the range parameters,  $b_{i}$, of the potential determined in Secs.\,\ref{subsec:equiv_pot} and \ref{sec:results_const_pot}. This includes checking the sensitivity of the scattering amplitudes with respect to variations of the range parameters. In Fig.~\ref{fig:Feff_diff_bgau}, we show the $\KN$ and $\pS$ scattering amplitudes in the $I=0$ channel generated by the potential $V^{\eqg}_{ij}(\bm{r},E)$ of Eq.\,\eqref{eq:Vequiv_full}, using different sets of range parameters: $b_\pS=b_\KN=$ 0.2, 0.6, and 1.0 fm. By comparison with Fig.\,\ref{fig:F_Veff_IHW} and the optimized values of $b_{\pS}=0.80$ fm and $b_{\KN}=0.43$ fm, it is evident that the amplitudes are very sensitive to the range parameters. Although the deviations in Fig.~\ref{fig:Feff_diff_bgau} can in principle be absorbed by tuning the adjustment term $\Delta V_{ij}$, it is certainly better justified instead to use the amplitudes in Fig.~\ref{fig:F_Veff_IHW} as a starting point. This validates the entire procedure in Sec.~\ref{subsec:equiv_pot} for the determination of the range parameters.

%
%

In order to interpret the distance scales associated with the range parameters\,\eqref{eq:bgau_I0} and \eqref{eq:range_I1}, it is instructive to consider a Yukawa-type potential as it would be realized in a boson-exchange picture. The procedure for determining the range parameters is the same as for the Gaussian case in Sec.\,\ref{subsec:equiv_pot}. The explicit form of the Yukawa potential which meets the requirements\,\eqref{eq:g_diag} and \eqref{eq:g_offdiag} is given by
\begin{align}
V^{\eqY}_{ij}(r,E) = \frac{e^{-r/( 1/2b^\text{Y}_i+1/2b^\text{Y}_j )}}{4\pi r ~b^\text{Y}_ib^\text{Y}_j} \sqrt{\frac{M_iM_j}{s\mu_i\mu_j}}\,V_{ij}(\rts),
\end{align}
with range parameters $b^\text{Y}_i$. With this potential, minimizing $\Delta F$ of Eq.\,\eqref{eq:delF_thre_couple} gives $b^\text{Y}_\pS = 0.66$ fm and $b^\text{Y}_\KN = 0.34$ fm. We have checked that this Yukawa potential reproduces the scattering amplitudes as well as the Gaussian potential in Fig.\,\ref{fig:F_Veff_IHW}.
Translating the Yukawa range parameters into mass scales of hypothetically exchanged ``bosons," $m^\text{Y}_i\sim 1/b^\text{Y}_i$, one finds in the $\KN$ channel
\begin{align}
m^\text{Y}_{\KN} \sim 580 \text{ MeV}~
\end{align}
and in the $\pS$ channel
\begin{align}
m^\text{Y}_{\pS} \sim 300 \text{ MeV}~.
\end{align}
These mass and distance scales are of natural order of magnitude. While $m^\text{Y}_{\pS}$ is reminiscent of two-pion exchange at intermediate range, the $\KN$ channel reflects dynamics at shorter distance involving a higher mass scale. In essence, several mechanisms presumably combine in determining the finite-range parameters. Consider, for example, the subtraction constants regularizing ultraviolet divergences in chiral SU(3) dynamics. These constants encode high-energy (short-distance) physics not resolved in the low-energy EFT, i.e., not treated explicitly within the active model space. These subtraction constants turn out to be different in sign and magnitude for the $\pS$ and $\KN$ $I=0$ channels:  $a_\pS \simeq +4.4 \times 10^{-3}$ and $a_\KN \simeq -2.4 \times 10^{-3}$ (at a renormalization scale $\mu = 1$ GeV)\,\cite{Ikeda:2012au}, effectively reducing the strong $\pS$ attraction\,\cite{Hyodo:2008xr} while slightly enhancing the $\KN$ attraction with respect to the leading-order driving interactions, as required by the detailed fits to the empirical $K^- p$ data base. Such differences in the subtraction constants are expected to be reflected also in the range parameters of the equivalent coupled-channel potential although there is no one-to-one correspondence. 

A further distinctive short-distance effect may be the (partial) Pauli blocking at the quark level in the $\pi^0\Sigma^0$ system, with an extra quark pair of identical flavor appearing in the $[u\bar{u}-d\bar{d}][uds]$ configuration. This mechanism selectively suppresses part of the $I=0$ $\pS$ attraction. Whether such effects might be at the origin of the different range parameters found in the $I=1$ and $I=0$ channels ($b^{I=1}_\pS \simeq 0.5$ fm as compared to $b^{I=0}_\pS \simeq 0.8$ fm) poses an interesting question. In the future one may expect that lattice QCD computations\,\cite{Aoki:2012tk} will also contribute to the determination of the range parameters in question.

\section{Normalization and orthogonality conditions with energy-dependent potential}
\label{app:E_dep}

Here we derive the normalization condition in Eq.\,\eqref{eq:norm_Gamow} together with the orthogonality relation of wave functions for an energy-dependent coupled-channel potential in a non-Hermitian system, following previous work  on the corresponding Hermitian system\,\cite{Formanek2004}. 
The derivation for a non-Hermitian single-channel system is given in Ref.\,\cite{Miyahara:2015bya}.

The strategy is to define the probability density $P_{E'E}$ and the current $\bm{J}_{E'E}$ for a pair of eigenstates (with energies $E$ and $E'$) that satisfy the continuity equation,
\begin{align}
\frac{\del}{\del t}P_{E'E}(\bm{r},t) = -\nabla\cdot \bm{J}_{E'E}(\bm{r},t).
\end{align}
For an energy-independent potential, this continuity equation implies the conservation of the norm and the orthogonality relation for the wave function. We first show the continuity equation for the eigenstates with energy-dependent potential, from which the normalization condition and the orthogonality relation are derived.

To this end, we start from the coupled-channel Schr\"odinger equation with an energy-dependent complex potential $V_{ij}\in \mathbb{C}$:
\begin{align}
i\frac{\del}{\del t}\Psi_i(\bm{r},t) &= \sum_j H_{ij} \Psi_j(\bm{r},t)~, \notag \\
H_{ij}&= H^{(0)}_i\delta_{ij}+ V_{ij}(\bm{r},i\frac{\del}{\del t})~,  \\
H^{(0)}_i& = -\frac{\nabla^2}{2\mu_i} + \Delta M_i~,\notag
\label{eq:Sch_usual}
\end{align}
with the reduced mass $\mu_{i}$, the mass difference $\Delta M_{i}$, and the multicomponent wave function $\Psi=(\Psi_1,\Psi_2,\ldots)^T$. The time-dependent eigenfunction of $H$ with energy $E$ is expressed as
\begin{align}
\Psi_{E,i}(\bm{r},t)=e^{-iEt}\psi_{E,i}(\bm{r})~, 
\end{align}
and the time-independent wave function $\psi_{E,i}$ satisfies
\begin{align}
&\sum_jH_{ij}\psi_{E,j}(\bm{r}) \notag \\
&= \sum_j\left[ H^{(0)}_i\delta_{ij}+ V_{ij}(\bm{r},E) \right]\psi_{E,j}(\bm{r}) \notag \\
&= E \psi_{E,i}(\bm{r})~.
\label{eq:tindep_Sch_usual}
\end{align}
To deal with the non-Hermitian systems, we introduce the Gamow vectors $\Psi^\dag_E$ and $\psi^\dag_E$\,\cite{Hokkyo:1965,Berggren:1968zz,non_hermitian} so that the current,
\begin{align}
{\bm j}^\g_E(\bm{r},t)&\equiv 
 -\sum_i\frac{i}{2\mu_i} \left[ \Psi_{E,i}^{\dag *}({\bm r},t)\nabla\Psi_{E,i}({\bm r},t) \right. \notag \\
&\quad \left.- \big\{ \nabla\Psi_{E,i}^{\dag *}({\bm r},t) \big\} \Psi_{E,i}({\bm r},t)   \right]~,
\label{eq:current_Gamow}
\end{align}
vanishes at $|\bm{r}|\to \infty$ (see Ref.\,\cite{Hokkyo:1965}).%
\footnote{In Ref.\,\cite{Hokkyo:1965}, from the definition of the Hermitian conjugate of $H$, the boundary condition for the current ${\bm j}^\g_E\to 0$ at $|\bm{r}|\to \infty$ leads to the expression of the adjoint Hamiltonian as the complex conjugate of the Hamiltonian, $H^\dag=H^*$. This then ensures the adjoint wave function $\psi_E^\dag=\psi_E^*$ as an eigenstate of $H^\dag$. On the other hand, when the potential is energy dependent, the relation $H^\dag=H^*$ does not follow from this boundary condition. Thus we define the Gamow state $\psi^{\dag}_{E}$ as the eigenstate of $H^*$.} This requirement is met for
\begin{align}
\Psi_{E,i}^\dag(\bm{r},t) &= e^{-iE^*t}\psi^\dag_{E,i}(\bm{r}), \notag \\
\psi^\dag_{E,i}(\bm{r}) &= \psi_{E,i}^*(\bm{r})~,
\label{eq:wavefnc_Gamow}
\end{align}
which satisfy the Schr\"odinger equation for $H^{*}$,
\begin{align}
i\frac{\del}{\del t}\Psi_{E,i}^\dag(\bm{r},t) &= \sum_j H_{ij}^* \Psi_{E,j}^\dag(\bm{r},t)~, \notag \\
\sum_j H_{ij}^*\psi_{E,j}^\dag(\bm{r}) &= \sum_j \left[H^{(0)}_i\delta_{ij} + V^*_{ij}(\bm{r},E) \right]\psi^\dag_{E,j}(\bm{r}) \notag \\
&= E^* \psi^\dag_{E,i}(\bm{r})~.
\label{eq:Sch_Gamow}
\end{align}
Namely, we adopt the eigenstate of $H^{*}$ as the Gamow state which satisfies the boundary condition for the current ${\bm j}^\g_E$.
The time-independent Schr\"odinger equation is just the complex conjugate of Eq.~\eqref{eq:tindep_Sch_usual}, whereas the time-dependent wave functions are not complex conjugate with respect to each other, $\Psi^\dag_{E,i}\neq\Psi^*_{E,i}$.

Next, we express the probability density $P^{\g}_{E'E}(\bm{r},t)$ and the current $\bm{J}^{\g}_{E'E}(\bm{r},t)$ in terms of the wave functions. 
In order to reduce to Eq.\,\eqref{eq:current_Gamow} when $E^{\prime}\to E$, the current is written
\begin{align}
{\bm J}^\g_{E'E}(\bm{r},t)& =-\sum_i\frac{i}{2\mu_i} \left[ \Psi_{E^\prime,i}^{\dag *}({\bm r},t)\nabla\Psi_{E,i}({\bm r},t) \right. \notag \\
&\quad \left.- \big\{ \nabla\Psi_{E^\prime,i}^{\dag *}({\bm r},t) \big\} \Psi_{E,i}({\bm r},t)   \right],
\end{align}
as a generalization of the standard form. On the other hand, given the Schr\"odinger equations\,\eqref{eq:Sch_usual} and \eqref{eq:Sch_Gamow}, the time derivative of a corresponding ansatz for the probability density becomes
\begin{align}
&\frac{\partial}{\partial t} \sum_i \left[ \Psi^\dag_{E',i}(\bm{r},t) \right]^* \Psi_{E,i}(\bm{r},t) \notag \\
&=-\nabla\cdot{\bm J}^\g_{E'E} + i \sum_{i,j} \Psi_{E^\prime,i}^{\dag*} \left[ V_{ij}(E^\prime) -V_{ij}(E) \right] \Psi_{E,j}~.
\label{eq:contgamow}
\end{align}
Because of the energy dependence of the potential, there appears the second term on the right-hand side of Eq.\,\eqref{eq:contgamow}. 
In a similar way as in Ref.\,\cite{Formanek2004}, we can rewrite this term in the form of a time derivative,
\begin{align}
\frac{\del}{\del t}& \left\{ \Psi_{E^\prime,i}^{\dag *}\left[ \frac{V_{ij}(E')-V_{ij}(E)}{E'-E}\right]\Psi_{E,j}  \right\} \notag \\
&= i \Psi_{E',i}^{\dag*} [ V_{ij}(E')-V_{ij}(E) ] \Psi_{E,j}~,
\end{align}
using the Schr\"odinger equations~\eqref{eq:Sch_usual}, \eqref{eq:tindep_Sch_usual}, and \eqref{eq:Sch_Gamow}. Hence, the additional term in Eq.~\eqref{eq:contgamow} can be regarded as a part of the probability density, 
\begin{align}
P^\g_{E'E}(\bm{r},t) &\equiv  \sum_{i,j} \Psi_{E',i}^{\dag *} \left[ \delta_{ij} - \frac{V_{ij}(E')-V_{ij}(E)}{E'-E} \right] \Psi_{E,j}~ .
\label{eq:Prob_def}
\end{align}
Therefore, the following generalized continuity equation holds:
\begin{align}
\frac{\del}{\del t}P^\g_{E'E} (\bm{r},t)
&= -\nabla\cdot\bm{J}^\g_{E'E}(\bm{r},t).
\label{eq:cont_Gamow}
\end{align}

Let us discuss several consequences of this continuity equation.
Given the probability density, we define the norm of a state with energy $E$ as
\begin{align}
N^\g_{E}(t) &= \lim_{E^{\prime}\to E}\int d^3r\ P^\g_{E^{\prime}E}(\bm{r},t) \notag \\
&= \sum_{i,j}\int d^3r\ \Psi_{E,i}^{\dag *}(\bm{r},t) \left[ \delta_{ij} - \frac{\del V_{ij}}{\del E}(E) \right] \Psi_{E,j} (\bm{r},t)\notag \\
&=  \sum_{i,j}  \int d^3r\ \psi_{E,i} (\bm{r})\left[ \delta_{ij} - \frac{\del V_{ij}}{\del E}(E) \right] \psi_{E,j}(\bm{r}) .
\label{eq:norm_Gamow_app}
\end{align}
The last expression demonstrates that the norm is time independent as it should be. This can also be shown by substituting Eq.\,\eqref{eq:cont_Gamow} with the boundary condition of $\bm{j}^\g_{E}\to 0$ at $|\bm{r}|\to \infty$. Setting $N^\g_{E}=1$, we obtain the normalization condition in Eq.\,\eqref{eq:norm_Gamow} for the eigenstates of the energy-dependent coupled-channel potential. Next we consider the orthogonality relation. In a non-Hermitian system, the current $\bm{J}^\g_{E^{\prime}E}$ vanishes at the boundary for a pair of suitably regularized resonance wave functions, while this is not the case for general eigenstates. The orthogonality relation can therefore be derived only for regularized eigenstates.\footnote{There is a correspondence at this point to the extended completeness relation in the complex scaling method, which incorporates the regularized resonant states as well as the bound and continuum states\,\cite{Aoyama2006}.}
Performing the spatial integration in Eq.~\eqref{eq:cont_Gamow} we obtain
\begin{align}
&\frac{\del}{\del t}\sum_{i,j}  \int d^3r\ \Psi_{E',i}^{\dag *} \left[ \delta_{ij} - \frac{V_{ij}(E')-V_{ij}(E)}{E'-E} \right] \Psi_{E,j} \notag \\
&=e^{-i(E-E')t}\sum_{i,j} \int d^3r\,\psi_{E',i} \left[ \delta_{ij} - \frac{V_{ij}(E')-V_{ij}(E)}{E'-E} \right] \psi_{E,j}  \notag \\
&= 0\ \ \ (E'\neq E)~,
\label{eq:ortho_Gamow}
\end{align}
which implies the orthogonality relation for the time-independent wave functions, $\psi_{E,j}$ and $\psi_{E',i}$, with $E'\neq E$.

It is mandatory to introduce the Gamow states for the treatment of non-Hermitian systems. Had we used $\Psi_{E^{\prime},i}^{*}$ instead of $\Psi_{E^{\prime},i}^{\dag *}$, the second term in Eq.~\eqref{eq:Prob_def} would be
\begin{align}
- \Psi_{E',i}^* \frac{V^*_{ij}(E')-V_{ij}(E)}{E'^*-E}\Psi_{E,j}~,
\end{align}
which diverges in the limit $E'\to E\in \mathbb{R}$ for a complex potential ${\rm Im}[V_{ij}]\neq0$. The corresponding current would be
\begin{align}
&\bm{J}_{E'E}(\bm{r},t) \notag \\
&= -\sum_i \frac{i}{2\mu_i}\Big[ \Psi_{E',i}^*(\bm{r},t)\nabla\Psi_{E,i}(\bm{r},t) \notag \\
&\phantom{ -\sum_i \frac{i}{2\mu_i}\Big[ aaa}  - \left\{ \nabla\Psi_{E',i}(\bm{r},t) \right\}^*\Psi_{E,i}(\bm{r},t) \Big] ~,
\end{align}
which cannot be regularized uniquely for resonance wave functions, in contrast to $\bm{j}_{E'E}^{\g}$: Norm conservation and the orthogonality relation rely crucially on the introduction of Gamow state vectors.


\end{document}